\documentclass[twocolumn]{aastex63}
\usepackage{booktabs}
\usepackage{filecontents,catchfile}
\usepackage{longtable}
\usepackage{cleveref}
\usepackage{CJKutf8}

\newcommand{\um}{\,\micron}
\newcommand{\NeII}{[Ne\,\textsc{ii}]}
\newcommand{\NeIII}{[Ne\,\textsc{iii}]}
\newcommand{\HII}{H\,\textsc{ii}}
\newcommand{\PAHFIT}{\textsc{Pahfit}}
\newcommand{\LIR}{\,L\ensuremath{_\mathrm{IR}}}
\newcommand{\Lsun}{\,L\ensuremath{_\mathrm{\odot}}}

\newcommand{\Lpahthree}{L\ensuremath{_\mathrm{PAH\,3.3}}}
\newcommand{\Lpahseventeen}{L\ensuremath{_\mathrm{PAH\,17}}}
\newcommand{\Ltotpah}{L\ensuremath{_\mathrm{\Sigma\,PAH}}}
\newcommand{\LNe}{L\ensuremath{_\mathrm{[Ne\,II] + [Ne\,III]}}}
\newcommand{\AKARI}{\textit{AKARI}}
\newcommand{\Spitzer}{\textit{Spitzer}}
\newcommand{\JWST}{\textit{JWST}}
\newcommand{\ISO}{\textit{ISO}}
\newcommand{\tauSi}{\ensuremath{\tau_\mathrm{Si}}}
\newcommand{\tauWater}{\ensuremath{\tau_\mathrm{H_{2}O}}}

\renewcommand{\thefootnote}{\fnsymbol{footnote}}

\graphicspath{{./}{figures/}}

\accepted{October 8, 2020}
\submitjournal{ApJ}

\shorttitle{All the PAHs}
\shortauthors{Lai et al.}

\begin{document}
\begin{CJK*}{UTF8}{bsmi}
\title{All the PAHs: an AKARI-Spitzer Cross Archival Spectroscopic Survey of Aromatic Emission in Galaxies}

\correspondingauthor{Thomas S.-Y. Lai}
\email{ThomasLai.astro@gmail.com}

\author[0000-0001-8490-6632]{Thomas S.-Y. Lai (賴劭愉)}
\affil{Ritter Astrophysical Research Center, University of Toledo, Toledo, OH 43606, USA}

\author{J.D.T. Smith}
\affiliation{Ritter Astrophysical Research Center, University of Toledo, Toledo, OH 43606, USA}

\author{Shunsuke Baba (馬場俊介)}
\affiliation{National Astronomical Observatory of Japan, National Institutes of Natural Sciences (NINS), 2-21-1 Osawa, Mitaka, Tokyo 181-8588, Japan}

\author{Henrik W.W. Spoon}
\affiliation{Cornell University, CRSR, Space Sciences Building, Ithaca, NY 14853, USA}

\author{Masatoshi Imanishi (今西昌俊)}
\affiliation{National Astronomical Observatory of Japan, National Institutes of Natural Sciences (NINS), 2-21-1 Osawa, Mitaka, Tokyo 181-8588, Japan}



\begin{abstract}


We present a large sample of 2.5--38\um\ galaxy spectra drawn from a cross-archival comparison in the \AKARI-\Spitzer\ Extragalactic Spectral Survey (ASESS), and investigate a subset of 113 star-forming galaxies with prominent polycyclic aromatic hydrocarbon (PAH) emission spanning a wide range of star formation properties. With \AKARI's extended 2.5--5\um\ wavelength coverage, we self-consistently model for the first time \emph{all} PAH emission bands using a modified version of \PAHFIT. We find \Lpahthree/\LIR\ $\sim\,0.1\%$ and the 3.3\um\ PAH feature contributes $\sim$1.5--3\% to the total PAH power --- somewhat less than earlier dust models have assumed.  We establish a calibration between 3.3\um\ PAH emission and star formation rate, but also find regimes where it loses reliability, including at high luminosity and low metallicity. The 3.4\um\ aliphatic emission and a broad plateau feature centered at 3.47\um\ are also modeled. As the shortest wavelength PAH feature, 3.3\um\ is susceptible to attenuation, leading to a factor of $\sim$\,3 differences in the inferred star formation rate at high obscuration with different assumed attenuation geometries. Surprisingly, \Lpahthree/\Ltotpah\ shows no sign of decline at high luminosities, and the low metallicity dwarf galaxy II~Zw~40 exhibits an unusually strong 3.3\um\ band; both results suggest either that the smallest PAHs are better able to survive under intense radiation fields than presumed, or that PAH emission is shifted to shorter wavelengths in intense and high energy radiation environments.  A photometric surrogate for 3.3\um\ PAH luminosity using \JWST/NIRCam is provided and found to be highly reliable at low redshift.


\end{abstract}


\section{Introduction} \label{sec:intro}
Galaxies with active star formation are dominated by strong emission features at 3.3, 6.2, 7.7, 8.6, 11.3, 12.7, 16.4, and 17\um\ complex generally attributed to Polycyclic Aromatic Hydrocarbons \citep[PAHs;][]{Leger1984, Allamandola1985, Smith2004, Werner2004}. These molecules typically consist of 20--1000 carbon atoms \citep{Leger1984, Tielens2008}, and have been found to be an abundant and energetically important component of interstellar dust, contributing $\sim5$--20\% of the total infrared emission from a galaxy \citep{Smith2007}. In addition, they are extremely effective in reprocessing radiative energy and coupling photon energy to the gas through photoelectric heating \citep{DHendecourt1987}. PAH grains of different sizes and ionization states vary widely in the efficiency with which they emit at different bands \citep{Draine2007}. However, the underlying mechanism that governs such variation had not yet been well understood.

Spanning the wavelength range of 2.4--45\um\, the Short Wavelength Spectrometer (SWS) on board the $\textit{Infrared Space Observatory}$ \citep[ISO: ][]{Kessler1996} offered for the first time the sensitivity to exploit the diagnostic power of many of the most luminous PAH bands \citep{Spoon2000, Peeters2002}. The advent of the sensitive Spitzer/IRS instrument \citep{Houck2004} permitted detailed explorations of PAH emission at wavelengths of 5–-20\um\, and provided crucial insights regarding the nature of PAH populations (i.e., size, charge state) within differing interstellar environments and on galaxy-wide scales \citep{Smith2007, Galliano2008, Gordon2008, Hunt2010, Sandstrom2012}. The metal content and radiation field of the ISM have been suggested as the primary drivers of variations in PAH emission. PAH abundance was found to decline rapidly as metallicity drops, vanishing entirely below 12+log(O/H)$\sim$8.1 \citep{Engelbracht2005, Draine2007, Smith2007, Sandstrom2012}. Also, \citet{Smith2007} found a very low PAH 7.7/11.3 ratio in galaxies that host an AGN \citep[see also][]{ODowd2009, Diamond-Stanic2010}. These deficits have been interpreted as preferential destruction of small PAHs by strong radiation fields, leading to an enhanced decrease of the shorter-wavelength band relative to the longer-wavelength band. In those studies, however, 3.3\um\ PAH emission was left unexplored, due to the limited spectral coverage of \Spitzer.

Putting the 3.3\um\ PAH in context with all other PAH bands at longer wavelengths is crucial. The 3.3\um\ C-H stretching feature is known to originate from the smallest members of the emitting PAH population \citep{Desert1990, Schutte1993, Draine2007}, making it the most sensitive probe to the size distribution at the grain/molecule interface given their larger surface-to-volume ratios.  Very recently, \citet{Maragkoudakis2020} utilized the NASA Ames PAH IR Spectroscopic Database \citep[PAHdb;][]{Bauschlicher2010} to compare 3.3\um\ PAH to other PAH bands and concluded that the 3.3\um\ band is most sensitive to the sizes of the smallest PAH molecules. Many theoretical and laboratory studies have shown that the ratio of 3.3/11.3 should be a powerful diagnostic to the size distribution as both bands originated from neutral PAHs \citep{Allamandola1999, Croiset2016, Maragkoudakis2018, Maragkoudakis2020}. 

The 3.3\um\ PAH emission has been observed by ground-based facilities in various galaxies, including starbursting and infrared luminous galaxies, as well as obscured AGNs associated with star-forming activity \citep{Moorwood1986, Imanishi2000, Imanishi2002, Rodriguez_Ardila2003, Spoon2003, Imanishi2004, Imanishi2006, Imanishi2006b, Risaliti2010}. Taking advantage of the redshifted spectra, \citet{Sajina2009} even observed the 3.3\um\ feature in galaxies at z\,$\sim\,$2 with \Spitzer/IRS. While \ISO\ also provided access to the 3.3\um\ PAH emission \citep{Spoon2000}, the interpretation of these spectral features was limited by sensitivity to sampling only bright, nearby sources \citep{Sturm2000, Lu2003, VanDiedenhoven2004}, and the 3.3\um\ PAH was typically only marginally detected. 

The 3.3\um\ PAH observed by the Infrared Camera \citep[IRC:][2.5--5\um]{Onaka2007} on board the \AKARI\ Space Telescope \citep{Murakami2007} has been applied to a suite of diagnostics to identify obscured AGNs in (U)LIRGs \citep{Imanishi2008, Imanishi2010, Lee2012, Ichikawa2014}. These authors found AGN-dominated sources have relatively low 3.3\um\ equivalent width (EQW), attributing it to either the destruction of small PAHs by radiation in the AGN environment or the enhanced integrated infrared luminosity (\mbox{8--1000\um}; \LIR) from the AGN activity. Recently, a starburst/AGN diagnostic technique using the 3.3\um\ PAH and the slope of the IRC continuum has also been reported \citep{Inami2018}. The fact that 3.3\um\ PAH is bright in star-forming galaxies while suppressed in star-forming sources hosting AGNs draws into question whether 3.3\um\ PAH can serve as a reliable star formation rate indicator \citep{Yano2016, Inami2018}. Several studies have shown a correlation between \Lpahthree\ and \LIR\ in local LIRGs, but also find a 3.3\um\ PAH deficit at the higher luminosity of ULIRGs \citep{Imanishi2010, Kim2012, Y2013,  Murata2017}. The measurement of the 3.3\um\ PAH in these studies, however, dealt only with the attenuation from a nearby 3.05\um\ water ice absorption band, but not the steeply rising continuum opacity at short mid-infrared (MIR) wavelengths near 3\um, which can be substantially larger. 

To investigate the role of attenuation and environment on the full MIR spectrum, and model all the prominent vibrational emission bands of PAH molecules in a widely selected sample of galaxies, we present the \AKARI-\Spitzer\ Extragalactic Spectral Survey (ASESS). ASESS is an exhaustive cross-archival match between the \AKARI/IRC dataset and the \Spitzer/IRS Infrared Database of Extragalactic Observables from \Spitzer\ (IDEOS) catalog \citep{Hernan-Caballero2016}. With a comprehensive spectral range of 2.5--38 $\mu m$, this survey gains access to $\textit{all}$ PAH bands in the MIR and to model them consistently using a systematic approach. Using a subset of the full \AKARI-\Spitzer\ cross-archival sample, we focus on a collection of star-forming galaxies that exhibit bright PAH emission over a broad range of total infrared luminosities (10$^{9.5-12.5}$ L$_\odot$). We present our sample selection in \S~\ref{sec:sample_selection}, followed by a detailed discussion of the spectral decomposition model in \S~\ref{sec:pahfit_decomposition}. We present our results in \S~\ref{sec:result} and discuss their interpretation in \S~\ref{sec:discussion}, summarizing these findings in \S~\ref{sec:summary}. Throughout this paper, a concordance cosmology of H$_\mathrm{0}$=75 km s$^{-1}$ Mpc$^{-1}$, $\Omega_\mathrm{M}$=0.3, and $\Omega_\mathrm{\Lambda}$=0.7 is adopted.

\begin{figure*}
    \centering
        \includegraphics[width=0.495\textwidth]{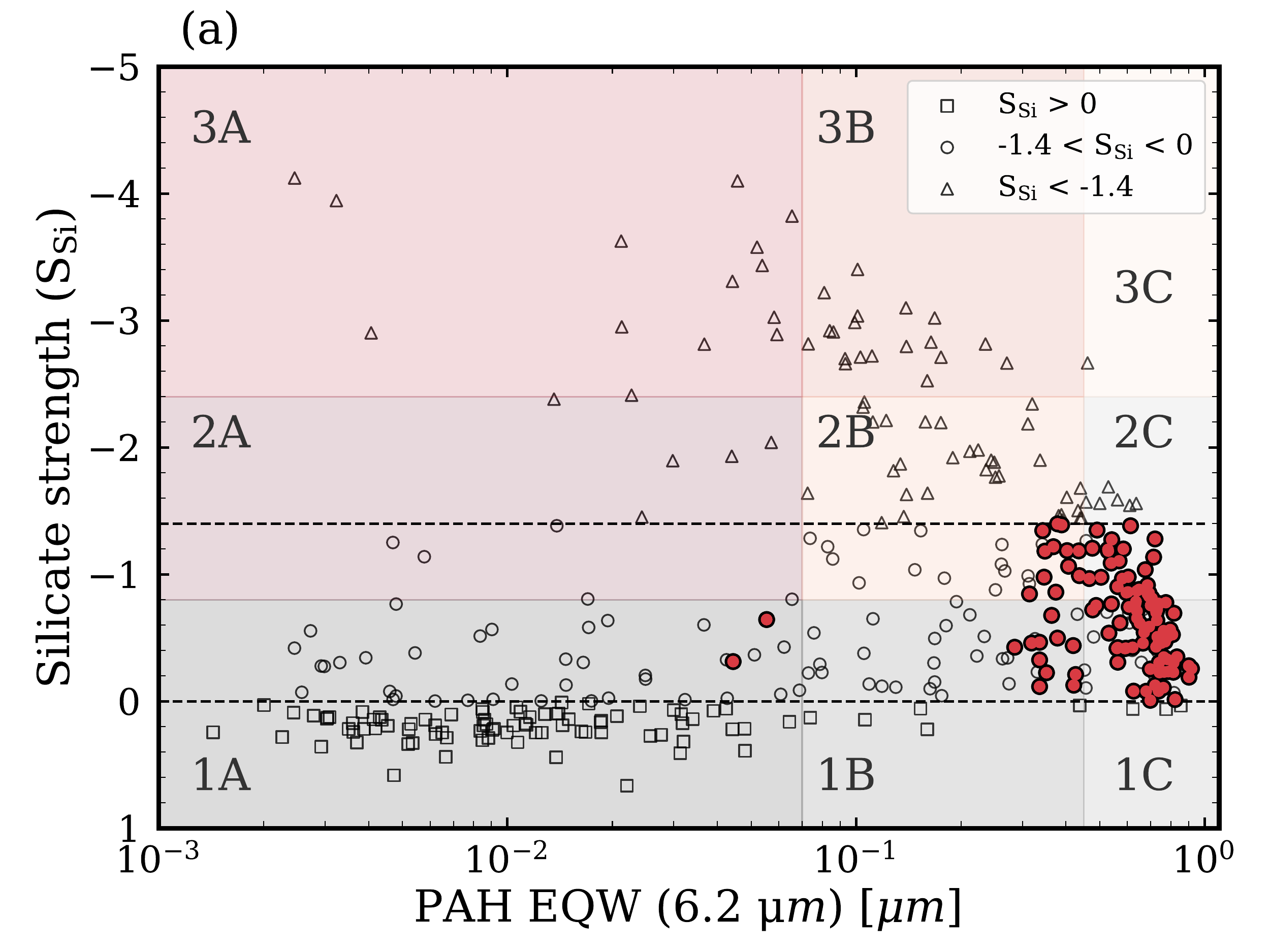} 
        \includegraphics[width=0.495\textwidth]{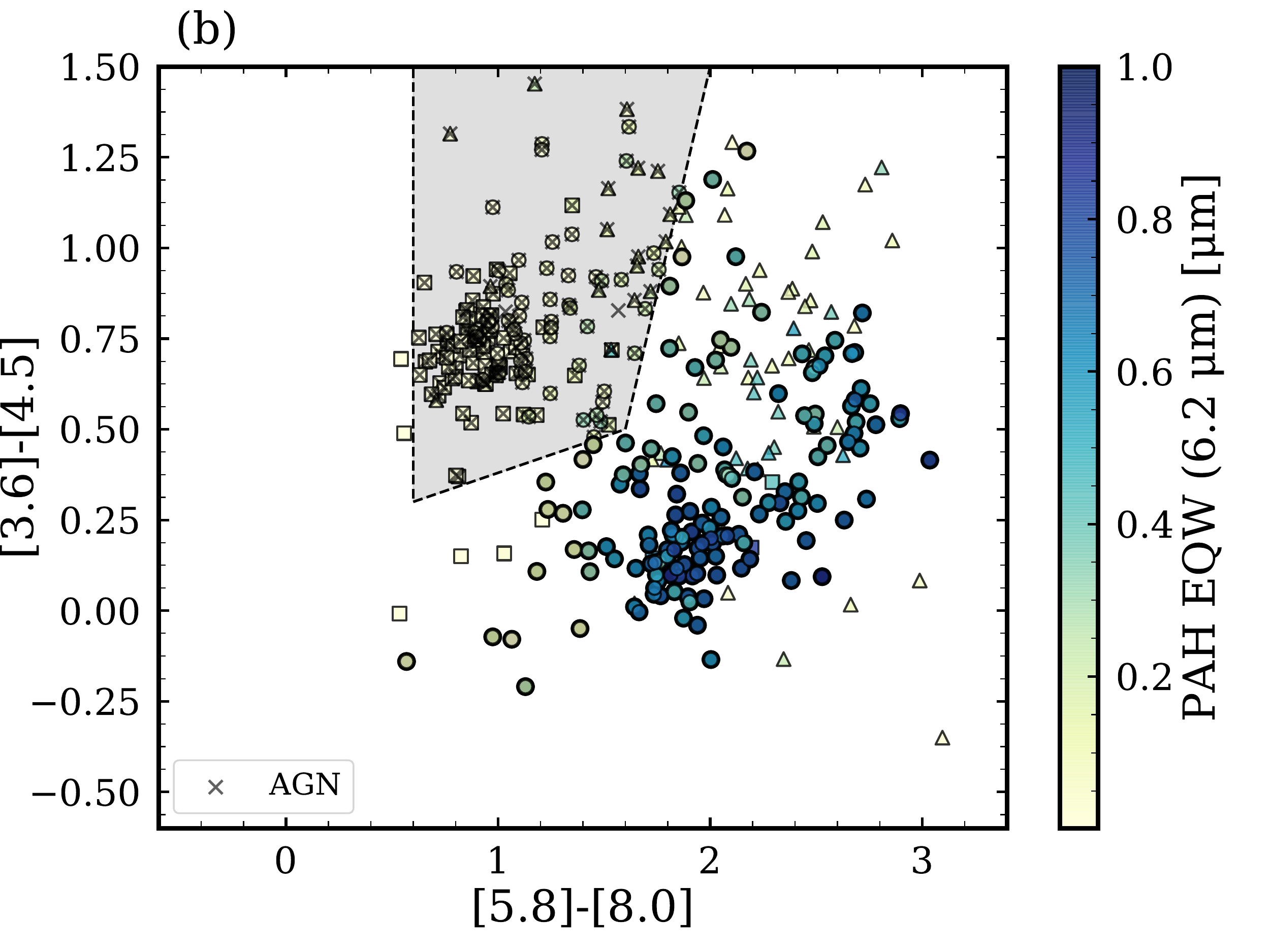} 
    \caption{Selection process of our bright-PAH sample. (a) Following the galaxy classification proposed by \citet{Spoon2007}, we present our parent sample based on the PAH 6.2 $\mathrm{\mu} m$ EQW and the 9.7 $\mathrm{\mu} m$ silicate strength. We first place a silicate cut (-1.4\,$<$ S$_\mathrm{Si}<$\,0) to include galaxies only subjected to moderate obscuration. (b) To exclude a number of low-obscuration AGNs, we further leave out galaxies lie within the ``Stern Wedge" (gray area). Consequently, with a spectral quality cut (3.3\um\ PAH $>$\,3$\sigma$ detection), our bright-PAH sample consists of 113 galaxies satisfying both criteria as shown in red in (a) (see \S~\ref{subsec:bright_PAH} for detail). The two galaxies in 1A are II~Zw~40 and IRAS 01004-2237.} 
    
    \label{fig:sample_selection} 
    
\end{figure*}

\section{Sample Selection, Observations, and Data reduction}
\label{sec:sample_selection}
With $\sim$100$\times$ better sensitivity than \ISO/SWS, the combined power of the \Spitzer\ with \AKARI\ allows us to extend deep MIR spectral coverage to more distant and less luminous sources. The ASESS sample is drawn from an exhaustive cross-archival comparison between the Infrared Database of Extragalactic Observables from \Spitzer\ (IDEOS) catalog \citep{Hernan-Caballero2016} and the public data provided by \AKARI\ Infrared Camera (IRC; \citealt{Onaka2007}) pointed spectroscopic observations compiled in the DARTS database Provided by ISAS, JAXA\footnote{https://darts.isas.jaxa.jp/astro/akari/pointing/}. 

The IRC offers two observing modes in the NIR: NG (high-resolution grism) and NP (low-resolution prism) mode. The grism mode covers a spectral range from 2.5--5$\mu m$ with R$\sim$120 at 3.6\um\ for point sources \citep{Onaka2007}, whereas the prism mode covers a spectral range from 1.5--5.5\um\ with low-resolution of R$\sim$20 at 3.6\um\ \citep{Ohyama2007}. In this cross-archival comparison, only the spectroscopic observations of 2220 galaxies with higher resolution NG/grism mode are considered. Also considered are 3361 IDEOS galaxies with \Spitzer/IRS low resolution (R\,$\simeq$\,60--125) staring mode observations covering the spectral range from 5.2--38\um\ \citep{Hernan-Caballero2016}.

In this study, we focus primarily on star-forming galaxies exhibiting strong PAH emission with modest obscuration. In this section, we describe the selection of our samples, followed by a detailed discussion of data reduction, \AKARI+\Spitzer\ spectral combination, and end with our selected subset that consists of exclusively star-forming galaxies.

\subsection{Parent Sample}
The parent sample is selected by matching sources from the IDEOS catalog to the IRC catalog. For each IRC spectrum, we create spatial profiles by averaging the spectral trace along the wavelength direction and fit up to two Gaussians. For the systems that consist of multiple nuclei within the IRC window, i.e., galaxy mergers, defined by an overlap of the Gaussian profiles within 2$\sigma$, are excluded. We further remove extended sources and spectra suffering from hot/cold pixels that are unable to recover by our profile fitting method (see \S~\ref{subsec:AKARI} for detail). Finally, we only include those spectra that have full coverage of 2.5--38\um\ for spectral decomposition, since a small number of IRS spectra have missing segments. These criteria result in a sample of 378 galaxies in our parent sample.

\subsection{IDEOS}
IDEOS provides the principle mid-infrared diagnostics for more than 3000 galaxies/galactic nuclei from spectroscopic measurements obtained by the IRS on the \Spitzer\ Space Telescope. The spectra have been drawn and vetted from the Combined Atlas of Spitzer-IRS Sources \citep[CASSIS;][]{Lebouteiller2011} a repository of 13500 homogeneously extracted, publication-quality, low-resolution spectra of all \Spitzer-IRS observations performed in staring-mode (as opposed to mapping mode). For more detail about the observation and data reduction of IDEOS, we refer the reader to a paper by \citet{Hernan-Caballero2016}.

\subsection{AKARI/IRC}
\label{subsec:AKARI}
Our 2.5--5\um\ spectroscopic sample is compiled from the archival data of the \AKARI\ \citep{Murakami2007} mission from the Infrared Camera (IRC), obtained by using the Astronomical Observing Template (AOT) IRC04 and IRCZ4 with the configuration b;Np, where b refers to the NG grism mode and Np refers to point source spectroscopy. While this is a slit-less mode, an aperture of $1\arcmin \times 1\arcmin$ is designed to block lights from nearby sources. To optimize the process of cosmic-ray removal, each target was observed by one to ten pointings, depending on the brightness of the source, and each pointing consists of eight or nine sub-frames. For each pointing, the total on-source exposure time was $\sim$ 6 minutes \citep{Ohyama2007}.

Our parent sample includes three phases of observations during the \AKARI\ mission -- phase 1 \& 2 were observations taken during the cryogenic phase, while phase 3 was post-cryogenic. The number of galaxies in phase 1 \& 2 and 3 are 50 and 328, respectively. Data reduction was carried out using \textit{IDL} packages prepared by the \AKARI\ team: ``IRC Spectroscopy Toolkit Version 20080528'' for phase 1 \& 2 data and ``IRC Spectroscopy Toolkit for Phase 3 data Version 20170225'' for phase 3 data. Both packages involve dark subtraction, linearity correction, flat-fielding, sub-frame combination, source detection, background/source subtraction, and wavelength and flux calibration. Furthermore, we applied a correction for second-order contamination as detailed in \citet{Baba2016, Baba2019}. We derived the spatial profile of each source from the 2D spectrum, using the 2$\sigma$ range for the spatial extraction width, which varies from 4-12 pixels (5$\farcs$85--17$\farcs$52). Extended galaxies that have a broad spatial profile with 2$\sigma$ range $>$12 pixels (17$\farcs$52) are excluded from our sample.

Several additional issues emerged in phase 3 after cryogen exhaustion: 1) the detector sensitivity was reduced by $\sim 70 \%$ compared to phase 2\footnote[2]{\AKARI\ observer's manual version 1.2}, and 2) the occurrence of \emph{hot} pixels increased to $\sim 5 \%$ of the total. These effects make 2D spectral images contaminated. To correct these contaminations, we follow the prescription from \citep{Horne1986} to construct a normalized spatial profile at each wavelength, and average these over wavelength to construct a single characteristic profile. The defect pixels, which typically shown as NaNs in the 2D spectral image, are then replaced by values scaled from the characteristic profile at the corresponding wavelength. To reach a higher S/N, we combine available spectra from different pointings (i.e., with similar primary ID but different sub-IDs) by calculating the weighted average.

\subsection{Spitzer-AKARI Spectra Stitching}
The $\sim$0.2\um\ gap in between the long-wavelength end of \AKARI/IRC and the start of \Spitzer/IRS complicates the process of combining the two spectral segments. For all 378 \AKARI-\Spitzer\ combined spectra in our parent sample, we scaled the flux of each segment to match their synthetic photometry to the \textit{WISE} photometry obtained from the profile-fitted magnitude in the ALLWISE Source Catalog \citep{Cutri2013}. The values were converted into fluxes in Jy based on the WISE calibration provided by \citet{Jarrett2011}. Only the W1 band is used to scale the IRC spectrum because the band transmission profile of W2 falls beyond the IRC spectral range. For scaling the IRS spectrum, we first follow the method of \citet{Wright2010} to correct the W4 fluxes, which appear to be $\sim$8--10\% too bright in the red sources, and further calculate the scaling factor using the average of W3 and W4 bands. For three sources lacking WISE photometry (IRAS 15250+3609, WKK 2031, and IC 4518 A), we used \textit{IRAC} photometry in a similar manner from the Spitzer Enhanced Imaging Products (SEIP) database to scale our spectra. For two additional sources (UGC 02369 and NGC 5990), where no photometry is provided in either database, we performed a linear extrapolation from feature-free regions on each side of the gap, scaling the IRC spectrum such that the two lines meet at gap center (5.1\um). We find the scaling factors of both the \AKARI\ and \Spitzer\ spectra lie mostly within 20\%, similar to the values reported by \citet{Baba2018}.

\subsection{Star-Forming Subset: Bright-PAH Sample} \label{subsec:bright_PAH}
Our parent sample consists of sources including (i) star-forming galaxies with prominent PAH emissions, (ii) heavily obscured AGNs, with strong 9.7 and 18\um\ silicate absorptions and PAH-diluting warm continuum, and (iii) low obscured AGN and Quasi-Stellar Objects (QSO) with smooth continuum, mainly dominated by dust re-emission with weaker PAH emissions. For the purpose of this study, we further narrow down our sample to galaxies exhibiting bright PAH emissions and dub them the ``bright-PAH sample". \citet{Spoon2007} provided a galaxy classification scheme based on the equivalent width (EQW) of the 6.2\um\ PAH and the silicate strength, S$_\mathrm{Si}$, defined by S$_\mathrm{Si}$ = ln(f$\mathrm{_{9.7 \mu m}/C_{9.7 \mu m})}$, where $f_\mathrm{9.7 \mu m}$ is the central flux intensity of the silicate absorption feature and C$_\mathrm{9.7 \mu m}$ is the estimated continuum flux intensity adopted for the local MIR spectrum. We follow this methodology and present our sample in Fig.~\ref{fig:sample_selection}(a) using the 6.2\um\ PAH EQW and the silicate strength both obtained from the IDEOS catalog. The average spectral template of each class will be presented in \S~\ref{sec:template}. Galaxies exhibiting strong PAH features are typically subjected to less silicate absorption. We further select only those galaxies with -1.4 $<$ S$_\mathrm{Si}$ $<$ 0 (circles), thus avoiding both deep obscuration (triangles) and silicate \emph{emission} (squares) contributed by inner AGN regions \citep[e.g.,][]{Fritz2006, Nenkova2008}. Galaxies satisfying the above criterion lie in between the two dashed lines in Fig.~\ref{fig:sample_selection}(a), but with this simple silicate cut, a modest number of lower-obscuration AGN sources remain.

To exclude these sources, we use the IRAC photometric AGN classification scheme of \citet{Stern2005}. In general, AGNs are redder in the [3.6]-[4.5] color and restricted to a narrower distribution of [5.8]-[8.0] color since those sources tend to have less prominent PAH emission diluted by the strong AGN continuum. We use synthetic photometry from IRAC response curves, excluding from our bright-PAH sample those IRAC-selected AGNs that lie inside the ``Stern Wedge" (shaded region in Fig.~\ref{fig:sample_selection}(b)), which are added with cross signs. The symbols used here are similar to Fig.~\ref{fig:sample_selection}(a) and are color-coded by strength of the 6.2\um\ PAH EQW. Galaxies exhibiting strong PAH emission tend to cluster at the bottom right corner as opposed to another galaxy cluster that lies inside the AGN region. Finally, we limit our bright-PAH sample to have robust ($>$\,3$\sigma$) detection of the 3.3\um\ PAH band. Our bright-PAH sample of 113 galaxies that meet all the imposed selection criteria are colored in red in Fig.~\ref{fig:sample_selection}(a). The two bright-PAH galaxies that lie within 1A are II~Zw~40 (See \S \ref{subsec:IIZw40} for detail) and IRAS~01004-2237, which hosts a moderately obscured AGN \citep{Imanishi2007} leading to a weak 6.2\um\ PAH emission. Even with this selection method, substantially buried AGNs which do not significantly contribute to the MIR continuum may still contribute to the total infrared luminosity, especially for galaxies with \LIR$>$10$^{12}$\Lsun\ \citep[][see also \S~\ref{sec:all_the_PAHs}]{Imanishi2008, Nardini2010, Petric2011, Alonso-Herrero2012, Imanishi2018, Imanishi2019}.

\begin{figure}
 	\includegraphics[width=1.0\columnwidth]{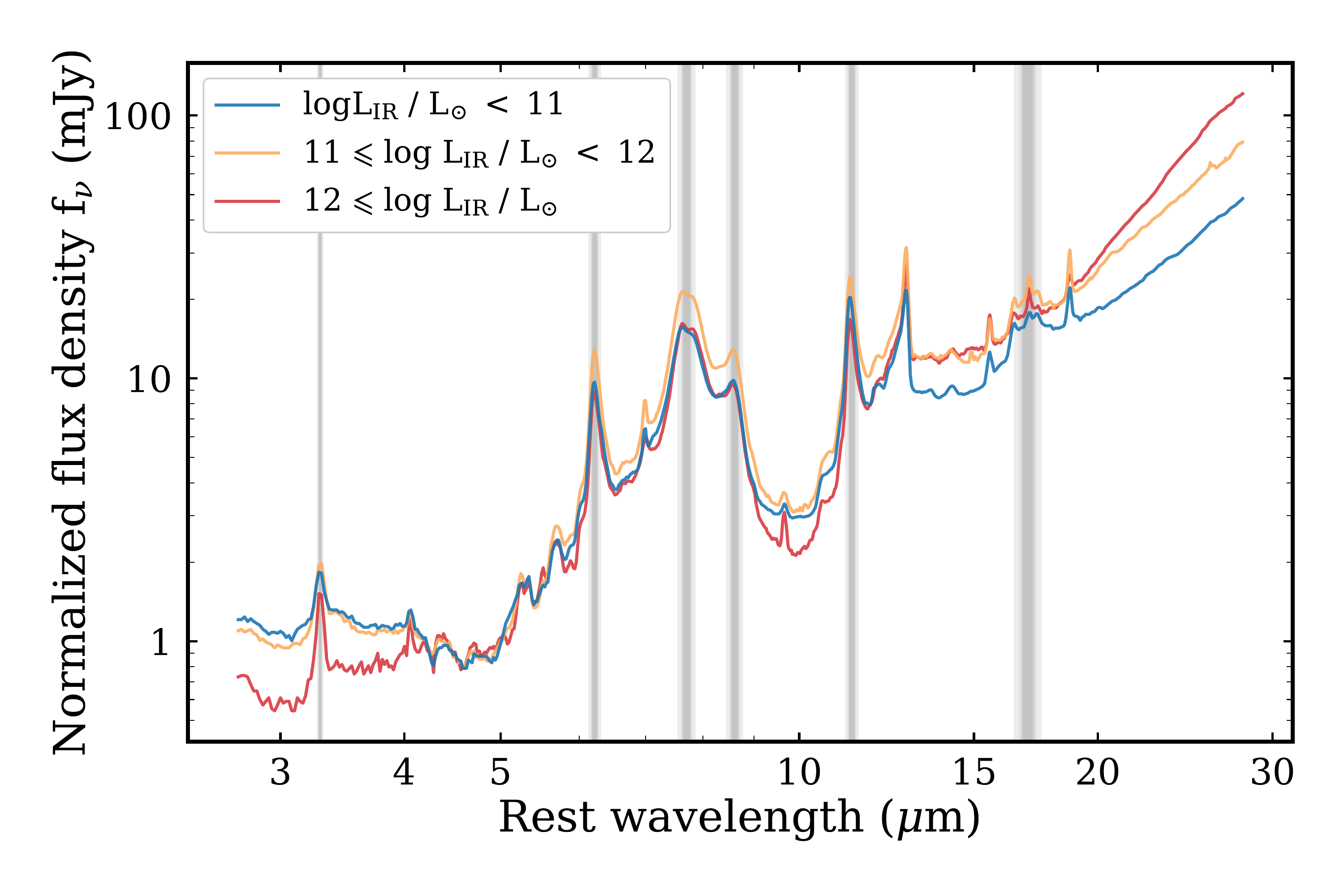}
    \caption{The averaged spectra of our bright-PAH sample binned by log\,\LIR. Spectra are normalized at 5\um\ and the width of the stripe indicates the FWHM of the main PAH features centered at 3.3, 6.2, 7.7, 8.6, 11.3, and 17\um. The numbers of galaxies used in the blue, yellow, and red spectra are 16, 73, and 23, respectively.}
    \label{fig:brightPAH_specs}
\end{figure}

To illustrate how infrared luminosity may affect the overall shape of the spectrum, we divide bright-PAH galaxies into three categories according to \LIR. Fig. \ref{fig:brightPAH_specs} shows these averaged spectra normalized at 5\um, with primary PAH emissions indicated. The \AKARI/IRC spectra of galaxies with log\,\LIR\,$<12$ show dominant stellar components, while galaxies with log\,\LIR\,$\geq 12$ have a steeper continuum slope towards red. Table~\ref{tab:bright_PAH_sample} summarizes basic information of galaxies in the bright-PAH sample, with Table~\ref{tab:bright_PAH_id} summarizes the observational details (See Appendix). The redshifts of our subset span a range 0 $\leq$ z $\leq$ 0.2 with a median of z=0.03.

\begin{deluxetable*}{lcrrcccc}
\tabletypesize{\footnotesize}
\tablecolumns{9}
\tablewidth{5pt}
\tablecaption{Bright-PAH Sample in ASESS}
\label{tab:bright_PAH_sample}
\tablehead{
 \colhead{Galaxy} & \colhead{z} &  \colhead{R.A.} &  \colhead{Dec.} & \colhead{EQW(6.2)} & \colhead{S$_\mathrm{Si}$} & \colhead{log L$_\mathrm{IR}$} & \colhead{Ref.} 
\vspace{-0.15cm}\\
\colhead{(IDEOS)} &  & \colhead{[deg.]} &  \colhead{[deg.]} & \colhead{[$\mu m$]} & & \colhead{[L$_{\odot}$]} &
\vspace{-0.15cm}\\
\colhead{(1)} &  \colhead{(2)} & \multicolumn{2}{c}{(3)} & \colhead{(4)} & \colhead{(5)} & \colhead{(6)} & \colhead{(7)} \vspace{-0.4cm}\\
}
\startdata
                 NGC0023 &  0.015 &   2.4726 &  25.9240 &    0.74 &            -0.23 &               11.05 &    2 \\
            ARP256 NED01 &  0.027 &   4.7119 & -10.3768 &    0.76 &            -0.35 &               11.42 &    4 \\
                 NGC0232 &  0.023 &  10.6908 & -23.5614 &    0.65 &            -0.62 &               11.40 &    4 \\
 2MASX J00480675-2848187 &  0.110 &  12.0282 & -28.8051 &    0.65 &            -0.88 &               12.15 &    3 \\
     MCG+12-02-001 NED02 &  0.016 &  13.5164 &  73.0850 &    0.79 &            -0.23 &               11.40 &    2 \\
          IRAS01003-2238 &  0.118 &  15.7083 & -22.3659 &    0.06 &            -0.64 &               12.26 &    4 \\
           MCG-03-04-014 &  0.035 &  17.5373 & -16.8527 &    0.78 &            -0.33 &               11.64 &    2 \\
       ESO244-G012 NED02 &  0.023 &  19.5345 & -44.4620 &    0.76 &            -0.56 &               11.39 &    4 \\
             CGCG436-030 &  0.031 &  20.0110 &  14.3618 &    0.39 &            -1.39 &               11.63 &    2 \\
             ESO353-G020 &  0.016 &  23.7135 & -36.1373 &    0.68 &            -0.92 &               10.99 &    2 \\
\enddata
\vspace{0.cm}
\tablecomments{Table \ref{tab:bright_PAH_sample} is published in its entirety in the machine-readable format. A portion is shown here for guidance regarding its form and content. Col. (1)--(5) are obtained from the IDEOS survey. Col. (1): Galaxy name. Col. (2): Redshift. Col. (3): Coordinate in J2000. Col. (4): The 6.2\um\ PAH equivalent width. Col. (5): Strength of the 9.7\um\ silicate feature, defined by the ratio of observed flux intensity (f$_{9.7\mu m}$) to continuum flux density (C$_{9.7\mu m}$) at 9.7\um, S$_\mathrm{Si}$ = ln$\mathrm{\frac{f_{9.7\mu m}}{C_{9.7\mu m}}}$. Col. (6): Logarithm of integrated infrared (8-1000\um) luminosity, in units of solar luminosity (L$_{\odot}$), derived from IRAS fluxes and calculated from L$_\mathrm{IR}$ = 2.1 $\times$ 10$^{39}$ $\times$ D(Mpc)$^2$ $\times$ (13.48$\times f_{12}$ + 5.16$\times f_{25}$ + 2.58$\times f_{60}$ + $f_{100}$) ergs s$^{-1}$ \citep{Sanders1996}. For fluxes in IRAS with upper limits, the prescriptions in \citet{Imanishi2008, Imanishi2010} are followed. Col. (7): Adopted  IRAS fluxes from the literature compilation (1: \citealt{Moshir1993}; 2: \citealt{Sanders2003}; 3: \citealt{Imanishi2008}; 4: \citealt{Imanishi2010}).
}
\end{deluxetable*}

\section{Decomposition of galaxy spectra from 2.5$-$38 $\mu m$}
\label{sec:pahfit_decomposition}
\begin{figure*}
 	\includegraphics[width=1.0\textwidth]{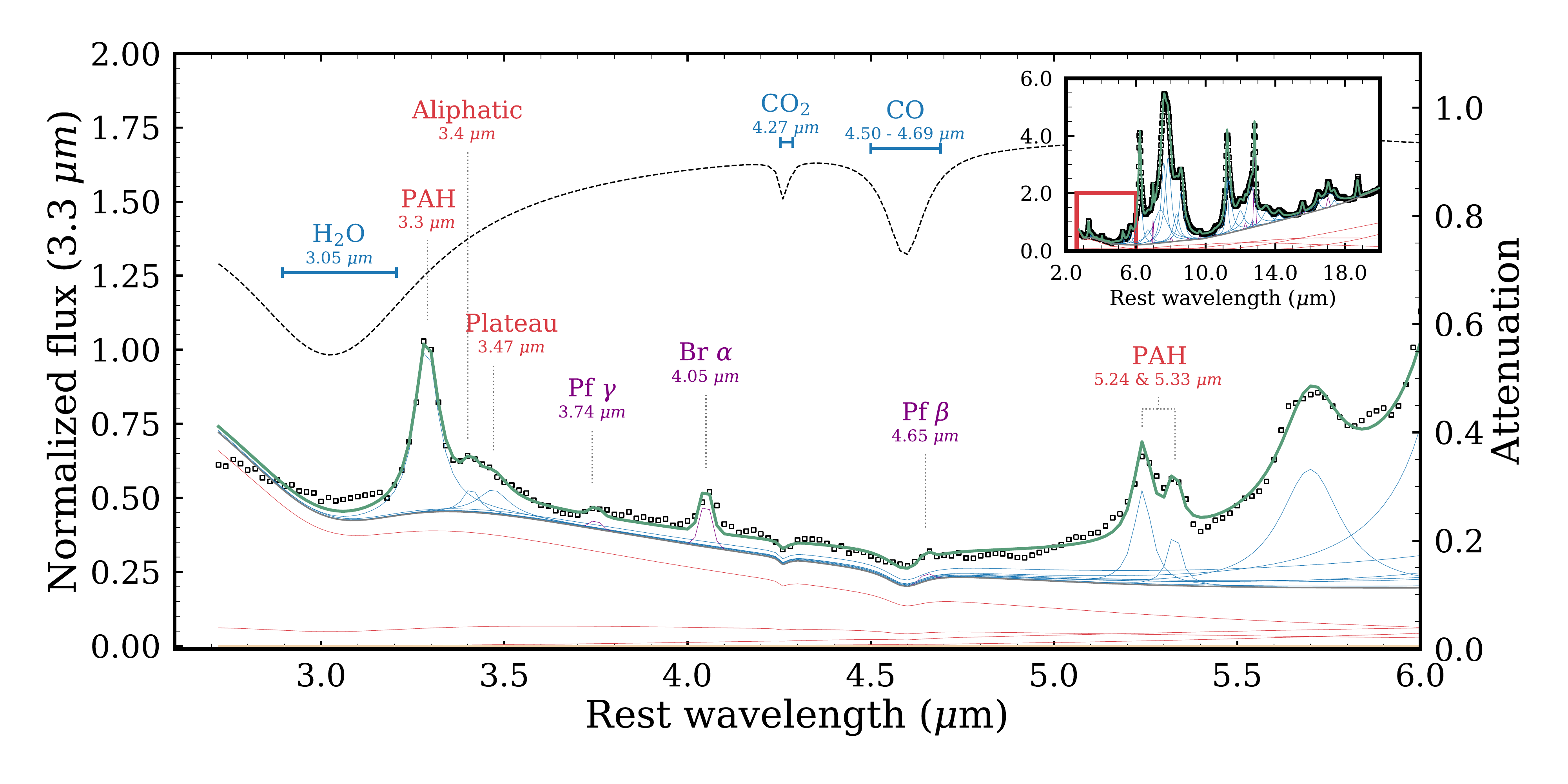}
    \caption{Detailed decomposition of the normalized type 1C spectrum ranging from 2.5--6\um. The newly added features in the enhanced \PAHFIT\ are labelled, including dust emission features (blue) of the 3.3\um\ PAH, 3.4\um\ aliphatic, and 3.47\um\ plateau, line emission features (purple) of the 3.74\um\ Pf$\gamma$, 4.05\um\ Br$\alpha$, and 4.65\um\ Pf$\beta$ (Pf$\beta$ are not obvious in this case). The dashed line indicates the attenuation caused by the fully mixed extinction with absorption features of H$_\mathrm{2}$O ice at 3.05\um, CO$_{2}$ at 4.27\um, and CO at 4.60\um. The full decomposition result is shown in the inset.}
    \label{fig:pahfit_example}
\end{figure*}

The infrared wavelengths from 2.5--5\um\ host a rich inventory of dust and line emission features, together with continuum and ice absorption. To assess how these features vary with the ISM environment, we use \PAHFIT\ to decompose the spectrum and to extract their integrated strength. \PAHFIT\ is a robust decomposition model first employed for estimating the strengths of weak, blended features of the low-resolution \Spitzer/IRS spectra in the MIR regime \citep[5.2--38 $\mu m$;][]{Smith2007}. IRS low-resolution spectra have a 5.2\um\ short wavelength cutoff, and therefore the 3.3\um\ PAH band was not included in \PAHFIT. Our main objective in extending \PAHFIT\ model coverage down to 2.5\um\ is to simultaneously model \emph{all} the PAH emission features with their underlying continua and absorption features. In this section, we first discuss the newly added emission components, followed by the new absorption components together with the adopted geometries. Fig.~\ref{fig:pahfit_example} shows the decomposition result of a typical star-forming galaxy.

\subsection{New Emission Components}
The profile of the 3.3\um\ PAH is set following the suggestions from \citet{Tokunaga1991} and \citet{VanDiedenhoven2004}. The 3.3\um\ PAH feature sometimes shows a broader wing towards longer wavelengths which is often attributed to a 3.4\um\ aliphatic band, together with an even broader underlying plateau centered at 3.47\um\ \citep[][see \S~\ref{sec:aliphatic}]{Hammonds2015}. The modest range of the redshift in our sample fills in the gap between the \AKARI/IRC and the \Spitzer/IRS spectra, which allows us to investigate the weaker PAH feature at $\sim$\,5.27\um\ that had been reported in \ISO\ spectra \citep[e.g.][]{Boersma2009}. We find the 5.27\um\ PAH feature can further be separated into two individual bands centered at 5.24 and 5.33\um. All the newly added/modified dust feature parameters can be found in Table~\ref{Tab:additional_features}(a).

To account for the unresolved recombination lines within the \AKARI/IRC spectra, we added Br$\beta$ at 2.63\um, Pf$\gamma$ at 3.74\um, Br$\alpha$ at 4.05\um, and Pf$\beta$ at 4.65\um\ to \PAHFIT. Since the Pf$\beta$ line at 4.65\um\ may be blended with an absorption band arising from CO at $\sim$ 4.67\um\ (see \citet{Spoon2003} for a high resolution spectrum), to avoid such ambiguity, we tied Pf$\beta$ with Br$\alpha$ using the theoretical line ratio of Pf$\beta$/Br$\alpha$=0.20 under the Case B condition, with n$_{e} \sim 10^{2}-10^{7}$ cm$^{-3}$ and T$_{e} \sim (3-30) \times 10^{3}$K \citep{Storey1995}. The newly added line feature parameters can be found in Table~\ref{Tab:additional_features}(b).

The original \PAHFIT\ model treats starlight continuum as a single temperature blackbody spectrum with T$_{\star}$=5000K. This assumption is sufficient for $\lambda>5$\um, but with wavelengths down to 2.5\um, additional features are needed to account for the non-stellar NIR excess that has an average temperature $\sim$ 1000K from the hot grains \citep{Lu2003} and the divergence of the mean starlight spectrum at $\lambda \sim$2--5\um. To address this and gain flexibility in the 2.5--4\um\ continuum fit, we add 800K and 1500K modified blackbody dust components, with the $\sim$1500K component representative of the dust sublimation temperature in AGNs \citep{Marchall2007, Nenkova2008}. Accurately separating stellar and non-stellar continuum sources is challenging given our spectral coverage and simplified treatment, so no physical significance is ascribed to individual continuum components.

\begin{table*}
	\footnotesize
	\centering
	\caption{Newly Added Features in \PAHFIT}
	\begin{tabular}{cccc}
		\textbf{(a) Dust feature parameters}\\
		\hline\hline
		$\lambda$($\mu$m) & $\gamma _{r}$ & FWHM($\mu$m) & Reference\\
        \hline
		3.29 & 0.013 & 0.043 & \citet{Tokunaga1991}\\
        3.40 & 0.009 & 0.031 & \citet{Hammonds2015}\\
        3.47 & 0.029 & 0.100 & \citet{Hammonds2015}\\
        5.24 & 0.011 & 0.058 & \\
        5.33 & 0.008 & 0.043 & \\
        \hline
        \multicolumn{4}{l}{%
            \begin{minipage}{11cm}%
            \footnotesize N\textsc{ote} -- $\gamma _{r}$ is the fractional FWHM, the ratio of FWHM to central wavelength ($\lambda$).
        \end{minipage}%
        }\\
        \\
        \textbf{(b) Line feature parameters}\\
		\hline\hline
        Line & $\lambda$($\mu$m) & FWHM($\mu$m)\\	
        \hline
        Br$\beta$ & 2.63 & 0.034\\
        Pf\,11 $^{*}$ & 2.87 & 0.034\\
        Pf$\epsilon$ $^{*}$ & 3.04 & 0.034\\
        Pf$\gamma$ & 3.74 & 0.034\\
        Br$\alpha$ & 4.05 & 0.034\\
        Hu\,13 $^{*}$ & 4.18 & 0.034\\
        He\,I $^{*}$ & 4.30 & 0.034\\
        Hu\,12 $^{*}$ & 4.38 & 0.034\\
        Pf$\beta$ & 4.65 & 0.034\\
        \hline
        \multicolumn{4}{l}{%
            \begin{minipage}{10cm}%
            \footnotesize $^{*}$ Additional emission lines identified in II~Zw~40
        \end{minipage}%
        }\\   
        \\
        \textbf{(c) Absorption feature parameters}\\
		\hline\hline        
        Name & $\lambda$($\mu$m) & FWHM($\mu$m) & Reference\\
        \hline
        H$_{2}$O & 3.05 & (0.406--0.528) & \\
        CO$_{2}$ & 4.27 & 0.033 & \citet{Gao2013}\\
        CO & 4.67 (4.50--4.69) & 0.100 & \citet{Gao2013}\\
        H$_{2}$O & 6.02 & 0.580 & \citet{Gao2013}\\
        \hline
        \multicolumn{4}{l}{%
            \begin{minipage}{10cm}%
            \footnotesize N\textsc{ote} -- Values in the parenthesis indicate the constraints in the model.
        \end{minipage}%
        }\\
        \\

    \end{tabular}
    \label{Tab:additional_features}
\end{table*}

\pagebreak
\subsection{New Absorption Features}
The expanded coverage of the \AKARI/IRC also includes several absorption features: a broad 3.05\um\ H$_{2}$O ice feature \citep[e.g.][]{Gibb2004}, 4.27\um\ CO$_{2}$ \citep{Spoon2000, Spoon2003}, and a very broad ($\sim$\,0.2\um) feature of CO absorption centered at 4.67\um\ \citep{Baba2018}. To precisely extract the 3.3\um\ PAH, it is necessary to differentiate the band from the nearby 3.05\um\ H$_{2}$O ice. Another H$_{2}$O ice absorption features at 6.02\um\ is also included in the model, with its peak optical depth tied to the 3.05\um\ H$_{2}$O ice. All of the newly added absorption features are approximated as Drude profiles with the widths and relative strengths of \citet{Gao2013} (see Table~\ref{Tab:additional_features}(c)), except for the FWHM of the 3.05\um\ H$_{2}$O ice. We found our sample have a profile on average 30\% wider than reported by Gao et al., with FWHM = 0.41\um. Since the profile of H$_{2}$O absorption is temperature dependent and its wing extends towards longer wavelength as temperature increases \citep{Spoon2002}, the upper limit of the FWHM is set to be 0.53\um, slightly less than the even broader ice profile (FWHM = 0.56\um) reported by \citet{Hammonds2015} when analyzing a large number of Galactic objects using \AKARI/IRC. We found no correlation between optical depths of H$_{2}$O and silicate, similar to what was reported in \citet{Doi2019}, so \tauWater\ and \tauSi\ were left uncoupled.

The CO ro-vibrational absorption profile (4.67 $\mu m$) has been used to model the column density and gas temperature of obscured AGNs \citep{Spoon2004, Shirahata2013, Baba2018}. As pointed by Baba et al., double-branched features of CO are observed, with the possible width of each branch extends to $\sim$\,0.2\um, depending on the CO gas temperature. Here, for simplicity, a Drude profile with a central wavelength constrained to 4.5\,$<\lambda<$\,4.69 is adopted to account for such deep and wide feature. Neither CO$_2$ nor CO absorption feature affects dust emission component (only Pf $\beta$ can be affected), so these simplifying treatments have limited impact on our results. 

\begin{figure}
 	\includegraphics[width=1.\columnwidth]{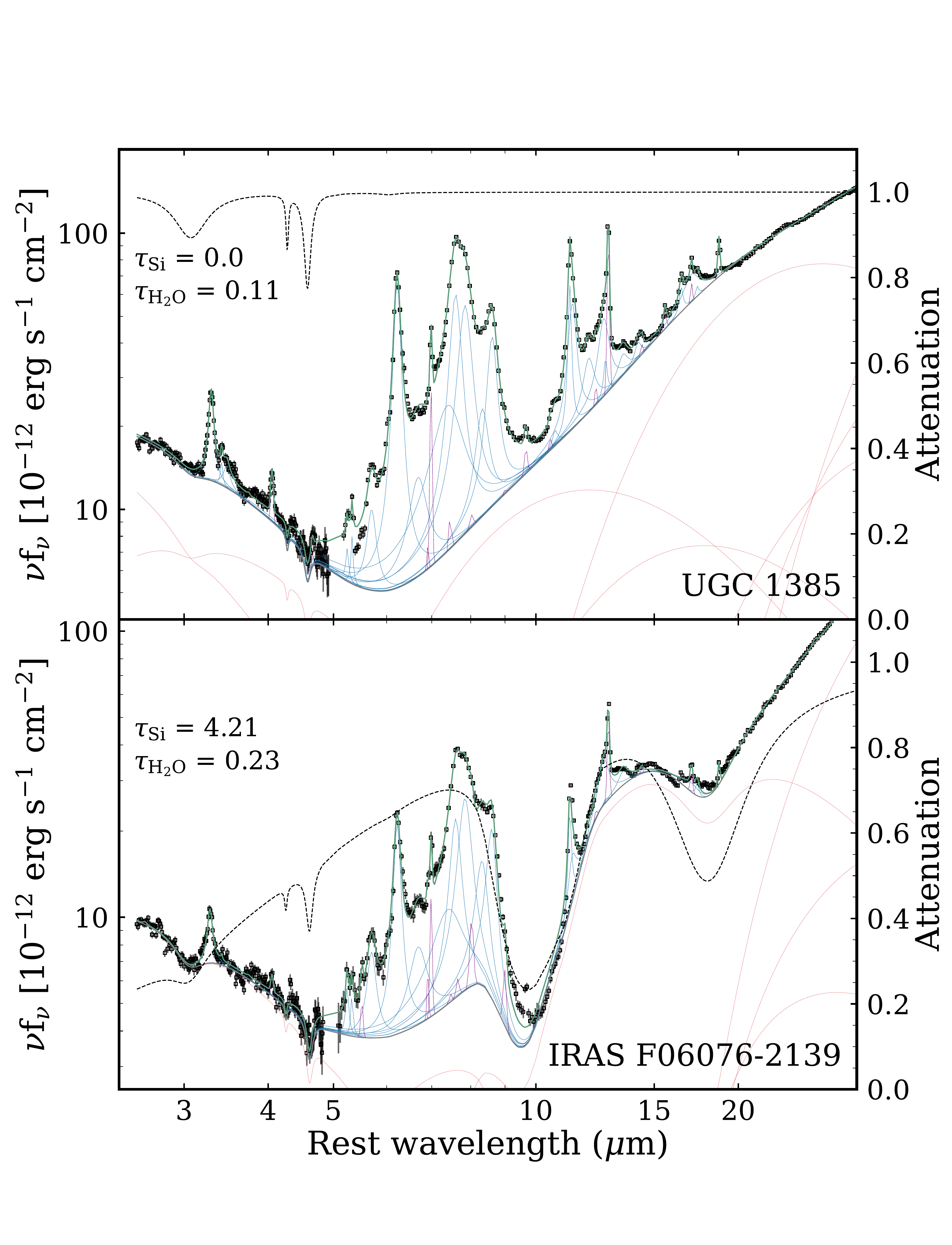}
    \caption{Two PAHFIT results of sample spectra that exhibit the minimum/maximum \tauSi\ (mixed) in our bright-PAH sample. Note that  \tauSi, \tauWater, and $\tau_\mathrm{CO+CO_2}$ vary independently.}
    \label{fig:galaxy_ext_curve}
\end{figure}

To demonstrate the full range (2.5--38\um) extinction profile that includes the newly added absorption features, we show PAHFIT results of the two example spectra (UGC~1385 and IRAS~F06076-2139) that contain the minimum and maximum \tauSi\ values in our bright-PAH sample in Fig.~\ref{fig:galaxy_ext_curve}. IRAS~F06076-2139 passed our selection criteria (\S~\ref{subsec:bright_PAH}) because its spline-based silicate strength is very close to but just below the cutoff \mbox{(S$_\mathrm{Si}$=-1.38)}.

\subsection{Attenuation in Different Geometries} \label{subsec:attenuation}
With a variety of smooth and spectrally resolved extinction components, the relative geometry of emission sources and obscuring material plays important roles in the interpretation of the MIR spectra of galaxies. For normal star-forming galaxies, it often considered appropriate to assume stars and dust are spatially well mixed, with both PAH+line and continuum emission subject to a similar range of attenuation\footnote{Here and throughout we use the term ``attenuation'' to refer to the combined effects of photon absorption and scattering, as well as the impact of geometry on these phenomena, reserving ``extinction'' for the geometry independent and wavelength dependent curves measured along individual lines of sight (typically in the MW).}. For galaxies powered by buried AGN or compact central starburst with only \HII-regions, however, the obscuration comes predominately from the central part of the system, while the PAH/line emissions excited by stars in the outer part are mostly not attenuated \citep{Imanishi2007, Imanishi2008}. To account for this latter scenario, we add a new geometry, hereafter the ``obscured continuum'' geometry, upon the fully mixed geometry that was already developed in \PAHFIT. For the extinction profile, we retain the custom curve developed with \PAHFIT\ \citep{Smith2007}, which derives from the Galactic center of \citet{Kemper2004}. With both the mixed and obscured continuum geometries, it is our attempt to bracket the range of realistic emission and absorption by these two simple schemes. In other words, the ``real" measurements should then be bracketed by these two extremes, and we interpret our results in both cases throughout this paper. We decide not to use the screen geometry as an alternative to the mixed geometry because the difference of the optical depths derived from these two schemes is simply a constant factor within our attenuation range, with no impact on our decomposition results. 

Including the additional obscured continuum geometry in the model is necessary since \PAHFIT\ was trained primarily on a sample of nearby galaxies from the SINGS survey \citep{Kennicutt2003}, in which the attenuation is moderate \citep[with mixed silicate opacity, $\tau_{\mathrm{Si}} $$\lesssim$ 0.35;][]{Smith2007}. Even our trimmed sample, however, contains galaxies with higher attenuation (see \S~\ref{subsec:bright_PAH}) with the median $\tau_{\mathrm{Si}}\!\sim\!1.1$. This amount of attenuation may affect our measurement of the 3.3\um\ PAH since it suffers the strongest effects of dust and ice attenuation (see \S~\ref{sec:3.3PAH} and \S~\ref{subsec:geometry_disc} for detailed discussion). 

\begin{figure}
 	\includegraphics[width=1.0\columnwidth]{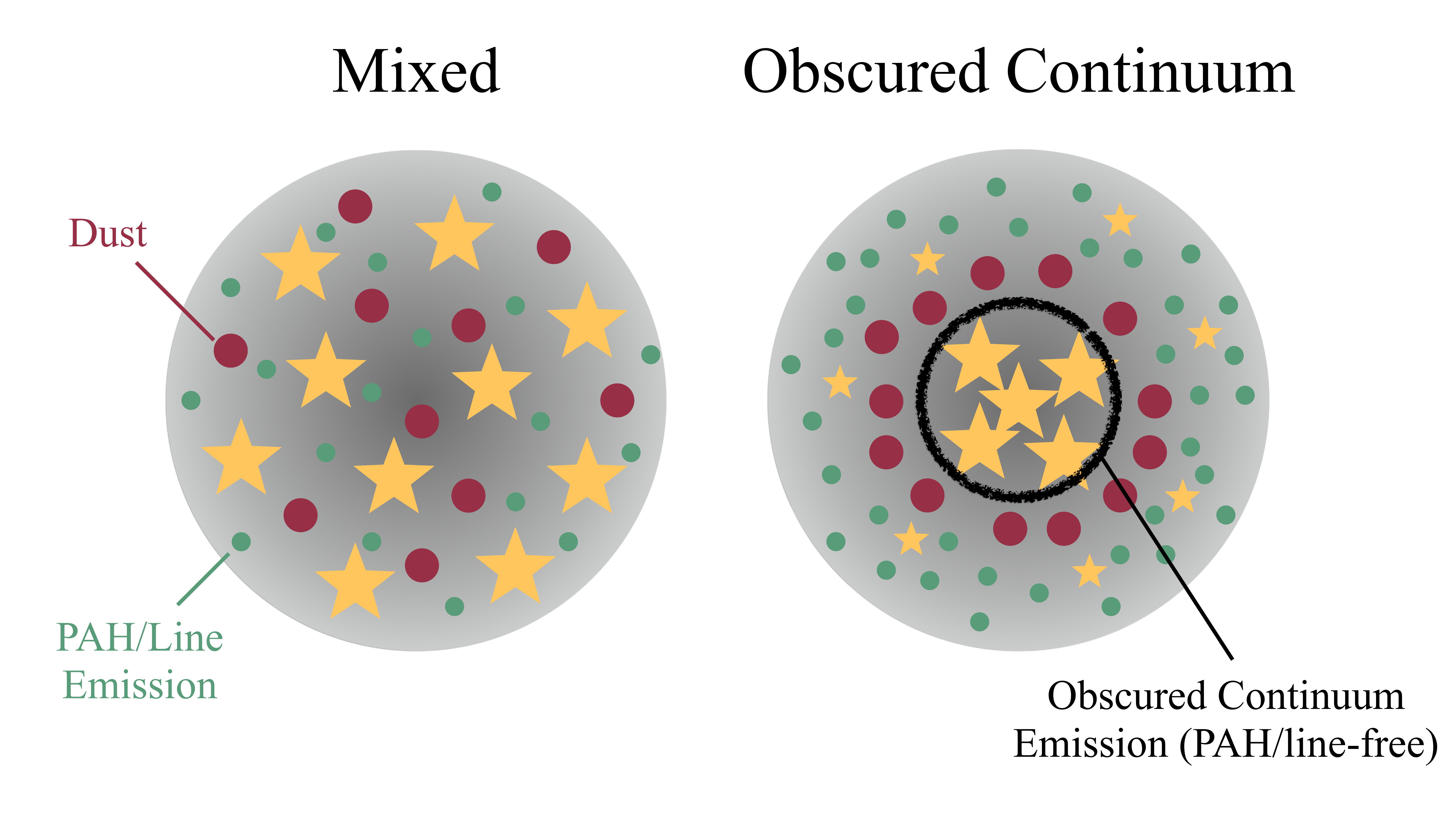}
    \caption{The two bracketing geometries considered in this study. In the mixed geometry, emitting sources and dust are fully mixed, whereas in the obscured continuum geometry, PAH and line emission occurs outside an obscured central continuum zone, and is not subjected to attenuation. 
    }
    \label{fig:geometry}
\end{figure}

In the new obscured continuum geometry, all dust and line emission features sit atop an attenuated continuum as illustrated schematically in Fig.~\ref{fig:geometry}, where both geometries in mixed and obscured continuum are demonstrated. While \PAHFIT\ by default outputs recovered fluxes corrected from dust attenuation, the measurements obtained from the obscured continuum geometry are closer to the observed fluxes measured directly by integrating the line/band profile above the assumed local continuum. Note that in both geometry schemes, optical depth acts in a similar fashion with a scaling factor of (1-$e^{-\tau_{\lambda}}$)/$\tau_{\lambda}$. 

\section{Results} \label{sec:result}
In this section, we first present a set of galaxy template spectra produced from ASESS and the spectral decomposition result of the galaxies in our bright-PAH sample. We then discuss the properties of 3.3\um\ PAH together with star formation rate calibration using 3.3\um\ and how different geometries may come into play. A special case of low-metallicity galaxy, II~Zw~40, is discussed. We then study the relationship between the aromatic and aliphatic molecules using the 3.3 and 3.4\um\ bands and a size diagnostic using 3.3, 11.3, and 17\um\ PAHs, followed by a comprehensive comparison between 3.3\um\ PAH and other prominent PAH features. We end this section with an outlook for probing the 3.3\um\ PAH strength using the \JWST\ photometric filters. 

\subsection{The Composite Template Spectra} \label{sec:template}
\begin{figure}
 	\includegraphics[width=1.0\columnwidth]{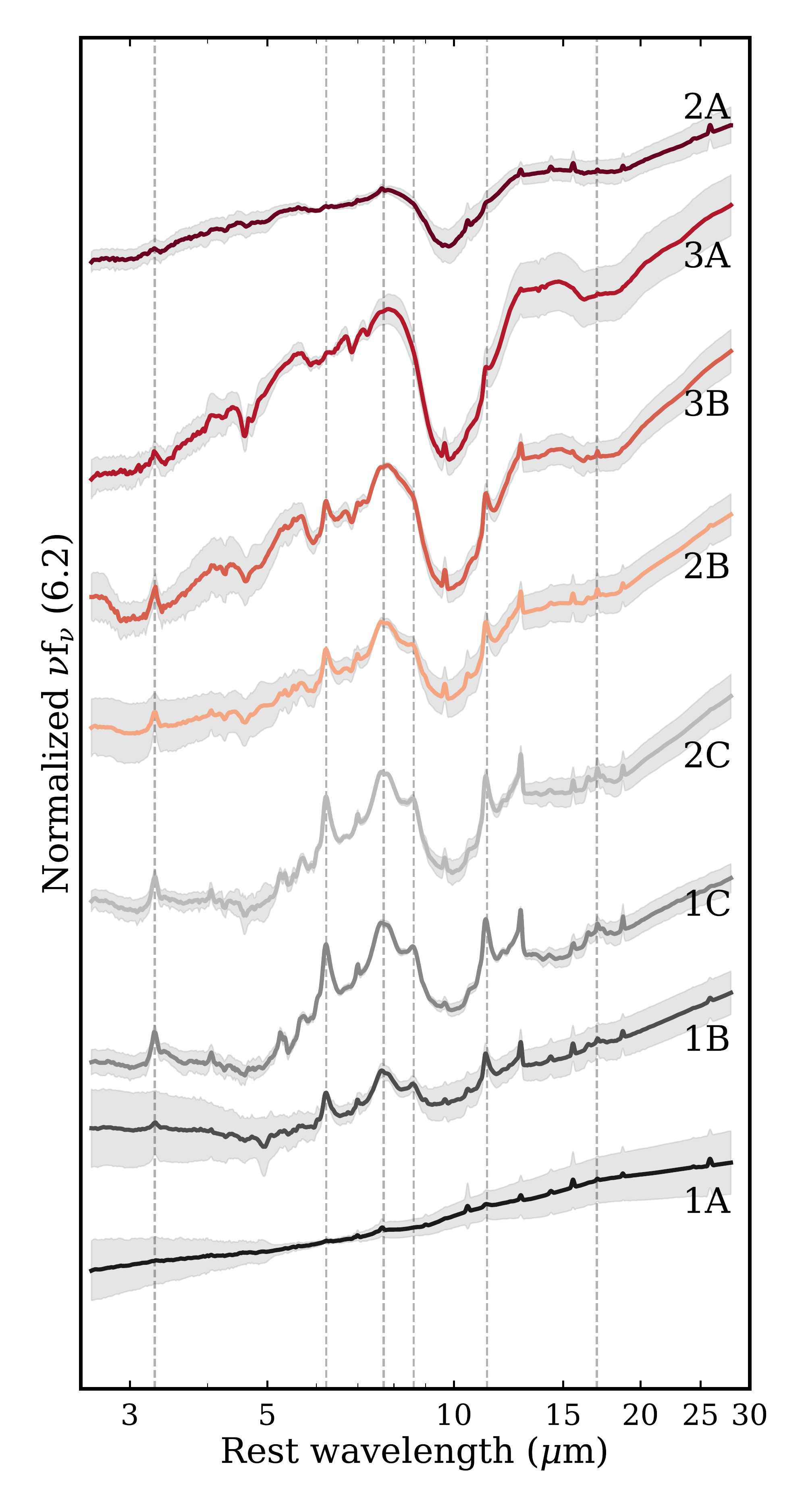}
    \caption{Galaxy templates of the eight classes defined in Fig.~\ref{fig:sample_selection}(a). All spectra are plotted in log scale and normalized to 6.2\um. The gray areas indicate the 1$\sigma$ dispersion. Vertical dashed lines denote the central wavelengths of the main PAH emission features at 3.3, 6.2, 7.7, 8.6, 11.3, and 17\um. Most galaxies in our moderate attenuation, bright-PAH sample are found within classes 1B, 2B, 1C, and 2C. Spectra are color-coded correspondingly to classes in Fig.~\ref{fig:sample_selection}(a). Template of class 3C is not provided since only one galaxy (NGC 4945) belongs to this class. The number of sources averaged for the templates range from 10 in 2A to 116 in 1A.}
    \label{fig:template}
\end{figure}

From the complete cross-archival sample of galaxies with compatible \Spitzer/IRS and \AKARI/IRC coverage (378 objects), we construct average spectral templates covering the complete spectral range of 2.6--28\um\ from the \citet{Spoon2007} galaxy classification
discussed in Fig.~\ref{fig:sample_selection}(a). Galaxy spectra in the eight classes are presented in Fig.~\ref{fig:template}, with each produced by the median of high-quality spectra (S/N $\geq$ 60 at 3.29\um) in the respective class.

The bulk of galaxies in our bright-PAH sample lies within classes  1B, 1C, 2B, or 2C, with more than half in 1C, as shown in Fig.~\ref{fig:sample_selection}(a). The 1C spectrum in Fig.~\ref{fig:template} is a typical star-forming galaxy, showing prominent PAH emission, with weak to moderate 9.7\um\ silicate absorption. As the spectrum shifts from class 1C to 2B, the 9.7\um\ silicate absorption becomes more prominent and the PAH emission bands become more suppressed. For classes 3A and 3B, the absorption features, including H$_{2}$O, CO$_{2}$, CO, and silicates all reach their maximum depths, while most PAH features are diminished. But as an exception, the 3.3\um\ PAH band remains apparent even when other strong PAH bands at 6.2, 7.7, and 8.6\um\ disappear in the 3A class. The behavior of the 3.3\um\ PAH in the most highly obscured systems will be a topic for a future study, but the impact of attenuation on this shortest wavelength PAH feature is considerable even at moderate levels of obscuration (see \S~\ref{subsec:PAH2Geoms}). The digital format of these templates can be found in CDS format in the online version of this publication.

\subsection{\PAHFIT\ Result}
\label{sec:result_pahfit}
In Fig.~\ref{fig:pahfit_example}, we demonstrate the capabilities of our enhanced \PAHFIT\ model on the composite, high-quality 1C spectrum (see Fig.~\ref{fig:template}). The data are shown in squares overlaid by the fitted profile in green. Dust features (blue) are modeled using Drude profile and emission lines (purple) are modeled with Gaussians. Newly added features are labelled. The \PAHFIT\ results of the bright-PAH sample in both the mixed and obscured continuum geometries can be found in Table~\ref{tab:PAHFIT_result_mx} and \ref{tab:PAHFIT_result_oc}, respectively.

\subsection{The 3.3\um\ PAH} \label{sec:3.3PAH}
Thanks to \Spitzer, studies of the PAH bands at wavelengths $>$ 5\um\ abound. However, investigations of 3.3\um\ PAH has been limited due to the lack of \Spitzer/IRS wavelength coverage, the hindrance by the atmospheric background emission for ground observation, and the low sensitivity at 3\um\ in \textit{ISO}. To put the 3.3\um\ PAH in context, we use the \AKARI-\Spitzer\ combined spectra obtained from the 113 galaxies in our bright-PAH sample and investigate how the 3.3\um\ PAH measurement varies under the assumption of different geometries, and as infrared luminosity changes.

\subsubsection{3.3\um\ PAH and Attenuation Geometry} \label{subsec:PAH2Geoms}
Among all the PAH emission features, the 3.3\um\ PAH has the shortest wavelength, and it sets itself apart from other prominent PAH emissions at longer wavelengths. Unlike the PAH band complexes at 6--9\um, for which measuring band strengths can be challenging due to the blending of multiple components, measuring the strength of the 3.3\um\ PAH seems to be straightforward. However, the coincidence with broad 3.05\um\ H$_{2}$O ice absorption \citep{Gibb2004}, together with the 3.4\um\ aliphatic feature, if present, affects the 3.3\um\ PAH measurement. In particular, the H$_{2}$O ice feature has an impact since its optical depth is not negligible even for some star-forming galaxies ---  the average H$_{2}$O opacity for our bright-PAH sample is $\tau_\mathrm{H_{2}O}=$0.20. Note that in this study, we only consider the 3.4\um\ aliphatic feature in emission, even though absorption feature can also be found in the ISM \citep{Pendleton2002, Doi2019}. For most standalone \AKARI/IRC studies, the 3.3\um\ PAH strength is measured by placing a linear continuum and integrating the band atop \citep{Imanishi2008, Imanishi2010, Ichikawa2014, Inami2018}. The result produced by this approach only takes into account the attenuation within the 2.5--5\um\ spectral range, but not the overall MIR attenuation predominately constrained by the 9.7\um\ absorption arising from the amorphous silicates and raised steeply towards the NIR at $\sim$\,3\um.

\begin{longrotatetable}

\begin{deluxetable*}{lllrrrrrrll}
\tablecaption{\PAHFIT\ Result of Bright-PAH Sample in the \emph{Mixed} Geometry}
\tablewidth{700pt}
\tabletypesize{\scriptsize}
\tablehead{
\colhead{Name} & 
\colhead{3.3 $\mu m$} & \colhead{3.4 $\mu m$} & 
\colhead{3.47 $\mu m$} & \colhead{6.2 $\mu m$} & 
\colhead{11.3 $\mu m$ complex} & \colhead{17 $\mu m$ complex} & \colhead{$\Sigma$ PAH} & \colhead{\NeII} & \colhead{\NeIII} & \colhead{SFR$_\mathrm{Ne}$}\\
\colhead{(1)} & \colhead{(2)} & 
\colhead{(3)} & \colhead{(4)} & 
\colhead{(5)} & \colhead{(6)} & 
\colhead{(7)} & \colhead{(8)} & 
\colhead{(9)} & \colhead{(10)} & \colhead{(11)}\\
} 
\startdata
                 NGC 0023 &    15.18$\pm$0.45 &         $<$1.62 &   6.09$\pm$0.48 &  11.08$\pm$0.04 &  12.82$\pm$0.03 &   6.23$\pm$0.04 &     10.61$\pm$0.02 &    15.45$\pm$0.09 &    2.53$\pm$0.07 &   19.58 \\
            ARP 256 NED01 &    42.33$\pm$1.56 &   5.07$\pm$1.19 &   7.64$\pm$0.76 &   18.65$\pm$0.1 &  15.89$\pm$0.09 &  10.01$\pm$0.17 &     15.84$\pm$0.05 &    38.51$\pm$0.27 &    9.24$\pm$0.26 &   51.08 \\
                 NGC 0232 &     31.9$\pm$1.53 &         $<$2.16 &   5.96$\pm$0.64 &  18.24$\pm$0.06 &  19.09$\pm$0.07 &  11.17$\pm$0.11 &     18.17$\pm$0.03 &    26.11$\pm$0.13 &    4.07$\pm$0.14 &   32.92 \\
  2MASX J00480675-2848187 &  175.03$\pm$13.59 &   8.12$\pm$2.29 &        $<$15.09 &   53.0$\pm$0.72 &    50.1$\pm$0.9 &  22.49$\pm$1.91 &     49.92$\pm$0.36 &   110.75$\pm$2.79 &    23.3$\pm$2.01 &  144.39 \\
     MCG +12-02-001 NED02 &    28.94$\pm$0.76 &   3.94$\pm$0.22 &   5.32$\pm$0.25 &   14.7$\pm$0.05 &  11.54$\pm$0.03 &   6.84$\pm$0.07 &     12.58$\pm$0.02 &     34.08$\pm$0.1 &    8.81$\pm$0.15 &   45.68 \\
          IRAS 01003-2238 &    44.77$\pm$3.62 &             ... &             ... &  19.08$\pm$1.75 &             ... &     2.6$\pm$0.0 &      45.57$\pm$0.8 &    57.95$\pm$6.16 &   25.33$\pm$5.24 &   85.29 \\
           MCG -03-04-014 &    75.87$\pm$2.23 &  10.42$\pm$0.61 &   14.5$\pm$0.77 &  53.29$\pm$0.18 &  46.11$\pm$0.19 &  22.41$\pm$0.36 &     45.86$\pm$0.09 &    79.09$\pm$0.49 &    9.62$\pm$0.36 &   97.53 \\
     ESO 244- G 012 NED02 &    29.54$\pm$1.58 &   3.12$\pm$0.99 &    5.77$\pm$0.7 &  17.91$\pm$0.12 &  16.22$\pm$0.21 &    8.7$\pm$0.27 &     16.31$\pm$0.07 &    49.84$\pm$0.56 &    9.46$\pm$0.34 &   64.19 \\
             CGCG 436-030 &    84.29$\pm$2.89 &   6.44$\pm$1.48 &             ... &  23.86$\pm$0.13 &  17.79$\pm$0.13 &  15.53$\pm$0.29 &     23.64$\pm$0.07 &    49.95$\pm$0.73 &    10.92$\pm$0.4 &   65.44 \\
           ESO 353- G 020 &    21.74$\pm$0.89 &         $<$1.86 &   3.93$\pm$0.43 &   10.6$\pm$0.05 &  10.33$\pm$0.08 &   5.93$\pm$0.06 &     10.06$\pm$0.02 &     13.9$\pm$0.08 &    1.46$\pm$0.05 &   16.95 \\
                 NGC 0633 &     9.91$\pm$0.85 &         $<$1.32 &    3.3$\pm$0.39 &   6.36$\pm$0.03 &   6.59$\pm$0.02 &   3.36$\pm$0.04 &      6.08$\pm$0.02 &    10.46$\pm$0.08 &    0.97$\pm$0.04 &   12.64 \\
       III Zw 035 &    12.76$\pm$1.15 &             ... &         $<$1.56 &   3.91$\pm$0.13 &    5.7$\pm$0.18 &    3.06$\pm$0.2 &      5.22$\pm$0.08 &     6.13$\pm$0.45 &    0.84$\pm$0.19 &    7.64 \\
                 NGC 0695 &     68.6$\pm$1.69 &   6.67$\pm$1.93 &  18.39$\pm$1.33 &  53.74$\pm$0.26 &  53.63$\pm$0.16 &   28.9$\pm$0.22 &     49.43$\pm$0.11 &    67.03$\pm$0.59 &   16.45$\pm$0.35 &   89.18 \\
  2MASX J01515140-1830464 &  183.61$\pm$23.93 &             ... &        $<$35.64 &  60.15$\pm$1.93 &  60.82$\pm$2.46 &         $<$19.8 &     65.63$\pm$1.15 &  153.25$\pm$10.36 &         $<$24.87 &     ... \\
                UGC 01385 &     9.59$\pm$0.49 &         $<$0.87 &   2.26$\pm$0.23 &    5.5$\pm$0.02 &   5.71$\pm$0.02 &   3.88$\pm$0.04 &       5.3$\pm$0.01 &    12.99$\pm$0.06 &    1.64$\pm$0.06 &   16.06 \\
                 NGC 0838 &    18.81$\pm$0.56 &   1.59$\pm$0.18 &   4.22$\pm$0.21 &   13.7$\pm$0.09 &  11.89$\pm$0.07 &   6.26$\pm$0.08 &     11.79$\pm$0.04 &    16.75$\pm$0.19 &    3.94$\pm$0.13 &   22.16 \\
                UGC 01845 &     26.0$\pm$1.11 &         $<$1.41 &   6.47$\pm$0.37 &  15.78$\pm$0.06 &  13.77$\pm$0.09 &   7.07$\pm$0.06 &     14.01$\pm$0.02 &     17.0$\pm$0.09 &    2.06$\pm$0.06 &   20.96 \\
                 NGC 0992 &    16.51$\pm$0.24 &   2.54$\pm$0.16 &   4.58$\pm$0.19 &   13.7$\pm$0.03 &  12.28$\pm$0.04 &   7.12$\pm$0.05 &     11.83$\pm$0.01 &    17.05$\pm$0.08 &    3.66$\pm$0.05 &   22.29 \\
                UGC 02238 &      67.7$\pm$2.1 &   5.52$\pm$0.66 &  12.62$\pm$0.84 &  39.69$\pm$0.18 &  32.98$\pm$0.22 &   18.34$\pm$0.2 &     34.89$\pm$0.08 &    39.14$\pm$0.32 &     5.45$\pm$0.3 &   48.83 \\
          UGC 02369 NED01 &    25.64$\pm$2.28 &         $<$3.87 &         $<$3.51 &  17.11$\pm$0.21 &  12.29$\pm$0.07 &   9.98$\pm$0.18 &     14.21$\pm$0.06 &    36.49$\pm$0.25 &    5.87$\pm$0.24 &   46.17 \\
 SDSS J032322.86-075615.2 &  116.03$\pm$23.79 &             ... &             ... &  53.56$\pm$1.53 &  54.51$\pm$1.87 &   23.64$\pm$5.7 &     51.22$\pm$0.92 &   121.33$\pm$7.22 &         $<$24.69 &     ... \\
  2MASX J03544214+0037033 &   64.25$\pm$15.76 &             ... &        $<$32.58 &  40.92$\pm$1.73 &  29.83$\pm$2.07 &        $<$13.41 &     36.91$\pm$0.85 &    69.34$\pm$7.64 &         $<$20.82 &     ... \\
  2MASX J04121945-2830252 &   109.05$\pm$8.58 &             ... &             ... &  49.95$\pm$1.45 &  46.73$\pm$1.51 &   39.12$\pm$2.0 &     41.72$\pm$0.64 &   109.37$\pm$3.68 &  107.12$\pm$3.19 &  202.85 \\
                UGC 02982 &    23.37$\pm$0.43 &   2.29$\pm$0.31 &    5.6$\pm$0.42 &  18.25$\pm$0.06 &  17.02$\pm$0.06 &   9.41$\pm$0.05 &     17.44$\pm$0.03 &    20.71$\pm$0.14 &    3.76$\pm$0.06 &   26.54 \\
                 NGC 1614 &    40.64$\pm$1.29 &   2.99$\pm$0.68 &   9.03$\pm$0.54 &  28.22$\pm$0.11 &  21.86$\pm$0.08 &  12.83$\pm$0.13 &     26.25$\pm$0.05 &     70.91$\pm$0.0 &   12.91$\pm$0.18 &   90.89 \\
\enddata
\vspace{0.cm}
\tablecomments{Table \ref{tab:PAHFIT_result_mx} is published in its entirety in the machine-readable format. 3$\sigma$ upper limits are prefaced with `$<$'. Non-detections are designated as `...'.\\
(1): Galaxy name\\
(2)-(4): Integrated luminosity of emission features in the \AKARI/IRC wavelength coverage returned by \PAHFIT, in units of 10$^{7}$\,\Lsun.\\
(5)-(7): Integrated luminosity of primary PAH emissions, in units of 10$^{8}$\,\Lsun.\\ 
(8): Total integrated PAH luminosity, in units of 10$^{9}$\,\Lsun.\\
(9)-(10): Integrated luminosity of Neon line emissions, in units of 10$^{7}$\,\Lsun.\\
(11): Neon line inferred star formation rate, in units of M$_{\odot}$/yr. Only provided are galaxies with $>3\sigma$ detections in both \NeII\ and \NeIII.\\
}
\label{tab:PAHFIT_result_mx}
\end{deluxetable*}
\end{longrotatetable}

\begin{longrotatetable}
\begin{deluxetable*}{lllrrrrrrll}
\tablecaption{\PAHFIT\ Result of Bright-PAH Sample in the \emph{Obscured Continuum} Geometry}
\tablewidth{700pt}
\tabletypesize{\scriptsize}
\tablehead{
\colhead{Name} & 
\colhead{3.3 $\mu m$} & \colhead{3.4 $\mu m$} & 
\colhead{3.47 $\mu m$} & \colhead{6.2 $\mu m$} & 
\colhead{11.3 $\mu m$ complex} & \colhead{17 $\mu m$ complex} & \colhead{$\Sigma$ PAH} & \colhead{\NeII} & \colhead{\NeIII} & \colhead{SFR$_\mathrm{Ne}$}\\
\colhead{(1)} & \colhead{(2)} & 
\colhead{(3)} & \colhead{(4)} & 
\colhead{(5)} & \colhead{(6)} & 
\colhead{(7)} & \colhead{(8)} & 
\colhead{(9)} & \colhead{(10)} & \colhead{(11)}\\
} 
\startdata
                 NGC 0023 &   14.19$\pm$0.47 &  2.38$\pm$0.33 &   4.57$\pm$0.47 &  11.23$\pm$0.03 &  12.76$\pm$0.02 &   6.54$\pm$0.04 &      10.7$\pm$0.02 &   15.46$\pm$0.09 &    2.67$\pm$0.07 &    19.7 \\
            ARP 256 NED01 &   31.17$\pm$1.09 &  4.06$\pm$0.94 &   6.15$\pm$0.61 &  17.46$\pm$0.09 &  14.19$\pm$0.06 &   9.15$\pm$0.14 &     14.85$\pm$0.04 &   37.11$\pm$0.26 &    8.73$\pm$0.25 &   49.09 \\
                 NGC 0232 &   21.66$\pm$0.97 &        $<$1.56 &    4.3$\pm$0.46 &  16.53$\pm$0.05 &  15.74$\pm$0.04 &   9.55$\pm$0.08 &     16.41$\pm$0.03 &    24.5$\pm$0.12 &    3.59$\pm$0.15 &   30.71 \\
  2MASX J00480675-2848187 &   75.42$\pm$5.38 &  4.16$\pm$1.15 &         $<$7.89 &  43.39$\pm$0.48 &  35.43$\pm$0.32 &  16.72$\pm$1.31 &     40.95$\pm$0.28 &   98.26$\pm$2.42 &    18.69$\pm$1.7 &  126.57 \\
     MCG +12-02-001 NED02 &   21.46$\pm$0.57 &  3.25$\pm$0.17 &   4.45$\pm$0.22 &  13.96$\pm$0.05 &  10.63$\pm$0.03 &   6.45$\pm$0.07 &     11.98$\pm$0.02 &    33.17$\pm$0.1 &    8.44$\pm$0.14 &   44.36 \\
          IRAS 01003-2238 &   28.85$\pm$2.05 &            ... &             ... &  17.27$\pm$1.55 &             ... &         $<$2.61 &     41.36$\pm$0.77 &   54.13$\pm$5.74 &   21.87$\pm$4.74 &   78.37 \\
           MCG -03-04-014 &   63.05$\pm$1.54 &  9.14$\pm$0.51 &  13.23$\pm$0.65 &  51.17$\pm$0.14 &  43.28$\pm$0.13 &  21.43$\pm$0.33 &     44.16$\pm$0.08 &   77.48$\pm$0.48 &    9.26$\pm$0.35 &   95.42 \\
     ESO 244- G 012 NED02 &   24.43$\pm$1.11 &  2.73$\pm$0.86 &   5.05$\pm$0.59 &  17.12$\pm$0.11 &   14.93$\pm$0.1 &   8.11$\pm$0.23 &     15.63$\pm$0.06 &   48.56$\pm$0.52 &    9.08$\pm$0.33 &   62.43 \\
             CGCG 436-030 &   27.91$\pm$0.93 &   2.3$\pm$0.55 &             ... &   18.32$\pm$0.1 &   10.3$\pm$0.07 &  10.34$\pm$0.17 &     17.89$\pm$0.05 &    41.47$\pm$0.6 &    8.16$\pm$0.31 &   53.63 \\
           ESO 353- G 020 &   10.91$\pm$0.48 &        $<$1.05 &   1.96$\pm$0.25 &   8.84$\pm$0.03 &   7.15$\pm$0.02 &   4.41$\pm$0.03 &      8.39$\pm$0.01 &   12.31$\pm$0.06 &    1.18$\pm$0.04 &   14.91 \\
                 NGC 0633 &     9.42$\pm$0.8 &        $<$1.29 &   3.43$\pm$0.39 &   6.35$\pm$0.03 &   6.59$\pm$0.02 &   3.32$\pm$0.04 &      6.08$\pm$0.02 &   10.48$\pm$0.08 &    0.95$\pm$0.04 &   12.65 \\
       III Zw 035 &    5.16$\pm$0.46 &            ... &         $<$0.72 &   3.06$\pm$0.09 &   3.47$\pm$0.09 &   2.05$\pm$0.13 &       4.0$\pm$0.06 &     5.2$\pm$0.38 &              ... &    5.82 \\
                 NGC 0695 &   63.55$\pm$1.25 &  9.41$\pm$1.03 &  14.07$\pm$1.18 &  54.96$\pm$0.25 &  52.98$\pm$0.16 &   30.5$\pm$0.24 &      49.7$\pm$0.11 &   66.91$\pm$0.59 &   16.83$\pm$0.35 &   89.34 \\
  2MASX J01515140-1830464 &   64.86$\pm$8.31 &            ... &        $<$17.85 &  50.06$\pm$1.97 &   42.34$\pm$1.4 &        $<$17.16 &     57.24$\pm$1.05 &  137.35$\pm$8.98 &   25.64$\pm$7.74 &  176.55 \\
                UGC 01385 &    9.02$\pm$0.46 &        $<$0.87 &    2.2$\pm$0.23 &   5.48$\pm$0.02 &   5.67$\pm$0.02 &   3.81$\pm$0.04 &      5.23$\pm$0.01 &   12.96$\pm$0.06 &    1.58$\pm$0.06 &   15.99 \\
                 NGC 0838 &   17.48$\pm$0.52 &  1.57$\pm$0.17 &   4.11$\pm$0.21 &  13.62$\pm$0.08 &  11.89$\pm$0.07 &   6.26$\pm$0.08 &     11.76$\pm$0.04 &   16.75$\pm$0.19 &    3.94$\pm$0.13 &   22.15 \\
                UGC 01845 &   14.91$\pm$0.57 &        $<$0.93 &   4.11$\pm$0.23 &  13.63$\pm$0.03 &  10.38$\pm$0.02 &   5.54$\pm$0.03 &     12.19$\pm$0.02 &   15.52$\pm$0.07 &    1.78$\pm$0.05 &   19.05 \\
                 NGC 0992 &   13.94$\pm$0.19 &  2.22$\pm$0.13 &   4.23$\pm$0.17 &  13.13$\pm$0.03 &  11.29$\pm$0.02 &   6.24$\pm$0.04 &     11.48$\pm$0.01 &   16.62$\pm$0.08 &    3.61$\pm$0.05 &   21.76 \\
                UGC 02238 &   34.45$\pm$0.64 &  3.07$\pm$0.35 &   7.08$\pm$0.46 &  32.27$\pm$0.13 &  22.61$\pm$0.09 &  13.46$\pm$0.13 &     28.49$\pm$0.06 &   34.36$\pm$0.28 &    4.25$\pm$0.26 &   42.43 \\
          UGC 02369 NED01 &   25.36$\pm$2.21 &        $<$3.84 &         $<$3.18 &  17.07$\pm$0.09 &  12.29$\pm$0.07 &   9.97$\pm$0.18 &      14.2$\pm$0.05 &   36.49$\pm$0.25 &    5.87$\pm$0.24 &   46.17 \\
 SDSS J032322.86-075615.2 &  60.27$\pm$11.86 &            ... &             ... &  45.36$\pm$1.19 &  40.39$\pm$0.96 &   17.3$\pm$3.71 &      43.4$\pm$0.71 &  109.77$\pm$6.47 &         $<$19.38 &     ... \\
  2MASX J03544214+0037033 &   31.48$\pm$3.45 &            ... &             ... &  32.46$\pm$1.39 &  19.33$\pm$1.13 &   3.35$\pm$0.96 &      29.6$\pm$0.63 &    59.36$\pm$6.5 &         $<$16.77 &     ... \\
  2MASX J04121945-2830252 &   75.58$\pm$6.51 &            ... &             ... &  46.08$\pm$1.74 &  41.14$\pm$1.31 &   37.07$\pm$1.8 &      38.6$\pm$0.62 &  104.96$\pm$3.52 &   101.1$\pm$3.02 &  193.48 \\
                UGC 02982 &    22.2$\pm$0.43 &        $<$1.32 &    6.78$\pm$0.4 &  18.01$\pm$0.06 &  17.03$\pm$0.04 &   9.33$\pm$0.05 &     17.36$\pm$0.03 &   20.74$\pm$0.14 &    3.91$\pm$0.06 &   26.69 \\
                 NGC 1614 &   37.84$\pm$1.22 &  2.02$\pm$0.45 &   9.33$\pm$0.48 &  28.68$\pm$0.12 &  20.97$\pm$0.09 &  12.13$\pm$0.13 &     25.99$\pm$0.06 &    70.11$\pm$0.0 &   11.04$\pm$0.18 &    88.5 \\
\enddata
\vspace{0.cm}

\tablecomments{Columns are similar to Table \ref{tab:PAHFIT_result_mx}.
}
\label{tab:PAHFIT_result_oc}
\end{deluxetable*}
\end{longrotatetable}

\begin{figure}
 	\includegraphics[width=\columnwidth]{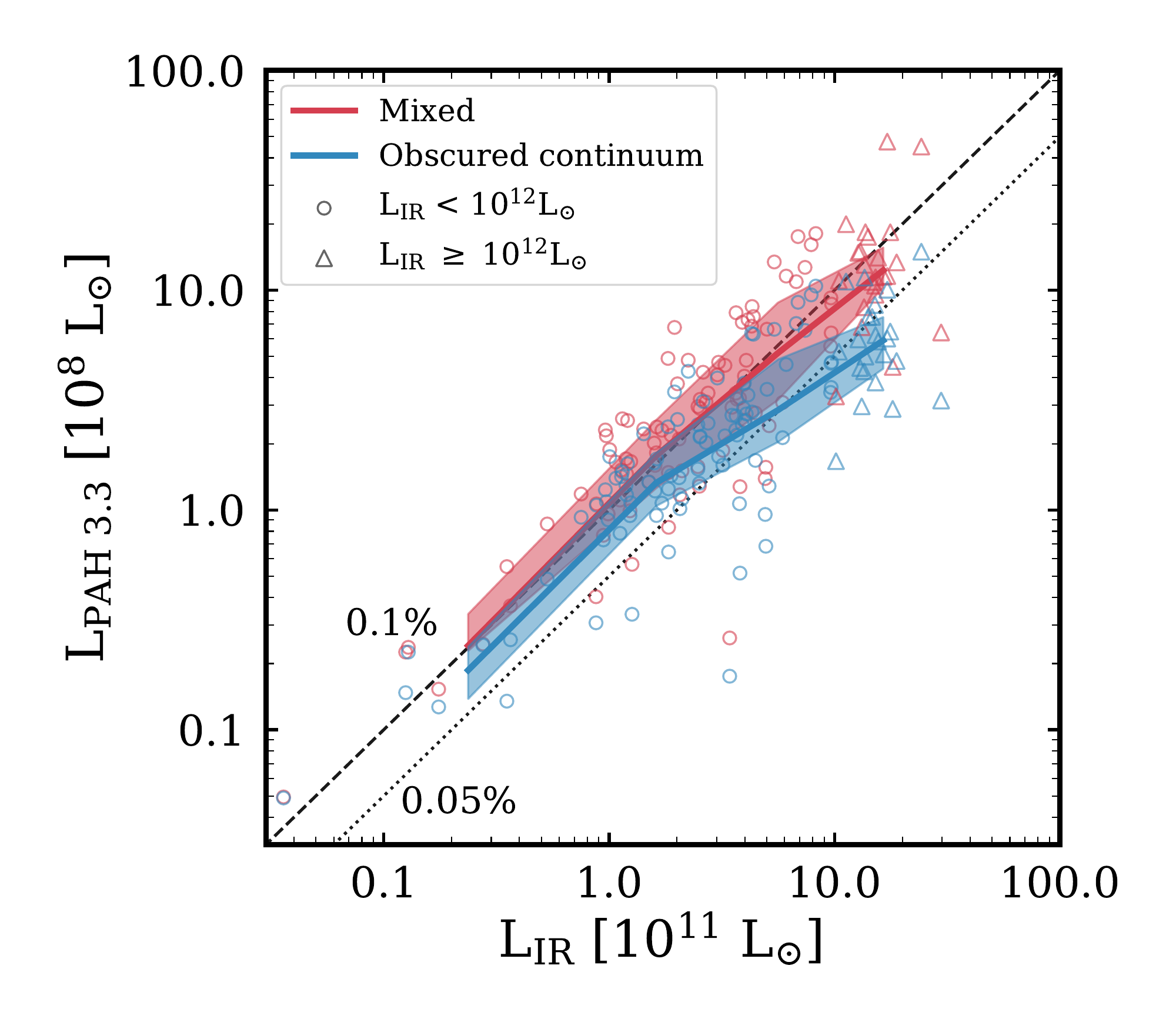}
    \caption{Comparison of the integrated infrared luminosity (L$\mathrm{_{IR}}$) and the 3.3\um\ PAH luminosity (L$\mathrm{_{PAH \ 3.3}}$) under the two different geometries shown in Fig.~\ref{fig:geometry}. Circles and triangles denote galaxies with \LIR $< 10^{12}$ L$_{\odot}$ and \LIR\ $\geq 10^{12}$ L$_{\odot}$ (ULIRGs), respectively. The solid lines are the median of the two binned distributions, with the 25--75\% ranges shaded. Both distributions fall between the dashed and dotted lines, representing \Lpahthree/\LIR\ = 0.1\% and 0.05\%, a typical range also found in \citet{Imanishi2010} and \citet{Y2013}. Note that II~Zw~40 is not included in this plot due to its exceptionally low 3.3\um\ PAH luminosity (\Lpahthree$\,\sim\,3 \times$10$^{5}$\Lsun).}
    \label{fig:L33_LIR}
\end{figure}

To explore how attenuation can alter inferred PAH strengths, we treated two different geometric scenarios as mentioned in \S~\ref{subsec:attenuation} -- fully mixed and obscured continuum. Given the steep rise of the extinction curve towards $\sim$\,3\um, we expect the 3.3\um\ PAH to suffer the greatest impacts of dust obscuration. Even with our bright-PAH sample, we found the 3.3\um\ PAH measured under the mixed geometry assumption can be boosted up to 5 times compared to the obscured continuum geometry. This ratio, r\,=\,\Lpahthree(mixed)/\Lpahthree(obscured continuum), can be approximated by a functional form: r\,=\,exp[(0.320\,$\pm$\,0.004)(\tauSi+\,\tauWater)], with optical depths measured in the mixed geometry. In our bright-PAH sample, the median of \tauSi+\,\tauWater\ is 1.31, indicating, in principle, a factor of 1.5 in between the measured strengths in the mixed and obscured continuum schemes, in which the limit of the ``true'' value is set.

\begin{figure*}
 	\includegraphics[width=1.0\textwidth]{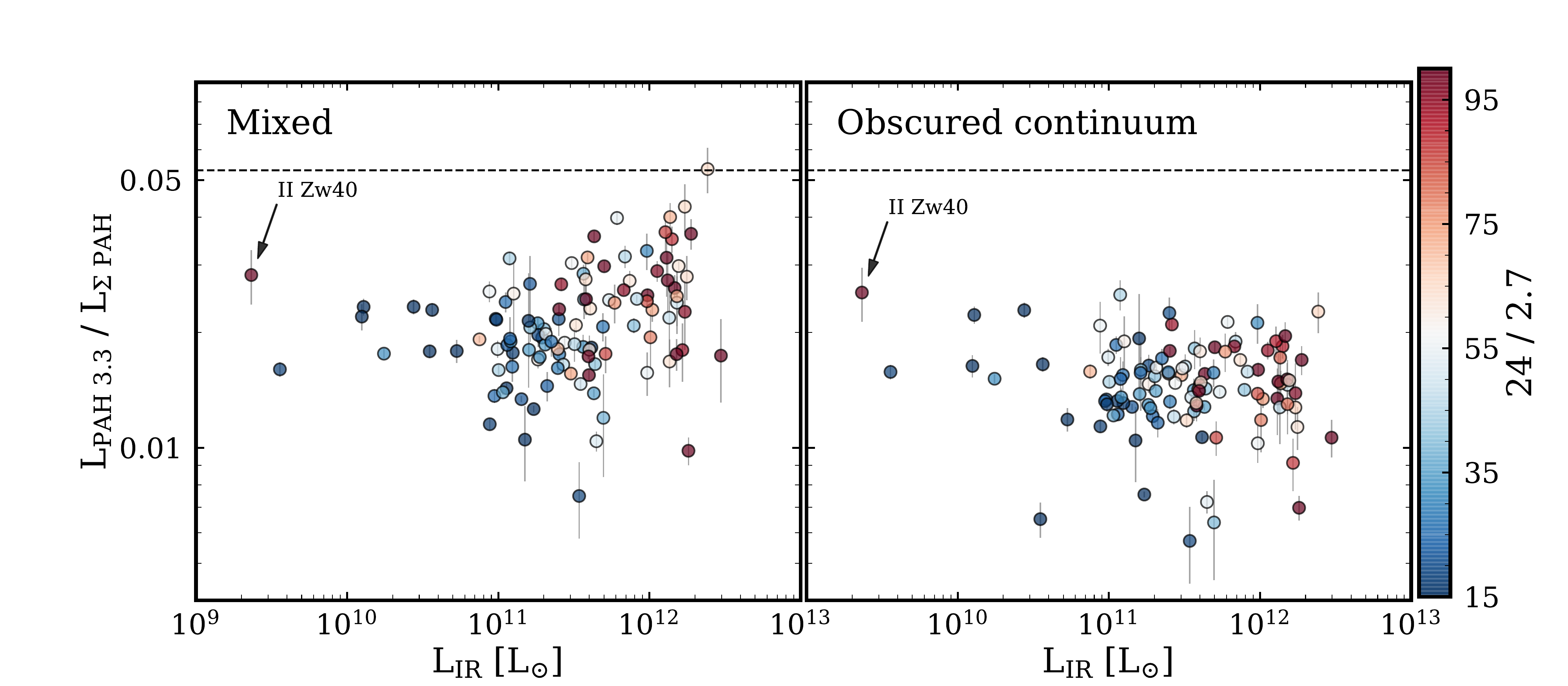}
    \caption{The strength of the fractional 3.3\um\ PAH against the infrared luminosity in two different adopted geometries. The fractional 3.3\um\ PAH strength increases with $\mathrm{L_{IR}}$ in the mixed geometry (\textit{left}) but remains constant in the obscured continuum geometry (\textit{right}). The data points are color-coded by the f$_{\nu}$(24\um)/f$_{\nu}$(2.7\um) ratio. The horizontal dashed lines indicate the nominal value adopted in the dust model of \citet{Draine2007}.}
    \label{fig:pah33_to_totpah}
\end{figure*}

\subsubsection{3.3\um\ PAH and \LIR}
The upper bound of the observed ratio of 3.3\um\ PAH to the integrated infrared luminosity (\Lpahthree/\LIR) for star-forming LIRGs has been found to be $\sim$ 10$^{-3}$ \citep{Mouri1990, Imanishi2010}. For ULIRGs (L$_\mathrm{IR} \geqslant 10^{12}$ \Lsun), however, a declining trend is seen, with  bolometric fractional luminosity of the uncorrected 3.3\um\ PAH (\Lpahthree/\LIR) dropping due to dust attenuation or the overall destruction of PAH molecules under intensive UV radiation \citep{Imanishi2010, Y2013}. Our results broadly agree with this trend. Fig.~\ref{fig:L33_LIR} shows binned 3.3\um\ PAH luminosity in two different geometries. In the obscured continuum geometry, the distribution lies in between the 0.1 and 0.05\% lines and gradually enters below the 0.05\% line at \LIR$\,\sim\,5 \times 10^{11}$ \Lsun. However, in the mixed geometry, rather than showing a 3.3\um\ PAH downturn, \Lpahthree/\LIR\ remains relatively constant over the whole range of \LIR. This is because the corrected strengths of the 3.3\um\ PAH in ULIRGs typically increase $\sim\,2 \times$ in the mixed geometry compared to the obscured continuum geometry. The difference between the two distributions suggests the high optical depth in ULIRGs likely causes the PAH deficit shown in the obscured continuum geometry.

\subsubsection{Fractional 3.3\um\ PAH}
Changes in PAH luminosity as a fraction of bolometric power can depend on both PAH power as well as variation in non-stochastic dust heating (AGN, etc.), but fractional PAH power relative to total PAH luminosity is more closely tied to properties of the PAH grains and their heating conditions. Fig.~\ref{fig:pah33_to_totpah} shows the relationship between the fractional 3.3\um\ PAH (\Lpahthree/\Ltotpah) and \LIR. The fractional 3.3\um\ PAH strength increases with \LIR\ in the fully mixed geometry. For galaxies with \LIR\,$<10^{12}$~\Lsun, the averaged fractional 3.3\um\ PAH is 2.06$\pm$0.59\%, whereas for \LIR\,$\geq\,10^{12}$~\Lsun, this value increases to 2.74$\pm$0.97\%. On the contrary, for the obscured continuum geometry, the fractional 3.3\um\ PAH power show little change with luminosity, with a similar average above and below $10^{12}$\Lsun\ of 1.50$\pm$0.38\%. We found galaxies with a steeper continuum slope ($f_\nu(24\,\mu m)/f_\nu(2.7\,\mu m)\,>\,60$) are associated with larger discrepancies among the two geometries because they are more likely dusty, obscured sources. The horizontal line in the figure indicates the nominal ratio of \Lpahthree\ to \Ltotpah\ in the \citet{Draine2007} dust model, in which the integrated strength of the 3.3\um\ PAH was increased by a factor of 2 to match the theoretical calculation. The fractional 3.3\um\ PAH obtained from the dust model is relatively high compared to our result, despite only some highly obscured ULIRGs are likely to have such a strong 3.3\um\ PAH emission measured in the mixed geometry. 

We note that low metallicity galaxy II~Zw~40 is a special case, which exhibits a steep 24/2.7 continuum slope, but the least \LIR\ among our sample. The \Lpahthree/\Ltotpah\ in II~Zw~40 in the mixed and obscured continuum geometries are 2.8\% and 2.5\%, respectively. It lies at the upper end of the distribution in the obscured continuum geometry, despite being being a blue compact, low-metallicity galaxy, where small grains are expected to be under-abundant (see \S~\ref{subsec:IIZw40}).

\subsection{3.3\um\ PAH as a Star Formation Rate Indicator}
\label{Sec:SFR}
One key to assessing the evolutionary history of galaxies is the link between their star formation rate and the physical condition in the ISM. Investigations into PAH emission as an indicator of star formation have been extensive \citep{Forster_Schreiber2004, Peeters2004a, Shipley2016, Xie2019}, but caution is needed when applying to low-metallicity galaxies, as PAH abundance is mostly driven by metallicity \citep{Calzetti2007}. By carefully treating galaxies with low-metallicity, \citet{Shipley2016} used a sample of star-forming galaxies and calibrated the extinction-corrected PAH bands at 6.2, 7.7, and 11.3\um\ to the SFR measured by H$\alpha$+24\um. More recently, \citet{Xie2019} provided a calibration using integrated 5--15\um\ PAH emission against the SFR measured by \NeII\ 12.8\um\ and \NeIII\ 15.6\um. These studies, however, used stronger PAH bands accessible from \textit{Spitzer} but not the 3.3\um\ PAH. Studies of the reliability of 3.3\um\ PAH as a SFR indicator are relatively scarce \citep{Kim2012, Y2013, Murata2017}. All these studies found 3.3\um\ PAH correlates with \LIR, but the correlation deviates at high luminosities (\LIR\ $\gtrsim 10^{12}$\Lsun), where \LIR\ may no longer be a direct SFR indicator but be contaminated by the contribution of AGN to the infrared luminosity. Alternatively, small PAHs traced by the 3.3\um\ PAH may be destroyed due to the energetic radiation field originating from the AGN in the host galaxy \citep{Genzel1998, Smith2007, ODowd2009, Diamond-Stanic2010, Wu2010}. In this study, we calibrate the SFR using the 3.3\um\ PAH and study how the results may differ for different assumptions of emission/attenuation geometry. It is important to note that this analysis only pertains to our pre-selected star-forming sample, and therefore does not pertain to galaxies that are metal-poor and/or AGN-dominated. For example, low-metallicity galaxy, II~Zw~40, shows a very different star-forming property as discussed in the next section.

\subsubsection{A New Calibration of SFR using 3.3\um\ PAH} \label{subsubsec:SFR_calibration}
Given the selection criteria described in \S~\ref{sec:sample_selection}, galaxies in our sample are primarily star-forming galaxies, with at most minor AGN contribution. To assess whether the 3.3\um\ PAH is a reliable SFR indicator is crucial and will benefit the study offered by the impending \JWST/NIRSpec with wavelength coverage ranging from 0.6--5\um\ with an unprecedented spectral resolution up to R=2700. 
Other accessible SFR indicators in our \mbox{2.5--38\um} range are the neon lines (\NeII\ 12.81\um\ and \NeIII\ 15.56 \um). To investigate the behavior of 3.3\um\ PAH emission as with SFR, we utilize the 102 galaxies from our bright-PAH subset with $>$3$\sigma$ detection in both \NeII\ and \NeIII. To estimate SFR inferred from \NeII\ and \NeIII\ (SFR$_\mathrm{Ne}$), we adopt the neon-based SFR relation of \citet{Zhuang2019}, an updated version of \citet{Ho2007} which explicitly includes the effect of metallicity. Here we adopt solar metallicity ($Z=Z_{\odot}$) for simplicity and normalize the SFR indicator to a Kroupa IMF \citep{Kroupa2001} with mass cutoffs at 0.1 and 100\,M$_{\odot}$.

\begin{figure}
 	\includegraphics[width=1.0\columnwidth]{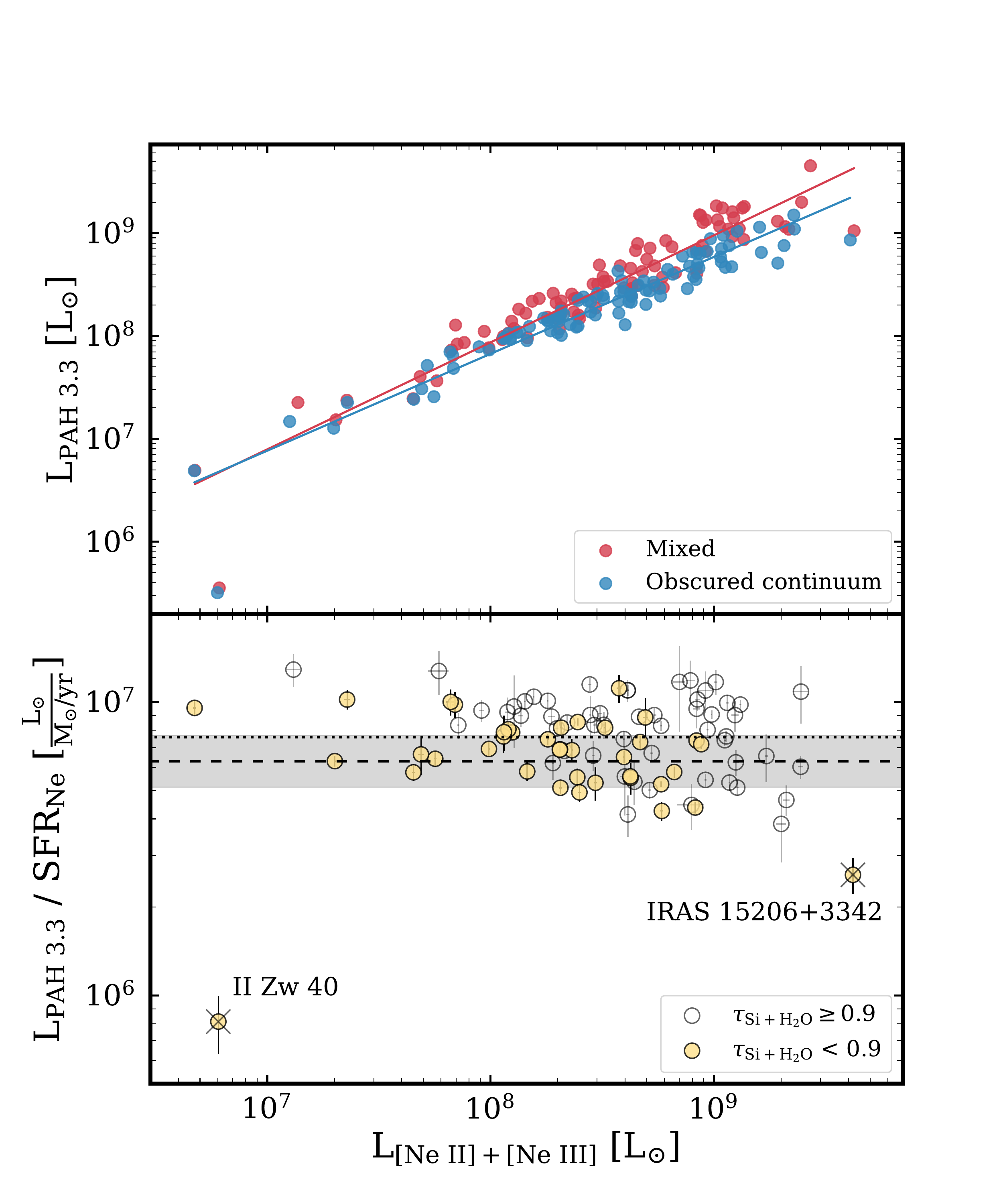}
    \caption{\textit{Top}: The relationship between the line and PAH luminosities \LNe\ and \Lpahthree, in both the mixed and obscured continuum geometries. The two color-coded lines are best fits for the corresponding geometry. \textit{Bottom}: The calibration of the SFR using 3.3\um\ PAH. Each data points is the weighted average of the values obtained from the mixed and obscured continuum geometries. Galaxies subject to only moderate optical depth (\tauSi+\tauWater $<$ 0.9) are shown as filled circles, and the dashed line and shading show the best weighted fit and 1$\sigma$ scatter (see Eq.~\ref{eqn:SFR}). The dotted line (at the upper end of the shading) indicates the average value from galaxies with (\tauSi+\tauWater $>$ 0.9). Two galaxies (II~Zw~40 and IRAS 15206+3342) exhibiting extremely low \Lpahthree/SNR ratios (below 3$\sigma$ from the calibrate correlation) are omitted from our calibration.}
    \label{fig:SFR_pah33_Ne}
\end{figure}

The top panel of Fig.~\ref{fig:SFR_pah33_Ne} shows the correlation between \LNe\ and \Lpahthree\ in the mixed and obscured continuum geometries, with scatters of 0.20 and 0.17 dex, respectively about the best fits. The bottom panel shows the relation of \LNe\ and the ratio of \Lpahthree\ to SFR$_\mathrm{Ne}$, calculated from the weighted average of the values obtained from both geometries. Aside from several outliers, the ratio of \Lpahthree\ to SFR$_\mathrm{Ne}$ shows no variation as \LNe\ spans $\sim$3 orders of magnitude, suggesting \Lpahthree\ and SFR$_\mathrm{Ne}$ are well correlated. The horizontal dashed line shows the best fit to the SFR calibration, which can be translated to a new calibration of SFR using the 3.3\um\ PAH:

\begin{equation}
\mathrm{log} \bigg(\mathrm{\frac{SFR}{M_{\odot}yr^{-1}} \bigg) = -(6.80 \pm 0.18) + log \bigg(\frac{L_{PAH \ 3.3}}{L_{\odot}} \bigg)}
\label{eqn:SFR}
\end{equation}

To avoid the effects of attenuation on this calibration, only galaxies with modest obscuration \tauSi+\tauWater\,$<$\,0.9 are used (see \S \ref{subsubsec:SFR_geometry} for detail). Galaxies with \tauSi+\tauWater\,$\geq$\,0.9 are not considered for SFR calibration since collectively they have the ratio of \Lpahthree/SFR$_\mathrm{Ne}$ that is $\sim$20\% higher (dotted line) than the value adopted. Also omitted are galaxies with peculiar low \Lpahthree/SFR$_\mathrm{Ne}$ ratios, below the 3$\sigma$ limit, namely II~Zw~40 and IRAS 15206+3342. We stress that our calibration applies only to star-forming, metal-rich galaxies. Metal-poor galaxies such as II~Zw~40 appears to have a ratio of \Lpahthree/SFR$_\mathrm{Ne}$ that is a factor of $\sim$\,8 lower than the average of our bright-PAH sample (see \S~\ref{subsec:IIZw40}). IRAS~15206+3342 lies very close to the AGN selection area in Fig.~\ref{fig:sample_selection}(b), and its nucleus is dominated by \HII\ regions and surrounded by shock-induced LINER ionization \citep{Arribas2002} that may suppress PAH emission. Clearly, PAHs lose their diagnostic power as a SFR indicator in environments such as low-metallicity or AGN/LINER-dominated regions.

\subsubsection{SFR Calibration in Different Geometries}
\label{subsubsec:SFR_geometry}
\begin{figure}
 	\includegraphics[width=1.0\columnwidth]{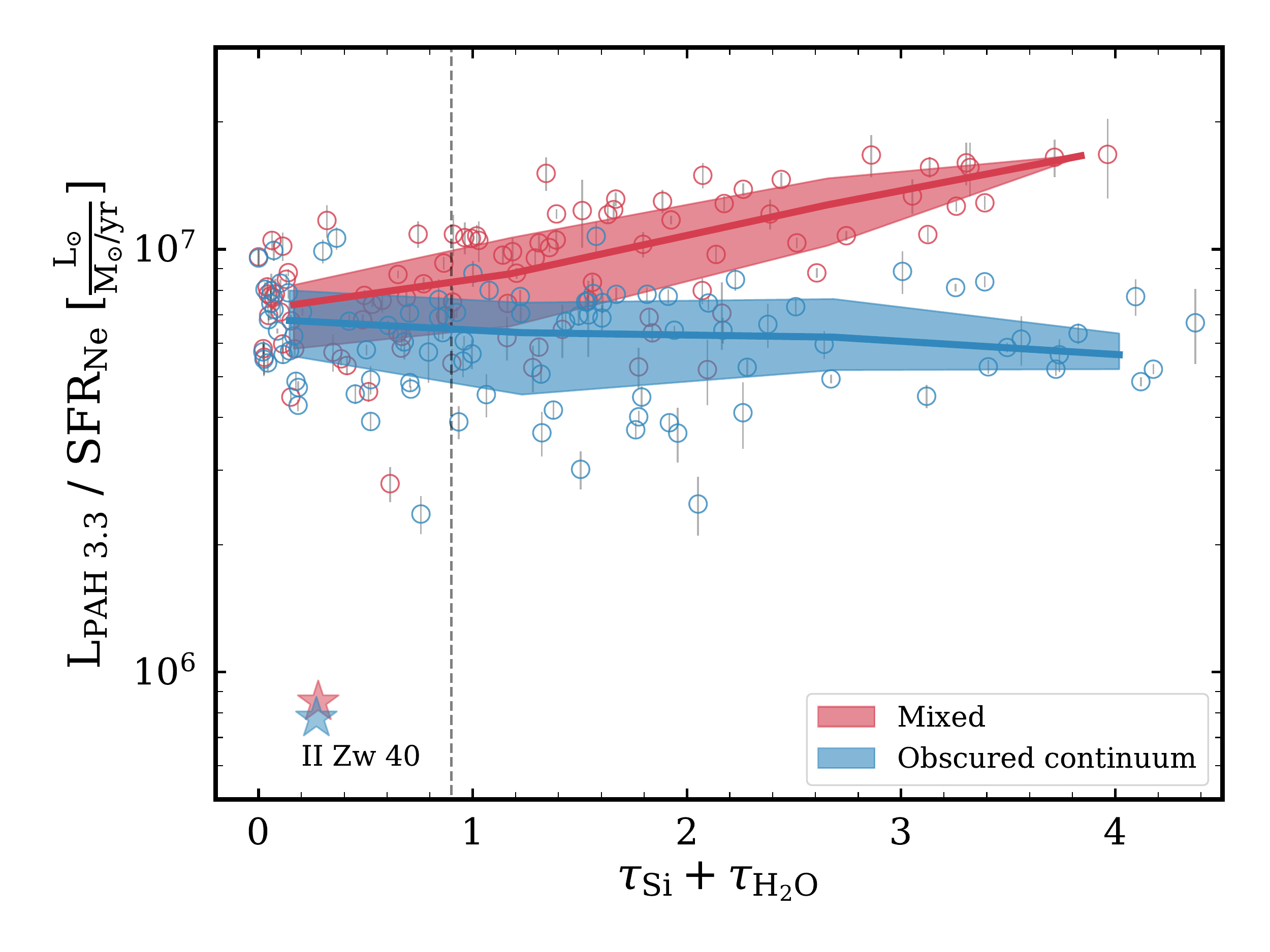}
    \caption{The dependence of the \Lpahthree/SFR ratio to the sum of silicate and water optical depths, in the mixed (red) and obscured continuum (blue) geometries. The trends are the median of the binned data with inner two quartiles shaded. Our SFR calibration is measured using galaxies with moderate obscuration ($\tau_\mathrm{Si}$+$\tau_\mathrm{H_{2}O}< 0.9$, left of the dashed line), where the two trends converge. The low metallicity outlier galaxy II~Zw~40 is indicated. }
    \label{fig:SFR_calibration}
\end{figure}

To explore how geometry affects 3.3\um\ PAH emission as a tracer for ongoing star formation, we examine how the ratio of \Lpahthree\ to SFR$_\mathrm{Ne}$ varies with respect to $\tau_\mathrm{Si}$+$\tau_\mathrm{H_{2}O}$ in both geometries. Fig.~\ref{fig:SFR_calibration} shows the two median trends in the binned data. Both trends exhibit no difference at low $\tau$ and progressively diverge as $\tau$ increases. The 3.3\um\ PAH luminosity per unit star formation rate stays rather consistent in the obscured continuum geometry within $0\,<\,\tau\,<\,4$, but monotonically increases together with $\tau$ in the mixed geometry. To test up to which $\tau$ value can the two populations be regarded as statistically similar, we perform a two-sample t-test, choosing a p-value significance level equal to 0.05, and find $\tau$=0.9 is the limit where the two populations can still count as identical. we therefore conclude the two trends exhibit no significant difference at $\tau <$ 0.9, indicated by the dashed line. Note that the ratio of \Lpahthree/\LIR\ for those galaxies with \tauSi+\tauWater $<$ 0.9 in the mixed and obscured continuum geometries are 0.11$\pm$0.05\% and 0.09$\pm$0.04\%, respectively. The offset between the two trends becomes as large as $\sim3\times$ in the most obscured cases in our bright PAH sample (\tauSi+\tauWater\,$\sim$\,4). Whether the two trends will continue diverging at even greater optical depths requires further study on more deeply embedded galaxies. 

\begin{figure*}
 	\includegraphics[width=1.0\textwidth]{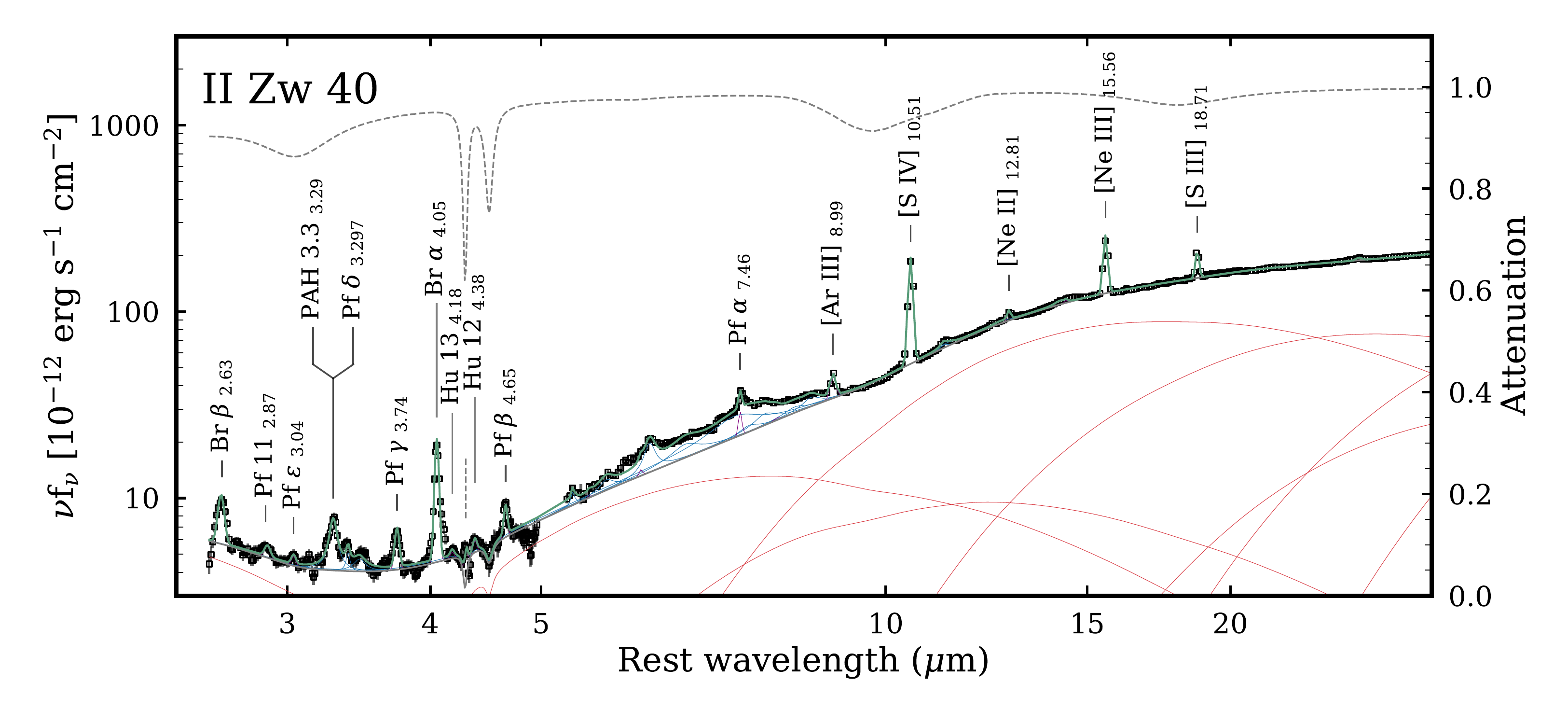}
    \caption{The extended \PAHFIT\ model of low metallicity galaxy II~Zw~40. The prominent 3.3\um\ PAH together with other emission lines and their corresponding wavelengths are labelled (with the dashed line indicating He I line at 4.30\um). Our 3.3\um\ PAH flux in this galaxy has been corrected by removing the contribution of the indistinguishable recombination line Pf~$\delta$ at 3.297\um\ that blends with the 3.3\um\ feature, contributing $\sim$\,21\% of the uncorrected 3.3\um\ PAH power.}
    \label{fig:IIZw40}
\end{figure*}

\subsection{Low Metallicity Case Study: II~Zw~40}
\label{subsec:IIZw40}
Large PAHs are expected to have a higher survival rate than small PAHs in a low-metallicity ISM \citep{Hunt2010}, since PAHs as a whole are subjected to enhanced photon interaction in the hard, intense interstellar radiation field which characterize as dust-poor environments \citep{Engelbracht2005, Galliano2005, Madden2006, Wu2006, Gordon2008}. However, a counter example has been found in SINGS galaxies, where \Lpahseventeen/\Ltotpah\ decreases with decreasing metallicity \citep{Smith2007}. Also, supporting such scenario, study by \citet{Sandstrom2012} found the size distribution of PAHs are apparently shifted to smaller sizes in the Small Magellanic Cloud relative to higher metallicity systems. These authors argued that the size shift towards smaller PAHs cannot be the result of ISM processing because small PAHs are more prone to photo-destruction under such conditions. Rather, PAHs at low-metallicity although experience more overall destruction, they have inherently smaller size distribution, making the PAH index $q_\mathrm{PAH}$, the fraction of PAHs relative to the total amount of dust, to have a positive correlation with the metallicity. Galaxies with 12+log$_{10}$(O/H)\,$<$\,8.1 typically have $q_\mathrm{PAH}$=1.0\%, a value that is at least three times less than that measured in galaxies with metallicity 12+log$_{10}$(O/H)\,$>$\,8.1 \citep{Drain2007b}. Given the many impact of low metal content on the conditions of the ISM, ascribing the PAH emission deficiency at low-Z to an exact mechanism may be challenging, but our sample galaxy II~Zw~40 provides a case for understanding the effect of metallicity on the formation/destruction of PAH molecules incorporating all the PAH emission bands.

II~Zw~40 is a nearby (D=10 Mpc), prototypical blue compact dwarf galaxy (BCD), containing a super star cluster and dominated by a giant \HII\ region with diameter $\sim$0.5 kpc \citep{Galliano2005}. Its metallicity is log(O/H)+12\,=\,8.09 \citep{Guseva2000}, or Z\,$\sim$\,1/4 Z$_{\odot}$. Fig.~\ref{fig:IIZw40} shows the extended \PAHFIT\ result of II~Zw~40. The overall PAH emission is highly suppressed in II~Zw~40, such that its ratio of \Ltotpah\ to \LIR\ is less than that measured in other sampled galaxies by a factor of $\sim$\,10. However, we find relatively strong 3.3\um\ PAH emission (\Lpahthree/\Ltotpah\,$\sim$\,2.8\%) with the absence of 17\um\ PAH, suggesting a size distribution shifted towards the smaller end.

Because the 5--9\um\ PAH features are weak compared to the continuum, and due to the limited sensitivity of PAHFIT to opacities \tauSi\,$\lesssim$\,0.2 \citep{Smith2007}, the total PAH power in II~Zw~40 is somewhat uncertain.  We estimate \tauSi\,$\simeq$\,0.2, consistent with the CO$_{2}$ and CO absorption bands near 4.5\um. As a simpler alternative with no dependence on attenuation or continuum modeling, we compare our result to a spline-based decomposition similar to the method employed in \citet{Peeters2002}. The estimated fractional 3.3\um\ PAH power with a spline-fitting technique is  \Lpahthree/\Ltotpah\,=\,6.6\%, suggesting that \Lpahthree/\Ltotpah\ from the full decomposition may be a lower limit. Additional evidence is found in the band ratios of isolated PAH features. Unlike ratios involving the integrated power of broad PAH features, whose strengths are often underestimated by spline-fitting methods  \citep[e.g.][see Table 6]{Smith2007}, the ratio between the 3.3 and 11.3\um\ PAH bands is robust against decomposition method, since these features are narrower and more isolated from other prominent PAH bands. The 3.3/11.3 ratio of 0.36 in II~Zw~40 is among the highest in the sample, and this ratio shows $<$\,20\% difference between the two estimation methods.  

II~Zw~40 has an unusually strong series of recombination lines. In addition to lines that have incorporated in \PAHFIT\ (see Table~\ref{Tab:additional_features}), five line features, including Pf 11, Pf$\epsilon$, Hu 13, He I, and Hu 12, are identified in the spectrum as shown in Fig.~\ref{fig:IIZw40}. The 3.3\um\ PAH strength may be overestimated since it is blended with the Pf$\delta$ line at 3.297\um. Because differentiating Pf$\delta$ line from the 3.3\um\ PAH is difficult with our spectral resolution, we correct the 3.3\um\ PAH strength in II~Zw~40 assuming the strength of Pf$\delta$ to be 69.71\% that of Pf$\gamma$ \citep{Hummer1987}. In such case, this results in a reduction of 3.3\um\ flux by 20.6\%. This is an unusually large correction compared to the typical value inferred from our 1C template spectrum of just 0.9\%. We note that this Pf$\delta$ correction of the 3.3\um\ PAH power has \emph{only} been applied to II~Zw~40 due to its exceptionally strong recombination lines exhibited in the spectrum. Also, this correction in II~Zw~40 has been applied throughout the results of this paper, and the corrected 3.3\um\ luminosity is in Table~\ref{tab:PAHFIT_result_mx} and Table~\ref{tab:PAHFIT_result_oc}.

\subsection{Aromatic Versus Aliphatic} \label{sec:aliphatic}
Understanding the abundance of aromatic and aliphatic components of the ISM can shed light on the evolution of dust grains and provide important constraints on dust models. The 3.3\um\ PAH emission is sometimes accompanied by a weak 3.4\um\ feature often attributed to the C-H vibrational modes in aliphatic hydrocarbons \citep{Joblin1996, Pendleton2002, Yang2016}. Anharmonicity of the aromatic C-H stretch has also been suggested \citep{Barker1987}, and more recently, based on high-resolution IR spectra of jet-cooled ``decorated'' PAHs, \citet{Maltseva2018} suggested that these hydrogenated and methylated PAHs are likely the carriers of the 3.4\um\ feature. In addition, the plateau feature centered at 3.47\um\ that extends up to $\sim$\,3.6\um\ (see Fig.~\ref{fig:pahfit_example}) is tightly correlated with 3.3\um\ PAH emission, suggesting the 3.47\um\ feature also arises from aromatic carriers \citep{Hammonds2015, Pilleri2015}. 

\begin{figure*}
 	\includegraphics[width=1.0\textwidth]{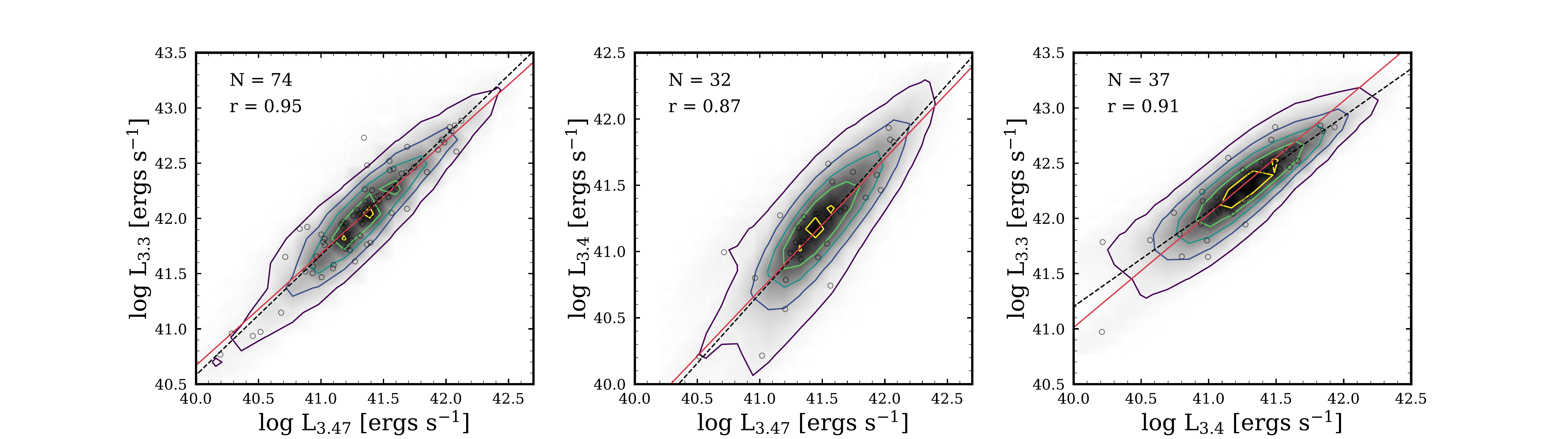}
    \caption{Correlations between the measured 3.3\um\ PAH, 3.4\um\ aliphatic emission, and 3.47\um\ plateau. Dispersions of the three correlations are 0.18, 0.25, and 0.19 dex, respectively. The dashed line denotes the fit of the data, with the red line indicating the proportional relationship. A tighter correlation between 3.3\um\ PAH and 3.47\um\ plateau is found, suggesting the plateau is dominated by the aromatic emission. The contours represent 90\%, 70\%, 50\%, 30\%, and 10\% enclosed probability.}
    \label{fig:aromatic_and_aliphatic}
\end{figure*}

We investigate the relationships among the 3.3\um\ aromatic emission, 3.4\um\ aliphatic emission, and the broad 3.47\um\ plateau in our bright-PAH sample. Figure~\ref{fig:aromatic_and_aliphatic} shows the correlations between all three combinations of the 3.3, 3.4, and 3.47\um\ band fluxes. In each panel, the correlation is calculated for galaxies with $>$\,3$\sigma$ detections in both bands. The sample size and the Pearson correlation coefficient are listed. We find the 3.3 and 3.47\um\ bands are best correlated (with r=0.95), supporting the claim that the plateau feature originates from aromatic bands \citet{Pilleri2015}. The 3.4\um\ aliphatic feature is relatively weak in our sample --- L$\mathrm{_{3.4}}$/L$\mathrm{_{3.3}}=0.10\,\pm\,0.04$ (either geometry).  This limits the sample size with robust 3.4\um\ detection given the sensitivity of \AKARI/IRC to $\sim$\,33\% of the galaxies in our bright-PAH sample. The 3.4\um\ aliphatic is less correlated with the 3.47\um\ plateau and the 3.3\um\ PAH band, with r=0.87 and r=0.91, respectively.

\begin{figure}
 	\includegraphics[width=1.0\columnwidth]{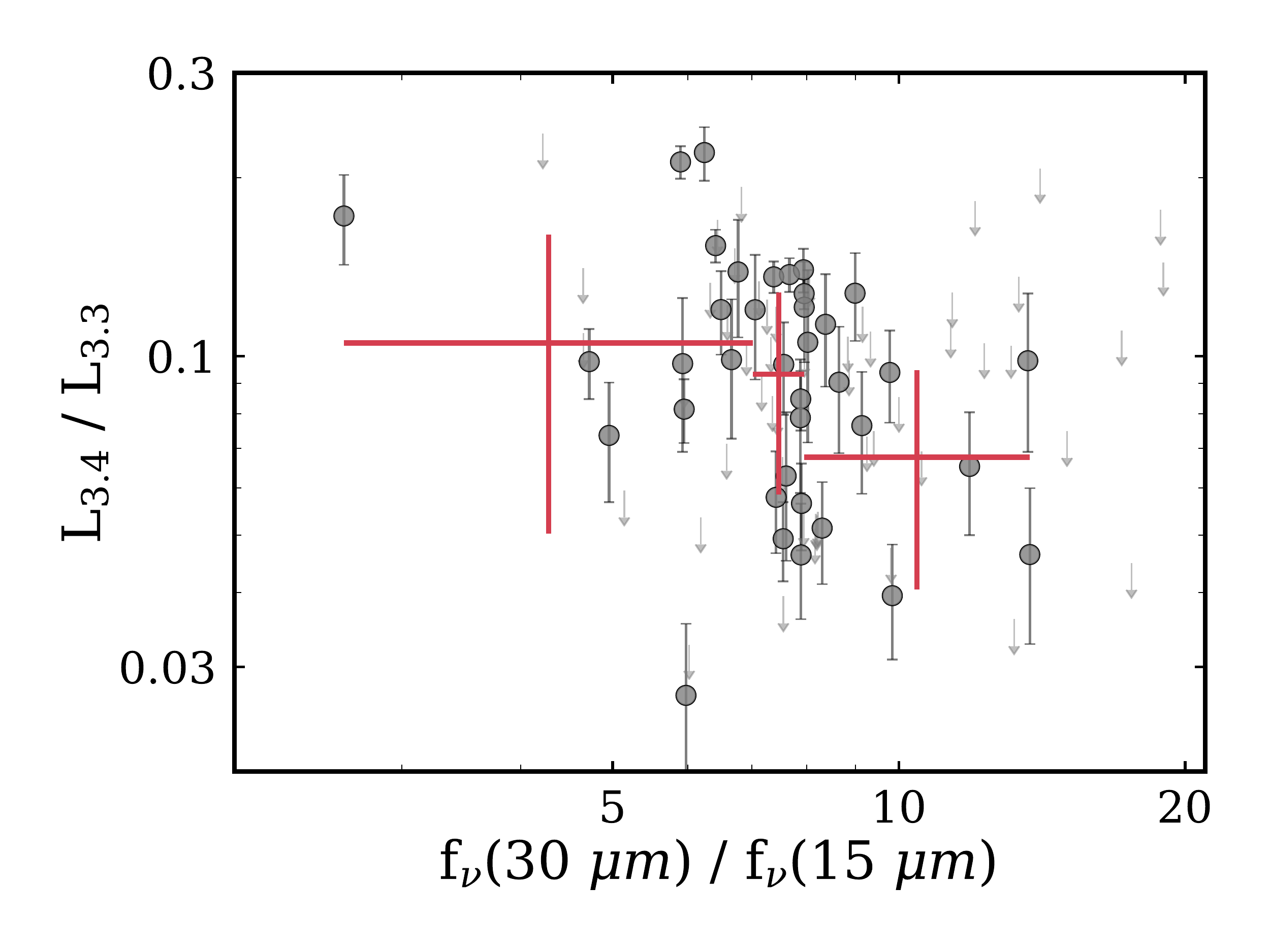}
    \caption{Correlation between L$_{3.4}$/L$_{3.3}$ (aliphatic to aromatic) and the monochromatic continuum at 15 and 30\um, which correlates with radiation field intensity. We divide our data into three equal-size sub-samples of the continuum slope, with the crosses indicating the weighted average. The error bars indicate the range of the continuum slope bins (horizontal) and the 1$\sigma$ scatter (vertical). We find a weak correlation with decreasing L$_{3.4}$/L$_{3.3}$ as  radiation fields increases. Galaxies with $<3\sigma$ detections of the L$_{3.4}$/L$_{3.3}$ ratio are displayed as limits.}
    \label{fig:34over33_30to15}
\end{figure}

\citet{Yamagishi2012} found that the 3.4/3.3 ratio decreases towards the center of starburst galaxy M82, suggesting that the relative abundance of the aliphatic band carriers drops towards regions with more intense star formation. The decline of the 3.4/3.3 ratio in PDRs is consistent with this picture, suggesting an efficient photochemical process leads to the destruction of the aliphatic subgroups present on the periphery of PAHs \citep{Joblin1996, Pilleri2015}. In Fig.~\ref{fig:34over33_30to15}, we compare the 3.4/3.3 ratio with the intensity of the radiation field inferred from the continuum slope between 15 and 30\um, where wavelengths are free from PAH and line emissions. The slope of the continuum is an often used surrogate for radiation field intensity, with steeper slopes implying stronger radiation in starburst galaxies \citep{Groves2008}. We find a mild correlation between the two. The 3.4/3.3 ratio ranges between $\sim$\,0.03--0.2 and drops by 40\% between the subsets with the weakest and strongest radiation fields. This agrees with the findings of \citet{Hammonds2015} and \citet{Pilleri2015}, who pointed out that aliphatics, with their fragile C-H bonds, are prone to efficient photo-destruction, resulting in higher relative abundance of the aromatics in intense radiation environments. 

\begin{figure*}
 	\includegraphics[width=\textwidth]{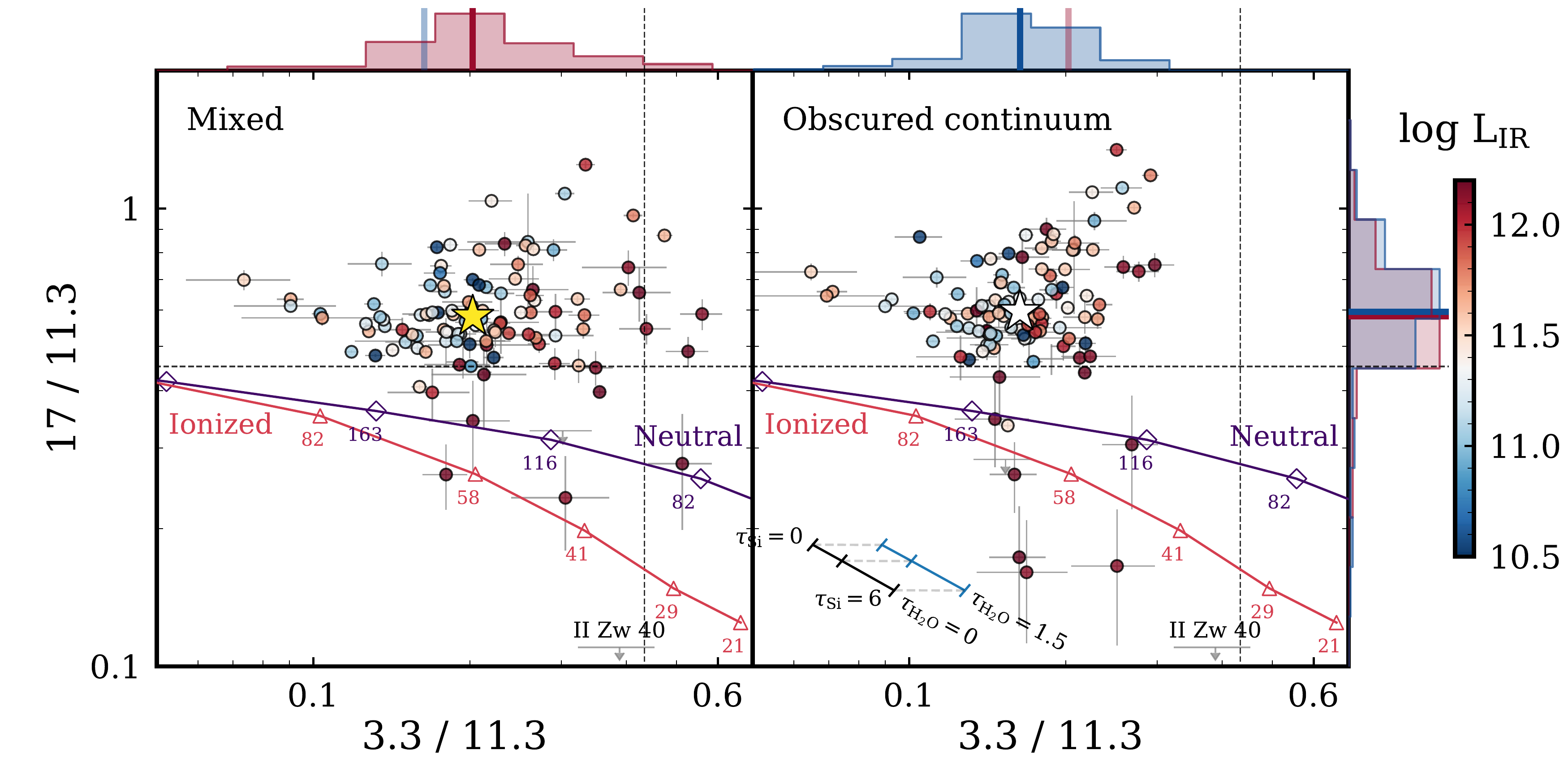}
    \caption{Size-sensitive PAH band ratios of 3.3/11.3 versus 17/11.3 in the mixed (left) and obscured continuum (right) geometries, color-coded with the integrated infrared luminosity. The marginalized distributions are also shown, with red and blue bars indicating the corresponding medians in the mixed and obscured continuum geometries, respectively, together with the yellow and white asterisks indicating the intersections of the bars. The dashed lines indicate the nominal values for individual PAH molecules from the model of \citet{Draine2007}, with tracks indicating neutral (purple) and ionized (red) PAHs with varying number of carbon atoms, N$_\mathrm{c}$. Detail discussion pertaining to the offset is in \S~\ref{sec:PAH_band}. Also shown are the attenuation tracks for these band ratios, parameterized by the silicate and water optical depths adopting the fully mixed dust geometry. 
    }
    \label{fig:pah33andpah17}
\end{figure*}

\subsection{Size diagnostics with the 3.3\um\ PAH} \label{sec:PAH_band}
The smallest PAH grains have the least heat capacities, and can therefore be stochastically heated to higher temperatures by single photon. As a result, small PAHs tend to emit efficiently at shorter wavelengths (3--9\um), while their larger counterparts tend to dominate at longer wavelengths \citep[15--20 $\mu m$;][]{Draine2007}. A widely used method for probing the grain size is to use the band intensity ratio 6.2/7.7 \citep{Draine2001}. However, since these two bands are so close in wavelength to each other, the range of grain sizes they probe is similar. With the full range of PAHs accessible in our study, we consider the relative strengths of the 3.3 and 17\um\ PAH bands as a probe of the size distribution, as these bands trace the smallest and largest ends of the PAH size distribution \citep{Schutte1993}. We note that In \PAHFIT, the 17\um\ complex is a combination of a group of PAH bands at 16.4, 17.0, 17.4 and 17.9\um, which is a broad PAH band feature thought to originate from PAH C-C-C banding modes \citep{Smith2004, Werner2004, Peeters2004b} and arise in PAHs with a few 1000 carbon atoms \citep{Draine2007}.

Fig.~\ref{fig:pah33andpah17} shows the ratios of both 3.3 and 17\um\ PAH normalized to the 11.3\um\ band for our two bracketing geometries, color-coded by \LIR. Overlaid are tracks of the model PAH band ratios from \citet{Draine2007} for individual neutral and ionized PAH grains, of varying carbon atom count (N$_\mathrm{c}$). The bulk of our sample lies at $\sim$\,80\,$<$\,N$_\mathrm{c}$\,$<$\,200. Galaxies with higher \LIR\ are likely to have a higher 3.3/11.3 ratio. We divide our sample into two groups and find galaxies with \LIR $<$ 10$^{12}$\Lsun\ and \LIR$ \geq 10^{12}$\Lsun\ have median of 3.3/11.3 as 0.20$^{+0.05}_{-0.04}$ and 0.29$^{+0.10}_{-0.07}$ in the mixed geometry, and 0.16$^{+0.03}_{-0.02}$ and 0.18$^{+0.04}_{-0.03}$ in the obscured continuum geometry, respectively (with the uncertainties indicating the inner two quartiles). Also shown are the dashed lines, indicating the nominal value derived from an averaged spectrum of PAH molecules with different N$_\mathrm{c}$'s, for a radiative heating intensity $U=1$, and assuming the size distribution from \citet{Weingartner2001} as adapted by \citet{Draine2007}\footnote{The average band ratios do not fall along the single-grain tracks due to large grains which produce weak PAH bands with moderate 17\um\ power (and hence high 17/11.3 ratios).}. Only a few number of galaxies with boosted 3.3\um\ PAH power in the mixed geometry can reach the nominal value in the x-axis, while no galaxy appear to have such high 3.3/11.3 in the obscured continuum geometry. In general, our 3.3\um\ PAH flux measurements are at least a factor of 2 less than the nominal value, as we have demonstrated in Fig.~\ref{fig:pah33_to_totpah}.

\begin{figure*}
 	\includegraphics[width=1.0\textwidth]{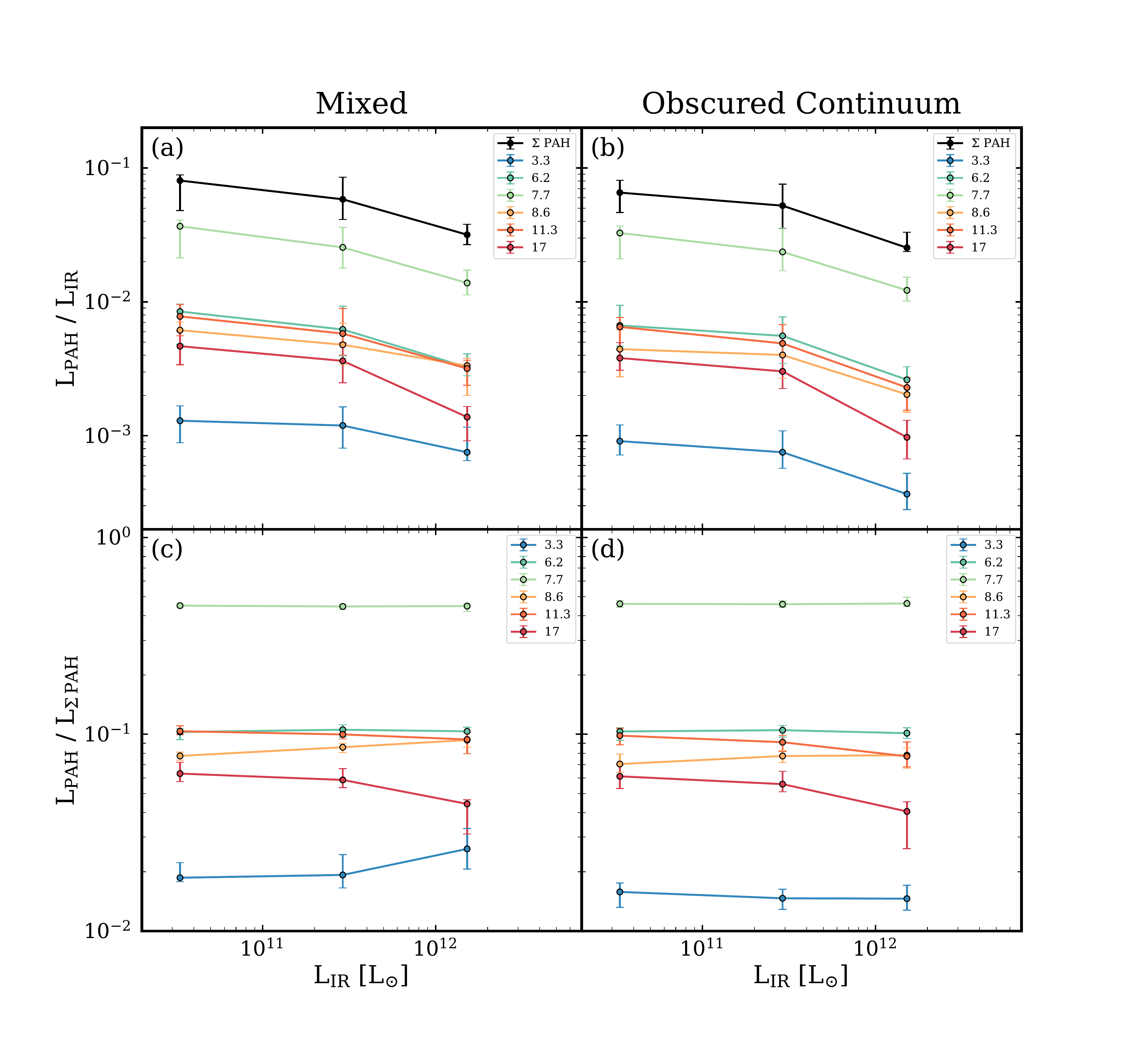}
    \caption{Changes of the main PAH band luminosity with respect to different \LIR\ under the two geometries. The trends show the median of the binned data in our bright-PAH sample, with the error bars indicating the inner two quartiles. While all L$_\mathrm{PAH}$/L$_\mathrm{IR}$ decline with increasing \LIR, diverse trends are shown in L$_\mathrm{PAH}$/\Ltotpah. }
    \label{fig:pah_trend}
\end{figure*}

More galaxies have higher 3.3/11.3 ratio in the mixed geometry than in the obscured continuum geometry suggest the 3.3\um\ PAH is boosted by attenuation correction, leading to a spread towards the right side in the fully mixed geometry. To study impact of the attenuation on migrating the points between the two adopted geometries, we use the PAH band profiles decomposed from the 1C spectrum and impose attenuation on those features to produce the attenuation tracks that depict the change of the band ratios with varying \tauSi\ and \tauWater. We show the black and blue lines indicate the attenuation trends assuming \tauWater=0 and \tauWater=1.5, respectively, with corresponding \tauSi\ shown in each tick. These $\tau$ values are all parameterized by assuming a mixed geometry. The optical depth of H$\mathrm{_{2}O}$ has impact solely on 3.3/11.3 since the 3.3\um\ PAH is the only band here that affected by \tauWater. The tracks indicate galaxies in the obscured continuum geometry are likely migrate towards the bottom right in the mixed geometry, depending on the levels the attenuation. We note that these tracks depict the normal behaviors of the galaxies, certain individuals may not follow such tracks, e.g., ULIRGs that are heavily obscured.

\subsection{PAH 3.3\um\ Emission in Context} \label{sec:all_the_PAHs}
A deficiency of PAH emission with increasing IR luminosity has been found in a variety of environments, including starburst galaxies and galaxies with moderate to strong AGN contribution; in particular, L$_\mathrm{PAH}$/\LIR\ drops significantly at ULIRG luminosities \citep{Wu2010, Pope2013, Stierwalt2014}. To assess whether a similar degree of decline is found between the 3.3\um\ PAH power, and how the emission geometry impacts the result, in Fig.~\ref{fig:pah_trend} (a) and (b), we show the relationship of \LIR\ and L$_\mathrm{PAH}$/\LIR\ in the two geometries, in three luminosity bins -- infrared galaxies (IRGs: \LIR\,$<$\,10$^{11}$\,\Lsun), LIRGs, and ULIRGs. Comparing \LIR\,$<10^{11}$\,\Lsun\ and \LIR\,$>10^{12}$\,\Lsun, we find the drop of the PAH power as a whole (black line) shows no difference in the two geometries, both decrease by 61\%. In general, the PAH band luminosities of individual bands also behave similarly in both geometries except for the 3.3\um\ PAH. While the drop of the \Lpahthree/\LIR\ is greater in the obscured continuum geometry (60\%) than in the mixed geometry (42\%), both drop less than \Ltotpah/\LIR\ (61\%). This suggests that the smallest PAHs are certainly not subjected to enhanced suppression, even though they are often considered the most vulnerable PAH population \citep{Allain1996}.

While all the PAH bands exhibit deficits as L$_\mathrm{PAH}$/\LIR\ declines with increasing \LIR, they show diverse trends in L$_\mathrm{PAH}$/\Ltotpah, as shown in Fig.~\ref{fig:pah_trend}(c) and (d). Both the 3.3 and 8.6\um\ PAH exhibit noticeable increases in fractional luminosity (by 40\% and 20\%, respectively), whereas the 17\um\ PAH decreases significantly by over 30\%. The increase in the fractional 8.6\um\ PAH is due to the fact that galaxies with higher infrared luminosity are likely exhibiting stronger silicate absorption, and the depth of such feature may affect the measurement of nearby 8.6\um\ PAH. The increase in fractional 3.3\um\ PAH power is rather interesting since it implies the size distribution shifts towards smaller end as \LIR\ increases. Even in the obscured continuum geometry 3.3\um\ PAH fractional power decreases only insignificantly while the 17\um\ PAH drops by 34\%. Note that the true, attenuation independent behavior is likely bracketed by the results in (c) and (d). Hence, we conclude there is no statistically significant downward trend in \Lpahthree/\Ltotpah\ as \LIR\ increases, suggesting rather non-intuitively that small PAH populations have an advantage in surviving in environments with strong global radiation fields, or are otherwise able to emit more effectively in exciting radiation (see \S~\ref{subsec:size_distribution} for details).

With our sample selection criteria using the AGN classification scheme of \citet[][see \S~\ref{subsec:bright_PAH}]{Stern2005}, a modest number of galaxies harboring deeply-embedded AGNs might remain in the sample \citep[e.g.][]{Petric2011}, especially at \LIR\,$>5 \times10^{12}$\Lsun \citep{Nardini2010, Alonso-Herrero2012}. Even though 20\% of our sample are ULIRGs (23 galaxies), none have \LIR\,$>5 \times10^{12}$\Lsun, with most below $3\times10^{12}$\Lsun, and only 2 galaxies in the range $=3$--$5\times10^{12}$\Lsun. We note that the possible inclusion of AGN in our sample could contribute to the drop of L$_\mathrm{PAH}$/\LIR\ in Fig.~\ref{fig:pah_trend}, but does not alter conclusions drawn from the absence of a decreasing trend of \Lpahthree/\Ltotpah\ with increasing \LIR\ (Fig.~\ref{fig:pah_trend} bottom panels).

\begin{deluxetable}{ccccc}
\tabletypesize{\footnotesize}
\tablecolumns{4}
\tablewidth{0pt}
\tablecaption{Best-fit Photometric Calibration Parameters for 3.3\um\ PAH using \JWST/NIRCam Medium Width Filters}
\label{tab:JWST_params}
\tablehead{
\colhead{z} &  \colhead{S} &  \colhead{$\mathrm{\sigma_S}$} & \colhead{$W_{300}$} & \colhead{$W_{360}$} \\
}
\startdata
0.00 & 10.78 & 0.44 & 0.35 & 0.65 \\
0.01 & 11.56 & 0.42 & 0.35 & 0.65 \\
0.02 & 12.88 & 0.52 & 0.34 & 0.66 \\
0.03 & 15.55 & 0.73 & 0.30 & 0.70 \\
\enddata
\vspace{0.cm}
\end{deluxetable}

\subsection{\JWST\ Photometric 3.3\um\ PAH Strengths} \label{sec:JWST_filter}
\begin{figure}
 	\includegraphics[width=\columnwidth]{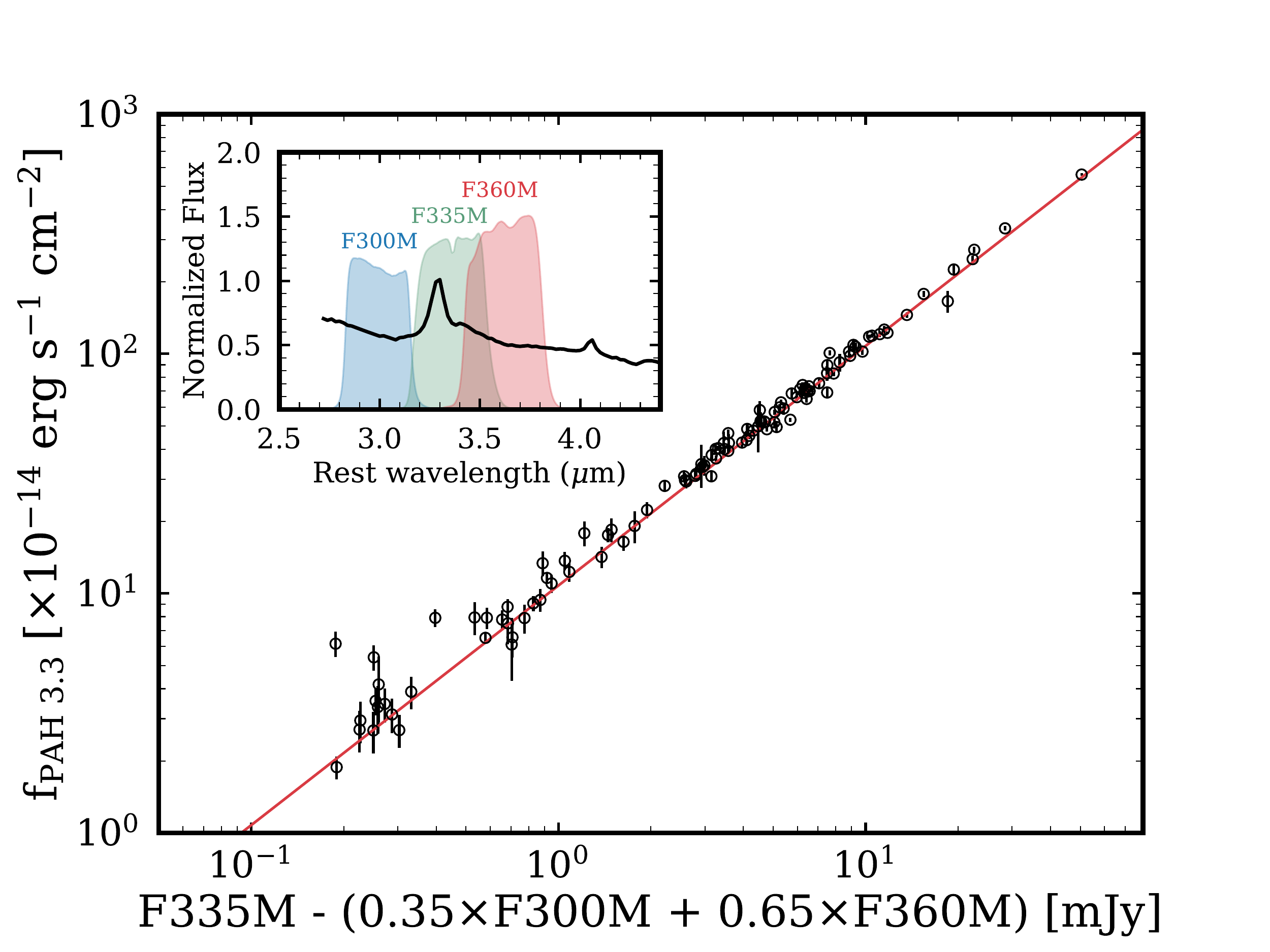}
    \caption{The comparison between full spectroscopic and \JWST/NIRCam synthetic photometry of the 3.3\um\ PAH feature. The response curves of the filters F300M, F335M, and F360M are inset. The x-axis is the optimized formula to narrow the dispersion at z=0 (see Table~\ref{tab:JWST_params}). 
    \label{fig:JWST_filter}}
\end{figure}

\JWST\ will provide routine photometric and spectroscopic access to the 3.3\um\ PAH feature, which has not been as extensively studied as the other brighter PAH emission features found at longer wavelengths. The three \JWST/NIRCam medium width filters F300M, F335M, and F360M\footnote[3]{https://jwst-docs.stsci.edu/near-infrared-camera/nircam-instrumentation/nircam-filters} are customized for detecting the 3.3\um\ PAH. To investigate the reliability of inferring 3.3\um\ PAH power from \JWST/NIRCam photometry, in Fig.~\ref{fig:JWST_filter} we compare synthetic photometry derived from our full spectra with the 3.3\um\ PAH strength recovered by \PAHFIT\ in the obscured continuum geometry (i.e. without attenuation correction). The synthetic photometry is calculated using the response curves of the respective \JWST/NIRCam medium filters with the rest-frame 1C spectrum. The x-axis represents the result of the synthetic photometry of the 3.3\um\ PAH band, derived from a subtraction of the continuum contribution to F335M. The continuum is represented by a weighted combination of F300M and F360M, with weighting coefficients optimized to provide the minimum scatter against the 3.3\um\ PAH band power measured in \PAHFIT. We only provide calibration up to z=0.03 since this calibration does not hold for higher redshifts as the 3.3\um\ band falls out of the F335M filter. The photometric calibration of the 3.3\um\ PAH feature using the photometric bands with respect to different redshifts can be parameterized by:

\begin{equation}
    \bigg(\frac{f_\mathrm{PAH \ 3.3}}{10^{-14} \ \mathrm{erg \ s^{-1} \ cm^{-2}}}\bigg) = (\mathrm{S} \ \pm \ \sigma_\mathrm{S}) \cdot x, 
\end{equation}
where
\begin{equation}
    x = \mathrm{F335M} - (\mathrm{W_{300}}\times\mathrm{F300M} + \mathrm{W_{360}}\times\mathrm{F360M}) \ \mathrm{[mJy]},
\end{equation}
$\mathrm{S}$ is the slope of the relation with an uncertainty of $\mathrm{\sigma_{S}}$, and $\mathrm{W_{300}}$ together with $\mathrm{W_{360}}$ are the weightings for the F300M and F360M filters. The corresponding best-fit parameters in different redshifts are listed in Table~\ref{tab:JWST_params}. Note that this calibration is derived from star-forming galaxies with moderate obscuration; heavily embedded galaxies may not be applicable. The reason for the local continuum at around 3.3\um\ to have a slightly lower weighting toward F300M may be due to the fact that F300M is subjected to the water ice absorption at 3.05\um\ that can induce a greater dispersion.

\begin{figure}
 	\includegraphics[width=\columnwidth]{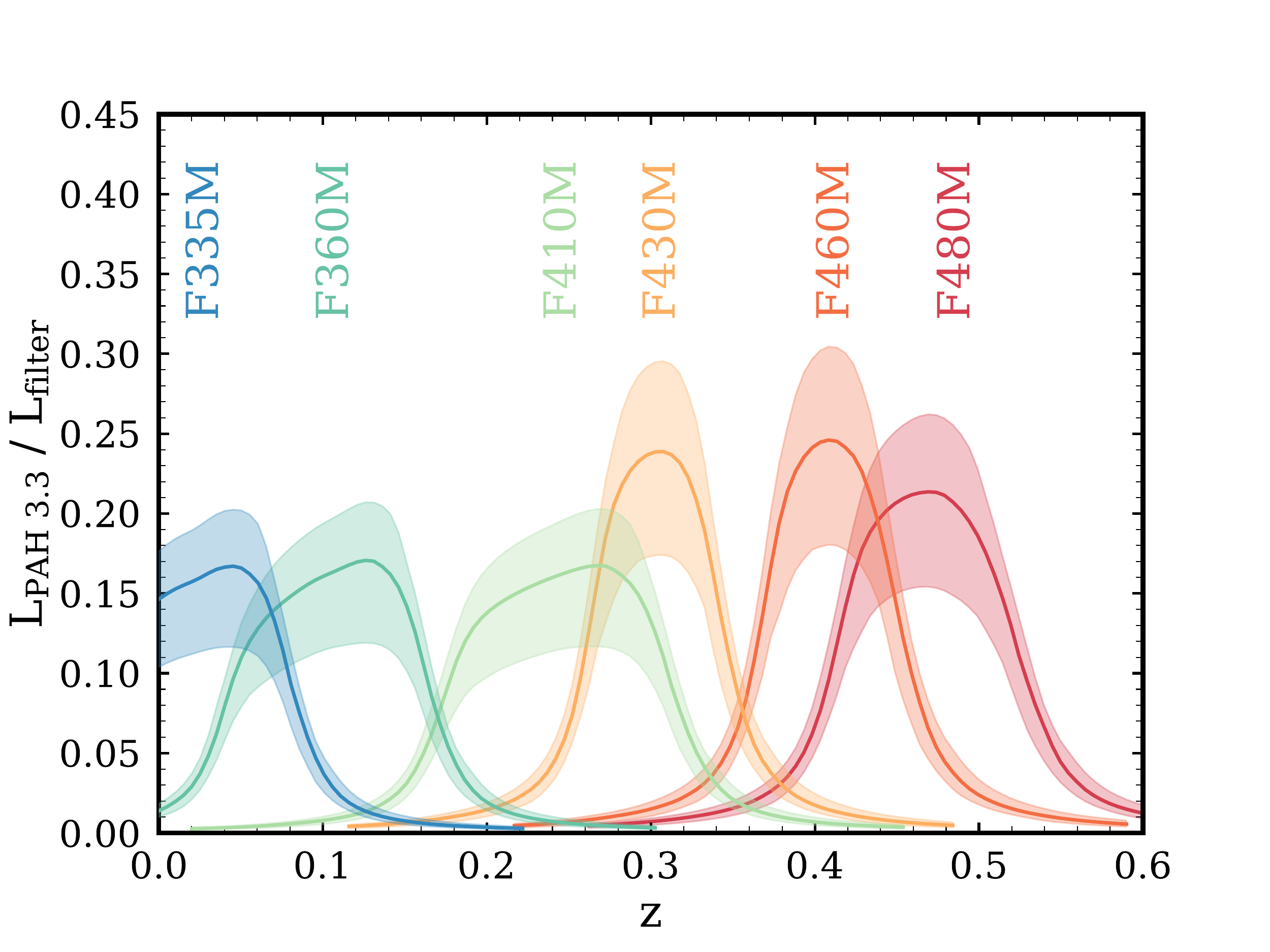}
    \caption{The fraction of the 3.3\um\ PAH that contributes to the corresponding \JWST/NIRCam medium filter with respect to the redshift. Each curve is the median of  spectra obtained from galaxies in class 1C. The shaded regions indicate the inner two quartiles of our 1C sample.}
    \label{fig:JWST_NIRCam_1C}
\end{figure}

Because the 3.3\um\ PAH has the shortest wavelength among all the PAH bands in the MIR, it is the final band to fall out of \JWST\ NIRCam or MIRI filters as redshift increases into the distant Universe, providing an advantage for the detection of resolved ISM features in galaxies up to z\,$\sim$\,7. Here we investigate the contribution of the 3.3\um\ PAH in different NIRCam filters with increasing redshift. In Fig.~\ref{fig:JWST_NIRCam_1C}, we show the contribution of the 3.3\um\ PAH in the respective NIRCam filters as a function of redshift. At z=0, 3.3\um\ PAH contributes $\sim$15\% to the F335M filter, and such value evolves with redshift. Each curve shows the median calculated from the star-forming galaxies, which were classified as 1C in Fig.~\ref{fig:sample_selection}(a), with shading region indicating the 25--75\% range. The three filters with longer central wavelengths, i.e. F430M, F460M and F480M, reach higher peaks since they are narrower compared to filters at shorter wavelengths. 

\section{Discussion} \label{sec:discussion}
With extended wavelength coverage down to 2.5\um, our cross-archival \Spitzer+\AKARI\ spectroscopic sample has opened a comprehensive view of the mid-infrared resolved dust emission features in star-forming galaxies. The features at 3.3--3.4\um, although not major energetic contributors, serve as an important diagnostic of the smallest dust grains and large aromatic and aliphatic molecules with 10's of carbon atoms. And yet as the shortest wavelength resolved dust features in the MIR, they suffer most significantly from the effects of dust attenuation, which can impact their interpretation. Here we discuss the effects of dust emission geometry on the intrepretation of 3.3\um\ PAH emission, and then consider the 3.3 and 17\um\ PAH features together as wide dynamic range tracers of opposite ends of the PAH size distribution. The relatively bright 3.3\um\ PAH emission together with the absence of 17\um\ PAH in II~Zw~40 leads to a discussion of possible mechanisms that conduce the stability of small dust grains under intense radiation fields.

\subsection{Impact of the Geometry: Mixed and Obscured Continuum} \label{subsec:geometry_disc}
Among all PAH bands, even considering the localized silicate absorption band, the 3.3\um\ feature suffers the most from attenuation. This is due both to the steep rise of continuum opacity in the extinction curve towards $\lambda$\,$\sim$\,3\um\ as well as 3.05\um\ H$_{2}$O ice absorption. Even at the limited attenuations of $\tau_\mathrm{mixed} \lesssim$ 3.0 of our PAH-bright sample, the interpretation of 3.3\um\ PAH behavior can depend on assumptions of emission/attenuation geometry, and accordingly we have explored the bracketing cases of \emph{fully mixed} and \emph{obscured continuum} geometries. 

For our adopted silicate attenuation profile, the optical depth at 3.3\um\ is $\sim$\,70\% of the peak of the primary silicate band at 9.7\um. While H$_{2}$O ice also has an impact at 3.3\um, the dominant attenuation component is the smooth continuum opacity associated with and constrained by silicate absorption. For the sample average $\tau_\mathrm{H_{2}O}$/$\tau_\mathrm{Si}$=0.25 in the mixed geometry case, H$_{2}$O ice only contributes an average $\sim$\,18\% of the total attenuation at 3.3\um. Among all PAH bands, 3.3\um\ PAH has the largest attenuation correction factor, $\sim$\,1.5, for an extinction curve characterized by typical parameters \tauSi=1 and \tauWater=0.2, followed by a factor of 1.2 for both the 8.6 and 11.3\um\ bands that flank the silicate feature.

Without prior knowledge of the geometry, it is challenging to constrain the \emph{intrinsic} power of 3.3\um\ PAH emission, especially for highly obscured galaxies. Even for high metallicity systems powered entirely by star formation, SFRs inferred from the 3.3\um\ PAH can differ by a factor of $\sim$\,3 compared to SFR$_\mathrm{Ne}$ in galaxies with $\tau_\mathrm{Si}$+$\tau_\mathrm{H_{2}O}\!\sim\!4$ (see Fig.~\ref{fig:SFR_calibration}). Statistically, only at $\tau _\mathrm{Si}$+$\tau_\mathrm{H_{2}O}$ $<$ 0.9 is there no significant difference in the ratios of 3.3\um\ PAH to neon-based SFR between the two geometries (see \S~\ref{subsubsec:SFR_geometry}). Another extreme beyond the mixed geometry is the screen geometry, which assumes attenuation arise from a monolithic foreground layer of dust. A similar optical depth in screen geometry results in more pronounced attenuation than in a mixed geometry. The ``convergent" attenuation $\tau$=0.90 in the mixed geometry is equivalent to $\tau$=0.42 in a screen geometry, and $\tau_\mathrm{screen}/\tau_\mathrm{mixed} \simeq 0.47$ is approximately constant within our attenuation range. It is important to note that \emph{the effective correction for this attenuation is nearly identical} between mixed and screen geometries, since both apply to all emission components (lines, continuum, and resolved dust emission bands). 

To conclude, applying both mixed and obscured continuum geometries permits constraining the impact of attenuation on our results. \Lpahthree/\LIR\ is most sensitive to the obscuration geometry, increasing by up to $\sim$\,5$\times$ for the most embedded sources in the mixed geometry. For \Lpahthree/\Ltotpah, since attenuation impacts all PAH bands to differing degrees, the difference is not as pronounced ---  at most a factor of $\sim$3$\times$. Similarly, as we compare \Lpahthree\ to SFR$_\mathrm{Ne}$, the strength of the neon lines are also subjected to the applied geometry, which results in a difference up to 3$\times$ at high obscuration in our sample. The aliphatic-to-aromatic ratio (3.4/3.3) is not significantly affected by the applied geometry due to wavelength proximity, with variation $\lesssim 20$\%.

\subsection{Impact of the extinction curve}
The impact of the selected extinction curve $A_\lambda$ on our study is minor compared to the geometry. In particular, the \emph{obscured continuum} geometry yield feature strengths which are largely independent of $A_\lambda$ (which only affects the underlying continuum). To investigate the impact on our \emph{mixed} geometry results, we compared PAH band ratios derived using the PAHFIT-default Galactic center extinction curve of \citet{Kemper2004}, to the newly published extinction curve of \citet{Hensley2020} measured from Galactic blue hypergiant Cyg OB2-12. The main difference between the two curves normalized at 9.7\um\ is that \citeauthor{Hensley2020} has a flatter (grayer) profile between 5-8\um\ with broader silicate features at 9.7 and 18\um. Despite these differences, A$_{3.3\um}$/A$_{9.7\um}$ is nearly identical between the two profiles. 

For the median galaxy in our sample with \tauSi=1.10 and \tauWater=0.20, applying the extinction curve of \citet{Hensley2020} reduces the fractional 3.3\um\ PAH power by $\sim$10\% due to the fact that the extinction-corrected PAH strengths between 5-8\um\ can be enhanced due to the flatter profile in that range. Our provided 3.3\um\ PAH star formation rate (Eq.~\ref{eqn:SFR}) is not affected by the choice of $A_\lambda$, since it is calibrated against galaxies with limited obscuration (\tauSi\ + \tauWater\ $<$ 0.9). Using the extinction curve of \citeauthor{Hensley2020} does not mitigate the offset between measured and modeled band ratios, as the ratio of 17/11.3 remains unchanged ($<$6\%) and the ratio of 3.3/11.3 drops only by $\sim$9\%. In the worst case of IRAS~F06076-2139 --- the most ``obscured" galaxy in our bright-PAH sample, with \tauSi=4.21 and \tauWater=0.23 in mixed geometry --- the drop of 3.3/11.3 is at most $\sim$30\%. Such a drop actually further \emph{increases} the discrepancies in Fig.~\ref{fig:pah33andpah17} between measurement and model. To gain a comprehensive understanding of how different extinction curves impact recovered PAH bands, further investigation on more embedded sources is needed.

\begin{figure}
 	\includegraphics[width=\columnwidth]{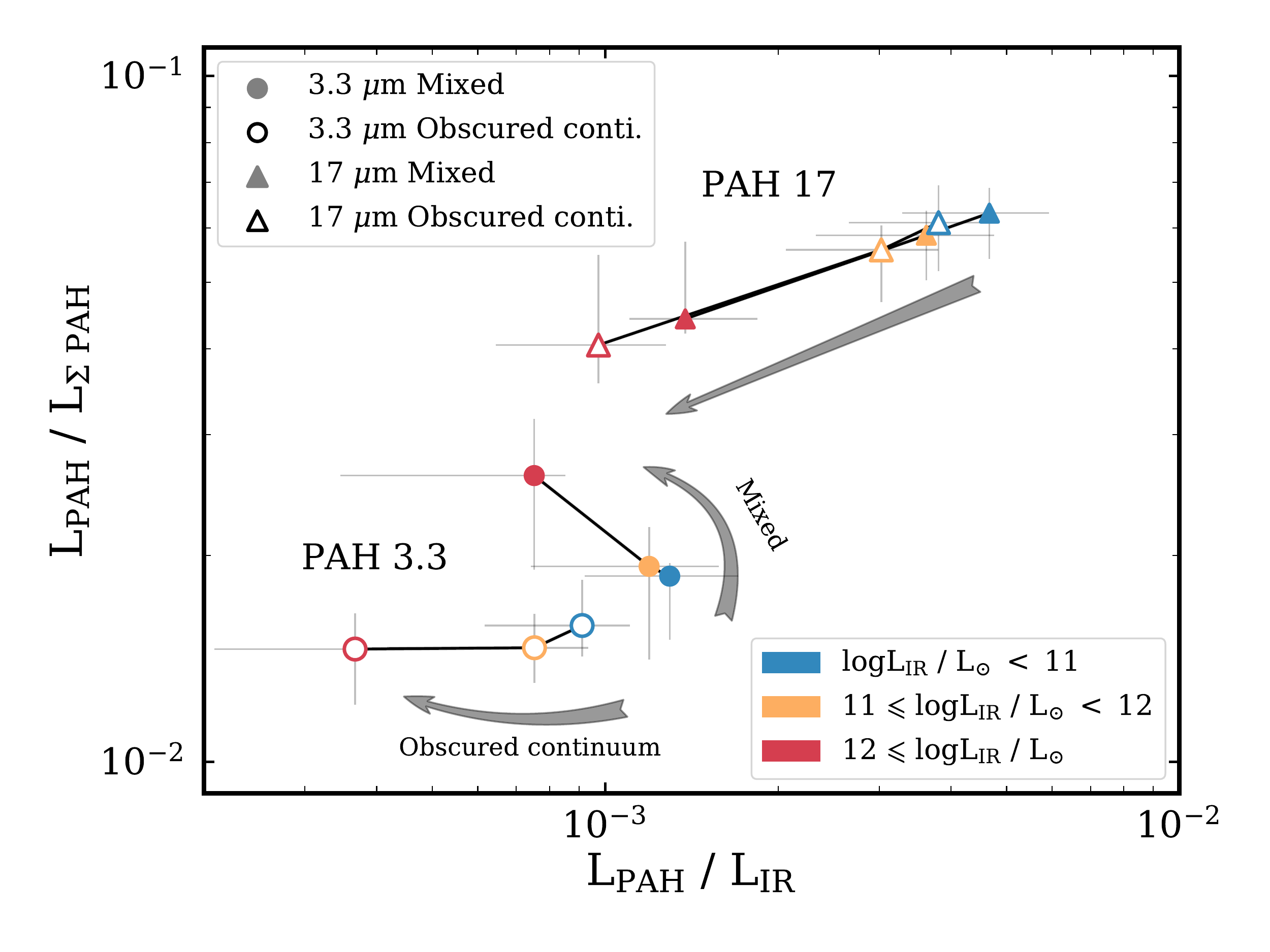}
    \caption{The change of the 3.3 (circle) and 17 (triangle)\um\ PAH in the L$\mathrm{_{PAH}}$/\LIR\ and L$\mathrm{_{PAH}}$/\Ltotpah\ space as infrared luminosity increases (directions of the arrows). Solid (empty) point indicates measurements in the mixed (obscured continuum) geometry. Apparently 17\um\ PAH band decreases in both geometries. While the behavior of 3.3\um\ PAH track depends on the adopted geometry, there is \emph{no} sign of decrease of the fractional 3.3\um\ PAH power as \LIR\ increases (see \S~\ref{subsec:size_distribution}).
    }
    \label{fig:PAH33and17}
\end{figure}

\subsection{The Size Distribution of PAHs} \label{subsec:size_distribution}
Among all the PAH emission bands, the 3.3\um\ band and 17\um\ complex trace PAH molecules at opposite ends of the PAH size distribution, probing the smallest and largest grains, respectively \citep{Schutte1993, Draine2007}. The variation between the two bands therefore allows us to probe the overall size shift in different interstellar environments across a wide range of sizes, from 10's to 1000's of carbon atoms. In Fig.~\ref{fig:pah_trend}, we demonstrated the behaviors of the dominant PAH emission features with increasing \LIR. Here in Fig.~\ref{fig:PAH33and17}, we specifically compare the changing contributions of the 3.3\um\ (circle) and 17\um\ (triangle) PAH bands for the entire sample. Both \Lpahthree/\LIR\ and \Lpahseventeen/\LIR\ decrease as \LIR\ increases (indicating by arrows all pointing left), regardless of the geometry. While the ratio of \Lpahseventeen\ to \LIR\ and \Ltotpah\ decreases in a similar fashion in both geometries, the behavior of 3.3\um\ band differs due to the larger impact of attenuation. In the fully mixed scheme, as \LIR\ increases, the trend of 3.3\um\ PAH emission tracks towards the upper left, suggesting an increase of the fractional 3.3\um\ PAH abundance. In contrast, the trend moves broadly towards the left in the obscured continuum scheme, with no sign of significant change in the fractional 3.3\um\ PAH power. A precise projection of the trend for an individual galaxy requires a better handle of the geometry --- a challenging task requiring a sophisticated radiative transfer model that is beyond the scope of this study. Nevertheless, these two bracketing trends allow us to narrow the possibilities ---  \Lpahthree/\Ltotpah\ either increases or is relatively unchanged as \LIR\ increases. 

The fact that the fractional 17\um\ PAH decreases together with the increase of the fractional 3.3\um\ PAH supports theories of top-down fragmentation \citep[e.g.][]{Jones2016}. Enhanced photo-destruction of PAHs may change the size distribution \citep{Allain1996}, leading to the breakdown of larger PAHs into smaller PAH segments. A scenario with the photochemical erosion process is also supported by the decline of the ratio of 3.4/3.3 with the increasing \LIR\ (\S~\ref{sec:aliphatic}). 

Another potential cause for shifting PAH power from longer to shorter wavelengths is the enhanced radiation field that makes larger PAHs ``act'' like smaller PAHs. In a high UV radiation density environment, the mean time between photon absorptions become shorter than the radiative cooling time for PAHs, allowing molecules to reach higher energy level by encountering multiple photons during the relaxation process, which in turn followed by short-wavelength emission that resembles emission of small PAHs \citep{Draine2007}. A detailed model of the PAH size distribution with respect to changing starlight fields will shed light on whether radiation could be the main driver of the enhancement of the fractional 3.3\um\ PAH in environments with high infrared luminosity or at low metallicity.

\subsection{The Stability of Small Grains} \label{subsec:grain_stability}
With a strong 3.3\um\ emission, together with non-detection of the 17\um\ PAH, II~Zw~40 casts considerable doubt on the suggestion that only larger PAHs are able to survive in harsh low-metallicity environments \citep{Madden2006, Gordon2008, Hunt2010}. \citet{Sandstrom2012} proposed an alternative scenario suggesting a size distribution shifted towards smaller PAHs in low-metallicity environment is due to the intrinsically enhanced formation of small grains rather than dust processing. That said, bottom-up formation process seemed be favored in low-metallicity environments as opposed to the top-down fragmentation process. The fact that strong 3.3\um\ PAH is detected in II~Zw~40 demonstrates that small grains are able to survive even in the harsh radiation environments of metal-poor starburst galaxies. 

Despite the fact that in general small PAHs are readily destroyed by UV photons, if the absorbed energy can be channeled through a relaxation process other than dissociation, photo-destruction can be avoided. PAH molecules have three relaxation channels: dissociation, IR vibrational emission, and recurrent fluorescence (RF). Dissociation becomes efficient only when excitation energy is higher than a certain level (depending on the size of the molecule). At lower energies, IR emission and recurrent fluorescence are the two main relaxation processes. The relaxation timescale for IR emission is on the order of $\sim$s, whereas recurrent fluorescence is a relatively fast radiative cooling process with relaxation time on the order of $\sim$ms \citep{Bernard2017}. According to \citet{Leger1988}, molecule consists of fewer carbon atoms produces RF photons more efficiently. Hence, it is worth noting that RF can be a potentially powerful mechanism to significantly increases the ability for small particles to survive under high UV radiation densities. Experimentally, such process has been confirmed in the laboratory that RF can be a main radiative cooling mechanism for PAH molecules by observing it through anthracene cations trapped in a compact electrostatic storage ring, the Mini-Ring \citep{Martin2013, Martin2015, Bernard2017}. Observationally, it has been proposed to conduce to the survival rate of the carriers of the extended red emission and diffuse interstellar bands \citep{Lai2017, Lai2020}. Therefore, RF may also play an important role in maintaining the abundance of small PAHs in the ISM, especially in environments with strong and high energy prevailing UV radiation fields such as II~Zw~40.

\section{Summary} 
\label{sec:summary}
We present a sample of 378 galaxies drawn from the \AKARI-\Spitzer\ Extragalactic Spectral Survey (ASESS), an exhaustive cross-archival comparison between the \AKARI/IRC dataset and the \Spitzer/IRS IDEOS extragalactic catalog, with combined wavelength coverage from 2.5--38\um. A set of composite template spectra adopted from the classification scheme of \citet{Spoon2007} is released. With extended wavelength coverage down to 2.5\um, we place the 3.3\um\ PAH feature in context of the complete MIR emission spectrum, and explore it as a powerful diagnostic of the smallest PAH molecules in a diverse range of galaxies. A subset of 113 PAH-bright galaxies with low-to-moderate dust attenuation were investigated using an extended version of the spectral decomposition tool \PAHFIT. This has enabled, for the first time, modeling \textit{all} PAH emission bands in the MIR simultaneously. Our major findings are summarized as follows: 

\smallskip

\begin{itemize}
\item An extended version of \PAHFIT\ with new features, including the aromatic and aliphatic bands, recombination lines, and several absorption components, found to be highly successful in modelling the complete MIR spectrum in a diverse sample of galaxies. We find the 5.27\um\ PAH feature can further be divided into two individual bands centered at 5.24 and 5.33\um.

\item A mild correlation exists between radiation heating intensity and the aliphatic-to-aromatic ratio (3.4/3.3), which suggests that aliphatics are more prone to photo-destruction under intense radiation environments. We also find that a broad ``plateau'' feature centered at 3.47\um\ is required in the model. This plateau typically carries 20\% of the 3.3\um\ PAH power, and is likely aromatic in nature, as it correlates better with 3.3\um\ PAH than with 3.4\um\ aliphatic emission.

\item As the shortest of the PAH vibrational emission features, we find the 3.3\um\ PAH is more susceptible to the effects of attenuation than any other PAH band. To bracket the impact of attenuation, two geometrical schemes are adopted: \emph{fully mixed geometry}, in which dust and emission components are mixed and \emph{obscured continuum geometry}, in which line and PAH emission lie atop an attenuated underlying continuum. We find that diagnostics including 3.3\um\ PAH emission diverge between these two geometries at moderate optical depths \tauSi+\tauWater\,$\gtrsim$\,1 (equivalent to 0.5 in screen geometry). 

\item The bolometric fraction \Lpahthree/\LIR\ is $\sim$\,0.1\%, although this varies between the two geometries by a median $\sim$\,1.5$\times$, and this discrepancy increases to $\sim$\,5$\times$ in the most obscured galaxies in our sample. The fractional 3.3\um\ PAH power depends less sensitively on attenuation geometry, and is typically \Lpahthree/\Ltotpah\,$\simeq$\,1.5--3\%.  

\item Ratios among the main non-ionized PAH bands at 3.3, 11.3, and 17\um\ follow well the expected tracks for individual neutral PAH molecules, but the ratio 3.3/11.3 is typically $2\times$ smaller than the nominal value adopted in the dust model of \citet{Draine2007}, indendent of the adopted extinction curve.  The ratio of 3.3\um\ to 11.3\um\ power rises mildly in galaxies with higher infrared luminosity, independent of the adopted geometry. 

\item We provide a new star formation rate calibration based on 3.3\um\ PAH luminosity, constructed from a carefully selected sample of low-attenuation star forming galaxies. However, we find that obscuration plays an important role in this use, with a factor of $\sim$\,3 divergence between different geometries at high obscuration, even in this pre-selected sample. Metallicity plays an even larger role, with SFR biased low by $\gtrsim8\times$ at $1/4\,Z_\sun$.

\item Although we confirm the significant drop in power of all PAH features as a fraction of bolometric infrared luminosity as \LIR\ increases from 10$^{9}$ to 10$^{12}$\Lsun, surprisingly, the 3.3\um\ feature, which traces the smallest and presumably most easily destroyed molecules, constitutes a flat or even \emph{rising} portion of total PAH emission, depending on the geometry adopted. In contrast, the 17\um\ feature, which traces the largest PAH grains with thousands of carbon atoms, drops most acutely at the highest infrared luminosities, independent of the assumed geometry.

\item We observed highly suppressed PAH emission in low metallicity galaxy II~Zw~40, with \Ltotpah/\LIR\ less than other sampled galaxies by a factor of $\sim$\,10. Surprisingly, however, 3.3\um\ power \emph{increases} relative to other PAH bands, with a complete absence of the 17\um\ PAH. This strongly suggests that caution is needed when using PAH bands as SFR indicators in moderate to low-metallicity environments, and that the smallest grains are more robust against photo-destruction in harsh radiation environments than previously believed. 

\item  We suggest that the likely mechanisms leading to strong fractional 3.3\um\ PAH power are either enhanced stability of small PAH molecules from efficient relaxation processes (e.g. recurrent fluorescence) or the migration of PAH power to shorter wavelengths under environments with intense and/or hard radiation fields.

\item Although 3.3\um\ PAH emission contributes just 15--25\% of the power in JWST/NIRCam medium band observations out to $z\,\sim\,0.5$, using the flanking bands for continuum subtraction permits recovery of the 3.3\um\ PAH power up to z=0.03 within 5\%.

\end{itemize}

\JWST\ will soon re-open the MIR window with more than an order of magnitude increase in sensitivity and spatial resolution, and the all-sky IR spectroscopic survey \textit{SPHEREx} will offer unprecedented access to more than 100 million galaxies in the 3\um\ wavelength regime. Together with comprehensive modeling of the complete suite of PAH emission, the diagnostic potential of these powerful bands will be fully realized, opening new windows into the study of star formation, radiation environment, metallicity, AGN processing, dust physics, and the physical properties of the ISM on sub-galactic scales and across much of cosmic time.
\\
\acknowledgments
We thank the anonymous referee for the insightful comments that helped to improve this manuscript. This research is based on observations with \AKARI, a JAXA project with the participation of ESA. This publication makes use of data products from the Wide-field Infrared Survey Explorer, which is a joint project of the University of California, Los Angeles, and the Jet Propulsion Laboratory/California Institute of Technology, funded by the National Aeronautics and Space Administration. TSYL gratefully acknowledges support from the University of Toledo Graduate Fellowship and the NASA ERS Grant and the host of ISAS, JAXA. HWWS would like to acknowledge support for the IDEOS project from NASA grant NNX16AF26G.
\\
\vspace{5mm}
\facilities{AKARI, Spitzer}


\software{Astropy \citep{Astropy2013, Astropy2018},
          Bokeh \citep{Bokeh2018},
          Matplotlib \citep{Hunter2007},
          Numpy \citep{VanderWalt2011},
          \PAHFIT\ \citep{Smith2007},
          SciPy \citep{Virtanen2020}
          }

\appendix
\section{Observation Detail for Bright-PAH galaxies}
Table~\ref{tab:bright_PAH_id} summarizes the observation for galaxies in bright-PAH sample.


\begin{deluxetable*}{llrrcr}
\tabletypesize{\footnotesize}
\tablecolumns{6}
\tablewidth{0pt}
\tablecaption{Observation for Bright-PAH galaxies}
\label{tab:bright_PAH_id}
\tablehead{
 \colhead{AKARI/IRC name} & \colhead{IDEOS name} &  \colhead{IRC ID} &  \colhead{IRC sub-ID} & \colhead{IRC Phase} & \colhead{IDEOS ID}
}

\startdata
              NGC 0023 &                  NGC 0023 &  1122291 &              [1] &      3 &  20348672\_0 \\
        MCG -02-01-051 &             ARP 256 NED01 &  1120101 &              [1] &      3 &  20313088\_0 \\
               NGC 232 &                  NGC 0232 &  1120102 &           [1, 2] &      3 &  20342016\_1 \\
       IRAS 00456-2904 &   2MASX J00480675-2848187 &  1100221 &              [1] &  1 \& 2 &  10109184\_0 \\
        MCG +12-02-001 &      MCG +12-02-001 NED02 &  5200048 &           [1, 2] &      3 &  20371712\_0 \\
       IRAS 01004-2237 &           IRAS 01003-2238 &  1120094 &        [1, 2, 3] &      3 &   4964608\_0 \\
        MCG -03-04-014 &            MCG -03-04-014 &  1122202 &     [1, 2, 3, 4] &      3 &  14835968\_0 \\
          ESO 244-G012 &        ESO 244-G012 NED02 &  1122223 &     [1, 2, 3, 4] &      3 &  20368384\_0 \\
       IRAS 01173+1405 &              CGCG 436-030 &  1120013 &           [1, 2] &      3 &  20356352\_0 \\
          ESO 353-G020 &              ESO 353-G020 &  1122308 &     [1, 2, 3, 4] &      3 &  20366080\_0 \\
\enddata
\vspace{0.cm}
\tablecomments{Table \ref{tab:bright_PAH_id} is published in its entirety in the machine-readable format. A portion is shown here for guidance regarding its form and content.}
\end{deluxetable*}

\bibliography{Lai_all_PAH}

\begin{thebibliography}{}
\expandafter\ifx\csname natexlab\endcsname\relax\def\natexlab#1{#1}\fi
\providecommand{\url}[1]{\href{#1}{#1}}

\bibitem[{{Allain} {et~al.}(1996){Allain}, {Leach}, \& {Sedlmayr}}]{Allain1996}
{Allain}, T., {Leach}, S., \& {Sedlmayr}, E. 1996, \aap, 305, 602

\bibitem[{{Allamandola} {et~al.}(1999){Allamandola}, {Hudgins}, \&
  {Sandford}}]{Allamandola1999}
{Allamandola}, L.~J., {Hudgins}, D.~M., \& {Sandford}, S.~A. 1999, \apjl, 511,
  L115

\bibitem[{{Allamandola} {et~al.}(1985){Allamandola}, {Tielens}, \&
  {Barker}}]{Allamandola1985}
{Allamandola}, L.~J., {Tielens}, A.~G.~G.~M., \& {Barker}, J.~R. 1985, \apjl,
  290, L25

\bibitem[{{Alonso-Herrero} {et~al.}(2012){Alonso-Herrero}, {Pereira-Santaella},
  {Rieke}, \& {Rigopoulou}}]{Alonso-Herrero2012}
{Alonso-Herrero}, A., {Pereira-Santaella}, M., {Rieke}, G.~H., \& {Rigopoulou},
  D. 2012, \apj, 744, 2

\bibitem[{{Arribas} \& {Colina}(2002)}]{Arribas2002}
{Arribas}, S., \& {Colina}, L. 2002, \apj, 573, 576

\bibitem[{{Astropy Collaboration} {et~al.}(2013){Astropy Collaboration},
  {Robitaille}, {Tollerud}, {Greenfield}, {Droettboom}, {Bray}, {Aldcroft},
  {Davis}, {Ginsburg}, {Price-Whelan}, {Kerzendorf}, {Conley}, {Crighton},
  {Barbary}, {Muna}, {Ferguson}, {Grollier}, {Parikh}, {Nair}, {Unther},
  {Deil}, {Woillez}, {Conseil}, {Kramer}, {Turner}, {Singer}, {Fox}, {Weaver},
  {Zabalza}, {Edwards}, {Azalee Bostroem}, {Burke}, {Casey}, {Crawford},
  {Dencheva}, {Ely}, {Jenness}, {Labrie}, {Lim}, {Pierfederici}, {Pontzen},
  {Ptak}, {Refsdal}, {Servillat}, \& {Streicher}}]{Astropy2013}
{Astropy Collaboration}, {Robitaille}, T.~P., {Tollerud}, E.~J., {et~al.} 2013,
  \aap, 558, A33

\bibitem[{{Astropy Collaboration} {et~al.}(2018){Astropy Collaboration},
  {Price-Whelan}, {Sip{\H{o}}cz}, {G{\"u}nther}, {Lim}, {Crawford}, {Conseil},
  {Shupe}, {Craig}, {Dencheva}, {Ginsburg}, {Vand erPlas}, {Bradley},
  {P{\'e}rez-Su{\'a}rez}, {de Val-Borro}, {Aldcroft}, {Cruz}, {Robitaille},
  {Tollerud}, {Ardelean}, {Babej}, {Bach}, {Bachetti}, {Bakanov}, {Bamford},
  {Barentsen}, {Barmby}, {Baumbach}, {Berry}, {Biscani}, {Boquien}, {Bostroem},
  {Bouma}, {Brammer}, {Bray}, {Breytenbach}, {Buddelmeijer}, {Burke},
  {Calderone}, {Cano Rodr{\'\i}guez}, {Cara}, {Cardoso}, {Cheedella}, {Copin},
  {Corrales}, {Crichton}, {D'Avella}, {Deil}, {Depagne}, {Dietrich}, {Donath},
  {Droettboom}, {Earl}, {Erben}, {Fabbro}, {Ferreira}, {Finethy}, {Fox},
  {Garrison}, {Gibbons}, {Goldstein}, {Gommers}, {Greco}, {Greenfield},
  {Groener}, {Grollier}, {Hagen}, {Hirst}, {Homeier}, {Horton}, {Hosseinzadeh},
  {Hu}, {Hunkeler}, {Ivezi{\'c}}, {Jain}, {Jenness}, {Kanarek}, {Kendrew},
  {Kern}, {Kerzendorf}, {Khvalko}, {King}, {Kirkby}, {Kulkarni}, {Kumar},
  {Lee}, {Lenz}, {Littlefair}, {Ma}, {Macleod}, {Mastropietro}, {McCully},
  {Montagnac}, {Morris}, {Mueller}, {Mumford}, {Muna}, {Murphy}, {Nelson},
  {Nguyen}, {Ninan}, {N{\"o}the}, {Ogaz}, {Oh}, {Parejko}, {Parley}, {Pascual},
  {Patil}, {Patil}, {Plunkett}, {Prochaska}, {Rastogi}, {Reddy Janga},
  {Sabater}, {Sakurikar}, {Seifert}, {Sherbert}, {Sherwood-Taylor}, {Shih},
  {Sick}, {Silbiger}, {Singanamalla}, {Singer}, {Sladen}, {Sooley},
  {Sornarajah}, {Streicher}, {Teuben}, {Thomas}, {Tremblay}, {Turner},
  {Terr{\'o}n}, {van Kerkwijk}, {de la Vega}, {Watkins}, {Weaver}, {Whitmore},
  {Woillez}, {Zabalza}, \& {Astropy Contributors}}]{Astropy2018}
{Astropy Collaboration}, {Price-Whelan}, A.~M., {Sip{\H{o}}cz}, B.~M., {et~al.}
  2018, \aj, 156, 123

\bibitem[{{Baba} {et~al.}(2018){Baba}, {Nakagawa}, {Isobe}, \&
  {Shirahata}}]{Baba2018}
{Baba}, S., {Nakagawa}, T., {Isobe}, N., \& {Shirahata}, M. 2018, \apj, 852, 83

\bibitem[{{Baba} {et~al.}(2019){Baba}, {Nakagawa}, {Usui}, {Yamagishi}, \&
  {Onaka}}]{Baba2019}
{Baba}, S., {Nakagawa}, T., {Usui}, F., {Yamagishi}, M., \& {Onaka}, T. 2019,
  \pasj, 71, 2

\bibitem[{{Baba} {et~al.}(2016){Baba}, {Nakagawa}, {Shirahata}, {Isobe},
  {Usui}, {Ohyama}, {Onaka}, {Yano}, \& {Kochi}}]{Baba2016}
{Baba}, S., {Nakagawa}, T., {Shirahata}, M., {et~al.} 2016, \pasj, 68, 27

\bibitem[{{Barker} {et~al.}(1987){Barker}, {Allamandola}, \&
  {Tielens}}]{Barker1987}
{Barker}, J.~R., {Allamandola}, L.~J., \& {Tielens}, A.~G.~G.~M. 1987, \apjl,
  315, L61

\bibitem[{{Bauschlicher} {et~al.}(2010){Bauschlicher}, {Boersma}, {Ricca},
  {Mattioda}, {Cami}, {Peeters}, {S{\'a}nchez de Armas}, {Puerta Saborido},
  {Hudgins}, \& {Allamandola}}]{Bauschlicher2010}
{Bauschlicher}, C.~W., J., {Boersma}, C., {Ricca}, A., {et~al.} 2010, \apjs,
  189, 341

\bibitem[{Bernard {et~al.}(2017)Bernard, Chen, Brédy, Ji, Ortéga, Matsumoto,
  \& Martin}]{Bernard2017}
Bernard, J., Chen, L., Brédy, R., {et~al.} 2017, Nuclear Instruments and
  Methods in Physics Research Section B: Beam Interactions with Materials and
  Atoms, 408, 21 , proceedings of the 18th International Conference on the
  Physics of Highly Charged Ions (HCI-2016), Kielce, Poland, 11-16 September
  2016.
\newblock
  \url{http://www.sciencedirect.com/science/article/pii/S0168583X17304020}

\bibitem[{{Boersma} {et~al.}(2009){Boersma}, {Mattioda}, {Bauschlicher},
  {Peeters}, {Tielens}, \& {Allamandola}}]{Boersma2009}
{Boersma}, C., {Mattioda}, A.~L., {Bauschlicher}, C.~W., J., {et~al.} 2009,
  \apj, 690, 1208

\bibitem[{{Bokeh Development Team}(2018)}]{Bokeh2018}
{Bokeh Development Team}. 2018, Bokeh: Python library for interactive
  visualization.
\newblock \url{https://bokeh.pydata.org/en/latest/}

\bibitem[{{Calzetti} {et~al.}(2007){Calzetti}, {Kennicutt}, {Engelbracht},
  {Leitherer}, {Draine}, {Kewley}, {Moustakas}, {Sosey}, {Dale}, {Gordon},
  {Helou}, {Hollenbach}, {Armus}, {Bendo}, {Bot}, {Buckalew}, {Jarrett}, {Li},
  {Meyer}, {Murphy}, {Prescott}, {Regan}, {Rieke}, {Roussel}, {Sheth}, {Smith},
  {Thornley}, \& {Walter}}]{Calzetti2007}
{Calzetti}, D., {Kennicutt}, R.~C., {Engelbracht}, C.~W., {et~al.} 2007, \apj,
  666, 870

\bibitem[{{Croiset} {et~al.}(2016){Croiset}, {Candian}, {Bern{\'e}}, \&
  {Tielens}}]{Croiset2016}
{Croiset}, B.~A., {Candian}, A., {Bern{\'e}}, O., \& {Tielens}, A.~G.~G.~M.
  2016, \aap, 590, A26

\bibitem[{{Cutri} \& {et al.}(2013)}]{Cutri2013}
{Cutri}, R.~M., \& {et al.} 2013, VizieR Online Data Catalog, II/328

\bibitem[{{Desert} {et~al.}(1990){Desert}, {Boulanger}, \&
  {Puget}}]{Desert1990}
{Desert}, F.~X., {Boulanger}, F., \& {Puget}, J.~L. 1990, \aap, 500, 313

\bibitem[{{D'Hendecourt} \& {Leger}(1987)}]{DHendecourt1987}
{D'Hendecourt}, L.~B., \& {Leger}, A. 1987, \aap, 180, L9

\bibitem[{{Diamond-Stanic} \& {Rieke}(2010)}]{Diamond-Stanic2010}
{Diamond-Stanic}, A.~M., \& {Rieke}, G.~H. 2010, \apj, 724, 140

\bibitem[{{Doi} {et~al.}(2019){Doi}, {Nakagawa}, {Isobe}, {Baba}, {Yano}, \&
  {Yamagishi}}]{Doi2019}
{Doi}, R., {Nakagawa}, T., {Isobe}, N., {et~al.} 2019, \pasj, 71, 26

\bibitem[{{Draine} \& {Li}(2001)}]{Draine2001}
{Draine}, B.~T., \& {Li}, A. 2001, \apj, 551, 807

\bibitem[{{Draine} \& {Li}(2007)}]{Draine2007}
---. 2007, \apj, 657, 810

\bibitem[{{Draine} {et~al.}(2007){Draine}, {Dale}, {Bendo}, {Gordon}, {Smith},
  {Armus}, {Engelbracht}, {Helou}, {Kennicutt}, {Li}, {Roussel}, {Walter},
  {Calzetti}, {Moustakas}, {Murphy}, {Rieke}, {Bot}, {Hollenbach}, {Sheth}, \&
  {Teplitz}}]{Drain2007b}
{Draine}, B.~T., {Dale}, D.~A., {Bendo}, G., {et~al.} 2007, \apj, 663, 866

\bibitem[{{Engelbracht} {et~al.}(2005){Engelbracht}, {Gordon}, {Rieke},
  {Werner}, {Dale}, \& {Latter}}]{Engelbracht2005}
{Engelbracht}, C.~W., {Gordon}, K.~D., {Rieke}, G.~H., {et~al.} 2005, \apjl,
  628, L29

\bibitem[{{F{\"o}rster Schreiber} {et~al.}(2004){F{\"o}rster Schreiber},
  {Roussel}, {Sauvage}, \& {Charmandaris}}]{Forster_Schreiber2004}
{F{\"o}rster Schreiber}, N.~M., {Roussel}, H., {Sauvage}, M., \&
  {Charmandaris}, V. 2004, \aap, 419, 501

\bibitem[{{Fritz} {et~al.}(2006){Fritz}, {Franceschini}, \&
  {Hatziminaoglou}}]{Fritz2006}
{Fritz}, J., {Franceschini}, A., \& {Hatziminaoglou}, E. 2006, \mnras, 366, 767

\bibitem[{{Galliano} {et~al.}(2005){Galliano}, {Madden}, {Jones}, {Wilson}, \&
  {Bernard}}]{Galliano2005}
{Galliano}, F., {Madden}, S.~C., {Jones}, A.~P., {Wilson}, C.~D., \& {Bernard},
  J.~P. 2005, \aap, 434, 867

\bibitem[{{Galliano} {et~al.}(2008){Galliano}, {Madden}, {Tielens}, {Peeters},
  \& {Jones}}]{Galliano2008}
{Galliano}, F., {Madden}, S.~C., {Tielens}, A. G.~G.~M., {Peeters}, E., \&
  {Jones}, A.~P. 2008, \apj, 679, 310

\bibitem[{{Gao} {et~al.}(2013){Gao}, {Jiang}, {Li}, \& {Xue}}]{Gao2013}
{Gao}, J., {Jiang}, B.~W., {Li}, A., \& {Xue}, M.~Y. 2013, \apj, 776, 7

\bibitem[{{Genzel} {et~al.}(1998){Genzel}, {Lutz}, {Sturm}, {Egami}, {Kunze},
  {Moorwood}, {Rigopoulou}, {Spoon}, {Sternberg}, {Tacconi-Garman}, {Tacconi},
  \& {Thatte}}]{Genzel1998}
{Genzel}, R., {Lutz}, D., {Sturm}, E., {et~al.} 1998, \apj, 498, 579

\bibitem[{{Gibb} {et~al.}(2004){Gibb}, {Whittet}, {Boogert}, \&
  {Tielens}}]{Gibb2004}
{Gibb}, E.~L., {Whittet}, D.~C.~B., {Boogert}, A.~C.~A., \& {Tielens},
  A.~G.~G.~M. 2004, \apjs, 151, 35

\bibitem[{{Gordon} {et~al.}(2008){Gordon}, {Engelbracht}, {Rieke}, {Misselt},
  {Smith}, \& {Kennicutt}}]{Gordon2008}
{Gordon}, K.~D., {Engelbracht}, C.~W., {Rieke}, G.~H., {et~al.} 2008, \apj,
  682, 336

\bibitem[{{Groves} {et~al.}(2008){Groves}, {Dopita}, {Sutherland}, {Kewley},
  {Fischera}, {Leitherer}, {Brandl}, \& {van Breugel}}]{Groves2008}
{Groves}, B., {Dopita}, M.~A., {Sutherland}, R.~S., {et~al.} 2008, \apjs, 176,
  438

\bibitem[{{Guseva} {et~al.}(2000){Guseva}, {Izotov}, \& {Thuan}}]{Guseva2000}
{Guseva}, N.~G., {Izotov}, Y.~I., \& {Thuan}, T.~X. 2000, \apj, 531, 776

\bibitem[{{Hammonds} {et~al.}(2015){Hammonds}, {Mori}, {Usui}, \&
  {Onaka}}]{Hammonds2015}
{Hammonds}, M., {Mori}, T., {Usui}, F., \& {Onaka}, T. 2015, \planss, 116, 73

\bibitem[{{Hensley} \& {Draine}(2020)}]{Hensley2020}
{Hensley}, B.~S., \& {Draine}, B.~T. 2020, arXiv e-prints, arXiv:2002.02457

\bibitem[{{Hern{\'a}n-Caballero} {et~al.}(2016){Hern{\'a}n-Caballero}, {Spoon},
  {Lebouteiller}, {Rupke}, \& {Barry}}]{Hernan-Caballero2016}
{Hern{\'a}n-Caballero}, A., {Spoon}, H.~W.~W., {Lebouteiller}, V., {Rupke},
  D.~S.~N., \& {Barry}, D.~P. 2016, \mnras, 455, 1796

\bibitem[{{Ho} \& {Keto}(2007)}]{Ho2007}
{Ho}, L.~C., \& {Keto}, E. 2007, \apj, 658, 314

\bibitem[{{Horne}(1986)}]{Horne1986}
{Horne}, K. 1986, \pasp, 98, 609

\bibitem[{{Houck} {et~al.}(2004){Houck}, {Roellig}, {van Cleve}, {Forrest},
  {Herter}, {Lawrence}, {Matthews}, {Reitsema}, {Soifer}, {Watson}, {Weedman},
  {Huisjen}, {Troeltzsch}, {Barry}, {Bernard-Salas}, {Blacken}, {Brandl},
  {Charmandaris}, {Devost}, {Gull}, {Hall}, {Henderson}, {Higdon}, {Pirger},
  {Schoenwald}, {Sloan}, {Uchida}, {Appleton}, {Armus}, {Burgdorf},
  {Fajardo-Acosta}, {Grillmair}, {Ingalls}, {Morris}, \& {Teplitz}}]{Houck2004}
{Houck}, J.~R., {Roellig}, T.~L., {van Cleve}, J., {et~al.} 2004, \apjs, 154,
  18

\bibitem[{{Hummer} \& {Storey}(1987)}]{Hummer1987}
{Hummer}, D.~G., \& {Storey}, P.~J. 1987, \mnras, 224, 801

\bibitem[{{Hunt} {et~al.}(2010){Hunt}, {Thuan}, {Izotov}, \&
  {Sauvage}}]{Hunt2010}
{Hunt}, L.~K., {Thuan}, T.~X., {Izotov}, Y.~I., \& {Sauvage}, M. 2010, \apj,
  712, 164

\bibitem[{Hunter(2007)}]{Hunter2007}
Hunter, J.~D. 2007, Computing in Science \& Engineering, 9, 90

\bibitem[{{Ichikawa} {et~al.}(2014){Ichikawa}, {Imanishi}, {Ueda}, {Nakagawa},
  {Shirahata}, {Kaneda}, \& {Oyabu}}]{Ichikawa2014}
{Ichikawa}, K., {Imanishi}, M., {Ueda}, Y., {et~al.} 2014, \apj, 794, 139

\bibitem[{{Imanishi}(2002)}]{Imanishi2002}
{Imanishi}, M. 2002, \apj, 569, 44

\bibitem[{{Imanishi}(2006)}]{Imanishi2006b}
---. 2006, \aj, 131, 2406

\bibitem[{{Imanishi} \& {Dudley}(2000)}]{Imanishi2000}
{Imanishi}, M., \& {Dudley}, C.~C. 2000, \apj, 545, 701

\bibitem[{{Imanishi} {et~al.}(2007){Imanishi}, {Dudley}, {Maiolino}, {Maloney},
  {Nakagawa}, \& {Risaliti}}]{Imanishi2007}
{Imanishi}, M., {Dudley}, C.~C., {Maiolino}, R., {et~al.} 2007, \apjs, 171, 72

\bibitem[{{Imanishi} {et~al.}(2006){Imanishi}, {Dudley}, \&
  {Maloney}}]{Imanishi2006}
{Imanishi}, M., {Dudley}, C.~C., \& {Maloney}, P.~R. 2006, \apj, 637, 114

\bibitem[{{Imanishi} {et~al.}(2008){Imanishi}, {Nakagawa}, {Ohyama},
  {Shirahata}, {Wada}, {Onaka}, \& {Oi}}]{Imanishi2008}
{Imanishi}, M., {Nakagawa}, T., {Ohyama}, Y., {et~al.} 2008, \pasj, 60, S489

\bibitem[{{Imanishi} {et~al.}(2010){Imanishi}, {Nakagawa}, {Shirahata},
  {Ohyama}, \& {Onaka}}]{Imanishi2010}
{Imanishi}, M., {Nakagawa}, T., {Shirahata}, M., {Ohyama}, Y., \& {Onaka}, T.
  2010, \apj, 721, 1233

\bibitem[{{Imanishi} {et~al.}(2018){Imanishi}, {Nakanishi}, \&
  {Izumi}}]{Imanishi2018}
{Imanishi}, M., {Nakanishi}, K., \& {Izumi}, T. 2018, \apj, 856, 143

\bibitem[{{Imanishi} {et~al.}(2019){Imanishi}, {Nakanishi}, \&
  {Izumi}}]{Imanishi2019}
---. 2019, \apjs, 241, 19

\bibitem[{{Imanishi} \& {Wada}(2004)}]{Imanishi2004}
{Imanishi}, M., \& {Wada}, K. 2004, \apj, 617, 214

\bibitem[{{Inami} {et~al.}(2018){Inami}, {Armus}, {Matsuhara}, {Charmandaris},
  {D{\'\i}az-Santos}, {Surace}, {Stierwalt}, {Ohyama}, {Howell}, {Marshall},
  {Evans}, {Linden}, \& {Mazzarella}}]{Inami2018}
{Inami}, H., {Armus}, L., {Matsuhara}, H., {et~al.} 2018, \aap, 617, A130

\bibitem[{{Jarrett} {et~al.}(2011){Jarrett}, {Cohen}, {Masci}, {Wright},
  {Stern}, {Benford}, {Blain}, {Carey}, {Cutri}, {Eisenhardt}, {Lonsdale},
  {Mainzer}, {Marsh}, {Padgett}, {Petty}, {Ressler}, {Skrutskie}, {Stanford},
  {Surace}, {Tsai}, {Wheelock}, \& {Yan}}]{Jarrett2011}
{Jarrett}, T.~H., {Cohen}, M., {Masci}, F., {et~al.} 2011, \apj, 735, 112

\bibitem[{{Joblin} {et~al.}(1996){Joblin}, {Tielens}, {Allamandola}, \&
  {Geballe}}]{Joblin1996}
{Joblin}, C., {Tielens}, A.~G.~G.~M., {Allamandola}, L.~J., \& {Geballe}, T.~R.
  1996, \apj, 458, 610

\bibitem[{{Jones}(2016)}]{Jones2016}
{Jones}, A.~P. 2016, Royal Society Open Science, 3, 160223

\bibitem[{{Kemper} {et~al.}(2004){Kemper}, {Vriend}, \& {Tielens}}]{Kemper2004}
{Kemper}, F., {Vriend}, W.~J., \& {Tielens}, A.~G.~G.~M. 2004, \apj, 609, 826

\bibitem[{{Kennicutt} {et~al.}(2003){Kennicutt}, {Armus}, {Bendo}, {Calzetti},
  {Dale}, {Draine}, {Engelbracht}, {Gordon}, {Grauer}, {Helou}, {Hollenbach},
  {Jarrett}, {Kewley}, {Leitherer}, {Li}, {Malhotra}, {Regan}, {Rieke},
  {Rieke}, {Roussel}, {Smith}, {Thornley}, \& {Walter}}]{Kennicutt2003}
{Kennicutt}, Robert~C., J., {Armus}, L., {Bendo}, G., {et~al.} 2003, \pasp,
  115, 928

\bibitem[{{Kessler} {et~al.}(1996){Kessler}, {Steinz}, {Anderegg}, {Clavel},
  {Drechsel}, {Estaria}, {Faelker}, {Riedinger}, {Robson}, {Taylor}, \&
  {Ximenez de Ferran}}]{Kessler1996}
{Kessler}, M.~F., {Steinz}, J.~A., {Anderegg}, M.~E., {et~al.} 1996, \aap, 500,
  493

\bibitem[{{Kim} {et~al.}(2012){Kim}, {Im}, {Lee}, {Lee}, {Jun}, {Nakagawa},
  {Matsuhara}, {Wada}, {Oyabu}, {Takagi}, {Inami}, {Ohyama}, {Yamada}, {Helou},
  {Armus}, \& {Shi}}]{Kim2012}
{Kim}, J.~H., {Im}, M., {Lee}, H.~M., {et~al.} 2012, \apj, 760, 120

\bibitem[{{Kroupa}(2001)}]{Kroupa2001}
{Kroupa}, P. 2001, \mnras, 322, 231

\bibitem[{{Lai} {et~al.}(2020){Lai}, {Witt}, {Alvarez}, \& {Cami}}]{Lai2020}
{Lai}, T. S.~Y., {Witt}, A.~N., {Alvarez}, C., \& {Cami}, J. 2020, \mnras, 492,
  5853

\bibitem[{{Lai} {et~al.}(2017){Lai}, {Witt}, \& {Crawford}}]{Lai2017}
{Lai}, T.~S.-Y., {Witt}, A.~N., \& {Crawford}, K. 2017, \mnras, 469, 4933

\bibitem[{{Lebouteiller} {et~al.}(2011){Lebouteiller}, {Barry}, {Spoon},
  {Bernard-Salas}, {Sloan}, {Houck}, \& {Weedman}}]{Lebouteiller2011}
{Lebouteiller}, V., {Barry}, D.~J., {Spoon}, H.~W.~W., {et~al.} 2011, \apjs,
  196, 8

\bibitem[{{Lee} {et~al.}(2012){Lee}, {Hwang}, {Lee}, {Kim}, \& {Lee}}]{Lee2012}
{Lee}, J.~C., {Hwang}, H.~S., {Lee}, M.~G., {Kim}, M., \& {Lee}, J.~H. 2012,
  \apj, 756, 95

\bibitem[{{Leger} {et~al.}(1988){Leger}, {D'Hendecourt}, \&
  {Boissel}}]{Leger1988}
{Leger}, A., {D'Hendecourt}, L., \& {Boissel}, P. 1988, Physical Review
  Letters, 60, 921

\bibitem[{{Leger} \& {Puget}(1984)}]{Leger1984}
{Leger}, A., \& {Puget}, J.~L. 1984, \aap, 500, 279

\bibitem[{{Lu} {et~al.}(2003){Lu}, {Helou}, {Werner}, {Dinerstein}, {Dale},
  {Silbermann}, {Malhotra}, {Beichman}, \& {Jarrett}}]{Lu2003}
{Lu}, N., {Helou}, G., {Werner}, M.~W., {et~al.} 2003, \apj, 588, 199

\bibitem[{{Madden} {et~al.}(2006){Madden}, {Galliano}, {Jones}, \&
  {Sauvage}}]{Madden2006}
{Madden}, S.~C., {Galliano}, F., {Jones}, A.~P., \& {Sauvage}, M. 2006, \aap,
  446, 877

\bibitem[{{Maltseva} {et~al.}(2018){Maltseva}, {Mackie}, {Candian},
  {Petrignani}, {Huang}, {Lee}, {Tielens}, {Oomens}, \& {Buma}}]{Maltseva2018}
{Maltseva}, E., {Mackie}, C.~J., {Candian}, A., {et~al.} 2018, \aap, 610, A65

\bibitem[{{Maragkoudakis} {et~al.}(2018){Maragkoudakis}, {Ivkovich}, {Peeters},
  {Stock}, {Hemachandra}, \& {Tielens}}]{Maragkoudakis2018}
{Maragkoudakis}, A., {Ivkovich}, N., {Peeters}, E., {et~al.} 2018, \mnras, 481,
  5370

\bibitem[{{Maragkoudakis} {et~al.}(2020){Maragkoudakis}, {Peeters}, \&
  {Ricca}}]{Maragkoudakis2020}
{Maragkoudakis}, A., {Peeters}, E., \& {Ricca}, A. 2020, \mnras,
  arXiv:2003.02823

\bibitem[{{Marshall} {et~al.}(2007){Marshall}, {Herter}, {Armus}, {Charmand
  aris}, {Spoon}, {Bernard-Salas}, \& {Houck}}]{Marchall2007}
{Marshall}, J.~A., {Herter}, T.~L., {Armus}, L., {et~al.} 2007, \apj, 670, 129

\bibitem[{Martin {et~al.}(2013)Martin, Bernard, Br\'edy, Concina, Joblin, Ji,
  Ortega, \& Chen}]{Martin2013}
Martin, S., Bernard, J., Br\'edy, R., {et~al.} 2013, Phys. Rev. Lett., 110,
  063003.
\newblock \url{https://link.aps.org/doi/10.1103/PhysRevLett.110.063003}

\bibitem[{Martin {et~al.}(2015)Martin, Ji, Bernard, Br\'edy, Concina, Allouche,
  Joblin, Ortega, Montagne, Cassimi, Ngono-Ravache, \& Chen}]{Martin2015}
Martin, S., Ji, M., Bernard, J., {et~al.} 2015, Phys. Rev. A, 92, 053425.
\newblock \url{https://link.aps.org/doi/10.1103/PhysRevA.92.053425}

\bibitem[{{Moorwood}(1986)}]{Moorwood1986}
{Moorwood}, A.~F.~M. 1986, \aap, 166, 4

\bibitem[{{Moshir} {et~al.}(1993){Moshir}, {Copan}, {Conrow}, {McCallon},
  {Hacking}, {Gregorich}, {Rohrbach}, {Melnyk}, {Rice}, \&
  {Fullmer}}]{Moshir1993}
{Moshir}, M., {Copan}, G., {Conrow}, T., {et~al.} 1993, VizieR Online Data
  Catalog, II/156A

\bibitem[{{Mouri} {et~al.}(1990){Mouri}, {Kawara}, {Taniguchi}, \&
  {Nishida}}]{Mouri1990}
{Mouri}, H., {Kawara}, K., {Taniguchi}, Y., \& {Nishida}, M. 1990, \apjl, 356,
  L39

\bibitem[{{Murakami} {et~al.}(2007){Murakami}, {Baba}, {Barthel}, {Clements},
  {Cohen}, {Doi}, {Enya}, {Figueredo}, {Fujishiro}, {Fujiwara}, {Fujiwara},
  {Garcia-Lario}, {Goto}, {Hasegawa}, {Hibi}, {Hirao}, {Hiromoto}, {Hong},
  {Imai}, {Ishigaki}, {Ishiguro}, {Ishihara}, {Ita}, {Jeong}, {Jeong},
  {Kaneda}, {Kataza}, {Kawada}, {Kawai}, {Kawamura}, {Kessler}, {Kester},
  {Kii}, {Kim}, {Kim}, {Kobayashi}, {Koo}, {Kwon}, {Lee}, {Lorente}, {Makiuti},
  {Matsuhara}, {Matsumoto}, {Matsuo}, {Matsuura}, {M{\"U}ller}, {Murakami},
  {Nagata}, {Nakagawa}, {Naoi}, {Narita}, {Noda}, {Oh}, {Ohnishi}, {Ohyama},
  {Okada}, {Okuda}, {Oliver}, {Onaka}, {Ootsubo}, {Oyabu}, {Pak}, {Park},
  {Pearson}, {Rowan-Robinson}, {Saito}, {Sakon}, {Salama}, {Sato}, {Savage},
  {Serjeant}, {Shibai}, {Shirahata}, {Sohn}, {Suzuki}, {Takagi}, {Takahashi},
  {Tanab{\'E}}, {Takeuchi}, {Takita}, {Thomson}, {Uemizu}, {Ueno}, {Usui},
  {Verdugo}, {Wada}, {Wang}, {Watabe}, {Watarai}, {White}, {Yamamura},
  {Yamauchi}, \& {Yasuda}}]{Murakami2007}
{Murakami}, H., {Baba}, H., {Barthel}, P., {et~al.} 2007, \pasj, 59, S369

\bibitem[{{Murata} {et~al.}(2017){Murata}, {Nakagawa}, {Matsuhara}, \&
  {Yano}}]{Murata2017}
{Murata}, K., {Nakagawa}, T., {Matsuhara}, H., \& {Yano}, K. 2017, arXiv
  e-prints, arXiv:1707.01652

\bibitem[{{Nardini} {et~al.}(2010){Nardini}, {Risaliti}, {Watabe}, {Salvati},
  \& {Sani}}]{Nardini2010}
{Nardini}, E., {Risaliti}, G., {Watabe}, Y., {Salvati}, M., \& {Sani}, E. 2010,
  \mnras, 405, 2505

\bibitem[{{Nenkova} {et~al.}(2008){Nenkova}, {Sirocky}, {Nikutta},
  {Ivezi{\'c}}, \& {Elitzur}}]{Nenkova2008}
{Nenkova}, M., {Sirocky}, M.~M., {Nikutta}, R., {Ivezi{\'c}}, {\v{Z}}., \&
  {Elitzur}, M. 2008, \apj, 685, 160

\bibitem[{{O'Dowd} {et~al.}(2009){O'Dowd}, {Schiminovich}, {Johnson}, {Treyer},
  {Martin}, {Wyder}, {Charlot}, {Heckman}, {Martins}, {Seibert}, \& {van der
  Hulst}}]{ODowd2009}
{O'Dowd}, M.~J., {Schiminovich}, D., {Johnson}, B.~D., {et~al.} 2009, \apj,
  705, 885

\bibitem[{{Ohyama} {et~al.}(2007){Ohyama}, {Onaka}, {Matsuhara}, {Wada}, {Kim},
  {Fujishiro}, {Uemizu}, {Sakon}, {Cohen}, {Ishigaki}, {Ishihara}, {Ita},
  {Kataza}, {Matsumoto}, {Murakami}, {Oyabu}, {Tanab{\'e}}, {Takagi}, {Ueno},
  {Usui}, {Watarai}, {Pearson}, {Takeyama}, {Yamamuro}, \&
  {Ikeda}}]{Ohyama2007}
{Ohyama}, Y., {Onaka}, T., {Matsuhara}, H., {et~al.} 2007, \pasj, 59, S411

\bibitem[{{Onaka} {et~al.}(2007){Onaka}, {Matsuhara}, {Wada}, {Fujishiro},
  {Fujiwara}, {Ishigaki}, {Ishihara}, {Ita}, {Kataza}, {Kim}, {Matsumoto},
  {Murakami}, {Ohyama}, {Oyabu}, {Sakon}, {Tanab{\'e}}, {Takagi}, {Uemizu},
  {Ueno}, {Usui}, {Watarai}, {Cohen}, {Enya}, {Ootsubo}, {Pearson}, {Takeyama},
  {Yamamuro}, \& {Ikeda}}]{Onaka2007}
{Onaka}, T., {Matsuhara}, H., {Wada}, T., {et~al.} 2007, \pasj, 59, S401

\bibitem[{{Peeters} {et~al.}(2002){Peeters}, {Hony}, {Van Kerckhoven},
  {Tielens}, {Allamandola}, {Hudgins}, \& {Bauschlicher}}]{Peeters2002}
{Peeters}, E., {Hony}, S., {Van Kerckhoven}, C., {et~al.} 2002, \aap, 390, 1089

\bibitem[{{Peeters} {et~al.}(2004{\natexlab{a}}){Peeters}, {Mattioda},
  {Hudgins}, \& {Allamand ola}}]{Peeters2004b}
{Peeters}, E., {Mattioda}, A.~L., {Hudgins}, D.~M., \& {Allamand ola}, L.~J.
  2004{\natexlab{a}}, \apjl, 617, L65

\bibitem[{{Peeters} {et~al.}(2004{\natexlab{b}}){Peeters}, {Spoon}, \&
  {Tielens}}]{Peeters2004a}
{Peeters}, E., {Spoon}, H.~W.~W., \& {Tielens}, A.~G.~G.~M. 2004{\natexlab{b}},
  \apj, 613, 986

\bibitem[{{Pendleton} \& {Allamandola}(2002)}]{Pendleton2002}
{Pendleton}, Y.~J., \& {Allamandola}, L.~J. 2002, \apjs, 138, 75

\bibitem[{{Petric} {et~al.}(2011){Petric}, {Armus}, {Howell}, {Chan},
  {Mazzarella}, {Evans}, {Surace}, {Sand ers}, {Appleton}, {Charmandaris},
  {D{\'\i}az-Santos}, {Frayer}, {Haan}, {Inami}, {Iwasawa}, {Kim}, {Madore},
  {Marshall}, {Spoon}, {Stierwalt}, {Sturm}, {U}, {Vavilkin}, \&
  {Veilleux}}]{Petric2011}
{Petric}, A.~O., {Armus}, L., {Howell}, J., {et~al.} 2011, \apj, 730, 28

\bibitem[{{Pilleri} {et~al.}(2015){Pilleri}, {Joblin}, {Boulanger}, \&
  {Onaka}}]{Pilleri2015}
{Pilleri}, P., {Joblin}, C., {Boulanger}, F., \& {Onaka}, T. 2015, \aap, 577,
  A16

\bibitem[{{Pope} {et~al.}(2013){Pope}, {Wagg}, {Frayer}, {Armus}, {Chary},
  {Daddi}, {Desai}, {Dickinson}, {Elbaz}, {Gabor}, \& {Kirkpatrick}}]{Pope2013}
{Pope}, A., {Wagg}, J., {Frayer}, D., {et~al.} 2013, \apj, 772, 92

\bibitem[{{Risaliti} {et~al.}(2010){Risaliti}, {Imanishi}, \&
  {Sani}}]{Risaliti2010}
{Risaliti}, G., {Imanishi}, M., \& {Sani}, E. 2010, \mnras, 401, 197

\bibitem[{{Rodr{\'\i}guez-Ardila} \& {Viegas}(2003)}]{Rodriguez_Ardila2003}
{Rodr{\'\i}guez-Ardila}, A., \& {Viegas}, S.~M. 2003, \mnras, 340, L33

\bibitem[{{Sajina} {et~al.}(2009){Sajina}, {Spoon}, {Yan}, {Imanishi}, {Fadda},
  \& {Elitzur}}]{Sajina2009}
{Sajina}, A., {Spoon}, H., {Yan}, L., {et~al.} 2009, \apj, 703, 270

\bibitem[{{Sanders} {et~al.}(2003){Sanders}, {Mazzarella}, {Kim}, {Surace}, \&
  {Soifer}}]{Sanders2003}
{Sanders}, D.~B., {Mazzarella}, J.~M., {Kim}, D.~C., {Surace}, J.~A., \&
  {Soifer}, B.~T. 2003, \aj, 126, 1607

\bibitem[{{Sanders} \& {Mirabel}(1996)}]{Sanders1996}
{Sanders}, D.~B., \& {Mirabel}, I.~F. 1996, \araa, 34, 749

\bibitem[{{Sandstrom} {et~al.}(2012){Sandstrom}, {Bolatto}, {Bot}, {Draine},
  {Ingalls}, {Israel}, {Jackson}, {Leroy}, {Li}, {Rubio}, {Simon}, {Smith},
  {Stanimirovi{\'c}}, {Tielens}, \& {van Loon}}]{Sandstrom2012}
{Sandstrom}, K.~M., {Bolatto}, A.~D., {Bot}, C., {et~al.} 2012, \apj, 744, 20

\bibitem[{{Schutte} {et~al.}(1993){Schutte}, {Tielens}, \&
  {Allamandola}}]{Schutte1993}
{Schutte}, W.~A., {Tielens}, A.~G.~G.~M., \& {Allamandola}, L.~J. 1993, \apj,
  415, 397

\bibitem[{{Shipley} {et~al.}(2016){Shipley}, {Papovich}, {Rieke}, {Brown}, \&
  {Moustakas}}]{Shipley2016}
{Shipley}, H.~V., {Papovich}, C., {Rieke}, G.~H., {Brown}, M. J.~I., \&
  {Moustakas}, J. 2016, \apj, 818, 60

\bibitem[{{Shirahata} {et~al.}(2013){Shirahata}, {Nakagawa}, {Usuda}, {Goto},
  {Suto}, \& {Geballe}}]{Shirahata2013}
{Shirahata}, M., {Nakagawa}, T., {Usuda}, T., {et~al.} 2013, \pasj, 65, 5

\bibitem[{{Smith} {et~al.}(2004){Smith}, {Dale}, {Armus}, {Draine},
  {Hollenbach}, {Roussel}, {Helou}, {Kennicutt}, {Li}, {Bendo}, {Calzetti},
  {Engelbracht}, {Gordon}, {Jarrett}, {Kewley}, {Leitherer}, {Malhotra},
  {Meyer}, {Murphy}, {Regan}, {Rieke}, {Rieke}, {Thornley}, {Walter}, \&
  {Wolfire}}]{Smith2004}
{Smith}, J.~D.~T., {Dale}, D.~A., {Armus}, L., {et~al.} 2004, \apjs, 154, 199

\bibitem[{{Smith} {et~al.}(2007){Smith}, {Draine}, {Dale}, {Moustakas},
  {Kennicutt}, {Helou}, {Armus}, {Roussel}, {Sheth}, {Bendo}, {Buckalew},
  {Calzetti}, {Engelbracht}, {Gordon}, {Hollenbach}, {Li}, {Malhotra},
  {Murphy}, \& {Walter}}]{Smith2007}
{Smith}, J.~D.~T., {Draine}, B.~T., {Dale}, D.~A., {et~al.} 2007, \apj, 656,
  770

\bibitem[{{Spoon} {et~al.}(2002){Spoon}, {Keane}, {Tielens}, {Lutz},
  {Moorwood}, \& {Laurent}}]{Spoon2002}
{Spoon}, H.~W.~W., {Keane}, J.~V., {Tielens}, A.~G.~G.~M., {et~al.} 2002, \aap,
  385, 1022

\bibitem[{{Spoon} {et~al.}(2000){Spoon}, {Koornneef}, {Moorwood}, {Lutz}, \&
  {Tielens}}]{Spoon2000}
{Spoon}, H.~W.~W., {Koornneef}, J., {Moorwood}, A.~F.~M., {Lutz}, D., \&
  {Tielens}, A.~G.~G.~M. 2000, \aap, 357, 898

\bibitem[{{Spoon} {et~al.}(2007){Spoon}, {Marshall}, {Houck}, {Elitzur}, {Hao},
  {Armus}, {Brandl}, \& {Charmandaris}}]{Spoon2007}
{Spoon}, H.~W.~W., {Marshall}, J.~A., {Houck}, J.~R., {et~al.} 2007, \apj, 654,
  L49

\bibitem[{{Spoon} {et~al.}(2003){Spoon}, {Moorwood}, {Pontoppidan}, {Cami},
  {Kregel}, {Lutz}, \& {Tielens}}]{Spoon2003}
{Spoon}, H.~W.~W., {Moorwood}, A.~F.~M., {Pontoppidan}, K.~M., {et~al.} 2003,
  \aap, 402, 499

\bibitem[{{Spoon} {et~al.}(2004){Spoon}, {Armus}, {Cami}, {Tielens}, {Chiar},
  {Peeters}, {Keane}, {Charmandaris}, {Appleton}, {Teplitz}, \&
  {Burgdorf}}]{Spoon2004}
{Spoon}, H.~W.~W., {Armus}, L., {Cami}, J., {et~al.} 2004, \apjs, 154, 184

\bibitem[{{Stern} {et~al.}(2005){Stern}, {Eisenhardt}, {Gorjian}, {Kochanek},
  {Caldwell}, {Eisenstein}, {Brodwin}, {Brown}, {Cool}, {Dey}, {Green},
  {Jannuzi}, {Murray}, {Pahre}, \& {Willner}}]{Stern2005}
{Stern}, D., {Eisenhardt}, P., {Gorjian}, V., {et~al.} 2005, \apj, 631, 163

\bibitem[{{Stierwalt} {et~al.}(2014){Stierwalt}, {Armus}, {Charmandaris},
  {Diaz-Santos}, {Marshall}, {Evans}, {Haan}, {Howell}, {Iwasawa}, {Kim},
  {Murphy}, {Rich}, {Spoon}, {Inami}, {Petric}, \& {U}}]{Stierwalt2014}
{Stierwalt}, S., {Armus}, L., {Charmandaris}, V., {et~al.} 2014, \apj, 790, 124

\bibitem[{{Storey} \& {Hummer}(1995)}]{Storey1995}
{Storey}, P.~J., \& {Hummer}, D.~G. 1995, \mnras, 272, 41

\bibitem[{{Sturm} {et~al.}(2000){Sturm}, {Lutz}, {Tran}, {Feuchtgruber},
  {Genzel}, {Kunze}, {Moorwood}, \& {Thornley}}]{Sturm2000}
{Sturm}, E., {Lutz}, D., {Tran}, D., {et~al.} 2000, \aap, 358, 481

\bibitem[{{Tielens}(2008)}]{Tielens2008}
{Tielens}, A.~G.~G.~M. 2008, \araa, 46, 289

\bibitem[{{Tokunaga} {et~al.}(1991){Tokunaga}, {Sellgren}, {Smith}, {Nagata},
  {Sakata}, \& {Nakada}}]{Tokunaga1991}
{Tokunaga}, A.~T., {Sellgren}, K., {Smith}, R.~G., {et~al.} 1991, \apj, 380,
  452

\bibitem[{{van der Walt} {et~al.}(2011){van der Walt}, {Colbert}, \&
  {Varoquaux}}]{VanderWalt2011}
{van der Walt}, S., {Colbert}, S.~C., \& {Varoquaux}, G. 2011, Computing in
  Science Engineering, 13, 22

\bibitem[{{van Diedenhoven} {et~al.}(2004){van Diedenhoven}, {Peeters}, {Van
  Kerckhoven}, {Hony}, {Hudgins}, {Allamandola}, \&
  {Tielens}}]{VanDiedenhoven2004}
{van Diedenhoven}, B., {Peeters}, E., {Van Kerckhoven}, C., {et~al.} 2004,
  \apj, 611, 928

\bibitem[{{Virtanen} {et~al.}(2020){Virtanen}, {Gommers}, {Oliphant},
  {Haberland}, {Reddy}, {Cournapeau}, {Burovski}, {Peterson}, {Weckesser},
  {Bright}, {van der Walt}, {Brett}, {Wilson}, {Jarrod Millman}, {Mayorov},
  {Nelson}, {Jones}, {Kern}, {Larson}, {Carey}, {Polat}, {Feng}, {Moore}, {Vand
  erPlas}, {Laxalde}, {Perktold}, {Cimrman}, {Henriksen}, {Quintero}, {Harris},
  {Archibald}, {Ribeiro}, {Pedregosa}, {van Mulbregt}, \&
  {Contributors}}]{Virtanen2020}
{Virtanen}, P., {Gommers}, R., {Oliphant}, T.~E., {et~al.} 2020, Nature
  Methods, 17, 261

\bibitem[{{Weingartner} \& {Draine}(2001)}]{Weingartner2001}
{Weingartner}, J.~C., \& {Draine}, B.~T. 2001, \apjs, 134, 263

\bibitem[{{Werner} {et~al.}(2004){Werner}, {Uchida}, {Sellgren}, {Marengo},
  {Gordon}, {Morris}, {Houck}, \& {Stansberry}}]{Werner2004}
{Werner}, M.~W., {Uchida}, K.~I., {Sellgren}, K., {et~al.} 2004, \apjs, 154,
  309

\bibitem[{{Wright} {et~al.}(2010){Wright}, {Eisenhardt}, {Mainzer}, {Ressler},
  {Cutri}, {Jarrett}, {Kirkpatrick}, {Padgett}, {McMillan}, {Skrutskie},
  {Stanford}, {Cohen}, {Walker}, {Mather}, {Leisawitz}, {Gautier}, {McLean},
  {Benford}, {Lonsdale}, {Blain}, {Mendez}, {Irace}, {Duval}, {Liu}, {Royer},
  {Heinrichsen}, {Howard}, {Shannon}, {Kendall}, {Walsh}, {Larsen}, {Cardon},
  {Schick}, {Schwalm}, {Abid}, {Fabinsky}, {Naes}, \& {Tsai}}]{Wright2010}
{Wright}, E.~L., {Eisenhardt}, P. R.~M., {Mainzer}, A.~K., {et~al.} 2010, \aj,
  140, 1868

\bibitem[{{Wu} {et~al.}(2006){Wu}, {Charmandaris}, {Hao}, {Brandl},
  {Bernard-Salas}, {Spoon}, \& {Houck}}]{Wu2006}
{Wu}, Y., {Charmandaris}, V., {Hao}, L., {et~al.} 2006, \apj, 639, 157

\bibitem[{{Wu} {et~al.}(2010){Wu}, {Helou}, {Armus}, {Cormier}, {Shi}, {Dale},
  {Dasyra}, {Smith}, {Papovich}, {Draine}, {Rahman}, {Stierwalt}, {Fadda},
  {Lagache}, \& {Wright}}]{Wu2010}
{Wu}, Y., {Helou}, G., {Armus}, L., {et~al.} 2010, \apj, 723, 895

\bibitem[{{Xie} \& {Ho}(2019)}]{Xie2019}
{Xie}, Y., \& {Ho}, L.~C. 2019, \apj, 884, 136

\bibitem[{{Yamada} {et~al.}(2013){Yamada}, {Oyabu}, {Kaneda}, {Yamagishi},
  {Ishihara}, {Kim}, \& {Im}}]{Y2013}
{Yamada}, R., {Oyabu}, S., {Kaneda}, H., {et~al.} 2013, \pasj, 65, 103

\bibitem[{{Yamagishi} {et~al.}(2012){Yamagishi}, {Kaneda}, {Ishihara}, {Kondo},
  {Onaka}, {Suzuki}, \& {Minh}}]{Yamagishi2012}
{Yamagishi}, M., {Kaneda}, H., {Ishihara}, D., {et~al.} 2012, \aap, 541, A10

\bibitem[{{Yang} {et~al.}(2016){Yang}, {Li}, {Glaser}, \& {Zhong}}]{Yang2016}
{Yang}, X.~J., {Li}, A., {Glaser}, R., \& {Zhong}, J.~X. 2016, \apj, 825, 22

\bibitem[{{Yano} {et~al.}(2016){Yano}, {Nakagawa}, {Isobe}, \&
  {Shirahata}}]{Yano2016}
{Yano}, K., {Nakagawa}, T., {Isobe}, N., \& {Shirahata}, M. 2016, \apj, 833,
  272

\bibitem[{{Zhuang} {et~al.}(2019){Zhuang}, {Ho}, \& {Shangguan}}]{Zhuang2019}
{Zhuang}, M.-Y., {Ho}, L.~C., \& {Shangguan}, J. 2019, \apj, 873, 103

\end{thebibliography}


\end{CJK*}
\end{document}